\documentclass[a4paper,11pt]{article}
\pdfoutput=1

\usepackage[a4paper,vmargin={30mm,30mm},hmargin={30mm,30mm}]{geometry}

\usepackage{lmodern}
\usepackage[T1]{fontenc}
\usepackage[utf8]{inputenc}

\usepackage{amsfonts}
\usepackage{amstext}
\usepackage{amsmath}
\usepackage{mathtools}

\usepackage[margin=\parindent,labelfont=bf]{caption}

\usepackage{pbox}

\usepackage[usenames,dvipsnames]{xcolor}

\usepackage{tikz}
\usetikzlibrary{decorations.markings}
\usetikzlibrary{arrows}

\numberwithin{equation}{section}
\numberwithin{figure}{section}

\usepackage[
pdfauthor={
  Rouven Frassek, 
  Nils Kanning, 
  Yumi Ko, 
  Matthias Staudacher
}, 
pdftitle={
  Bethe Ansatz for Yangian Invariants:
  Towards Super Yang-Mills Scattering Amplitudes
},
colorlinks=true,
linkcolor=RedViolet,
citecolor=Red,
urlcolor=Magenta,
unicode=true,
hypertexnames=false
]{hyperref}

\usepackage{cite}
\usepackage[parsep]{collref}

\DeclareMathOperator{\D}{d}
\DeclareMathOperator{\tr}{tr}
\DeclareMathOperator{\Min}{min}

\newcommand{\mon}{M}
\newcommand{\bvac}{|\Omega\rangle}

\newcommand{\osca}{\mathbf{a}}
\newcommand{\oscb}{\mathbf{b}}
\newcommand{\s}{\mathbf{s}}
\newcommand{\bs}{\bar{\mathbf{s}}}

\newcommand{\rap}{\theta}
\newcommand{\spec}{u}
\newcommand{\specp}{u'}
\newcommand{\brt}{u}
\newcommand{\magn}{x}
\newcommand{\inhdiff}{z}
\newcommand{\inh}{v}
\newcommand{\inht}{w}

\newcommand{\epb}{j}
\newcommand{\epe}{i}

\newcommand{\graph}{\mathbf{G}}
\newcommand{\inhtset}{\mathbf{w}}
\newcommand{\brtset}{\mathbf{u}}
\newcommand{\magnset}{\mathbf{x}}
\newcommand{\repset}{\boldsymbol{\Lambda}}
\newcommand{\rapset}{\boldsymbol{\theta}}
\newcommand{\stateset}{\boldsymbol{\alpha}}

\newcommand{\sites}{L}
\newcommand{\lines}{N}
\newcommand{\brts}{P}
\newcommand{\dsites}{K}

\begin{document} 

\baselineskip 5mm

\begin{titlepage}

  \hfill
  \pbox{5cm}{
    \texttt{HU-Mathematik-2013-14}\\
    \texttt{HU-EP-13/34}\\
    \texttt{AEI-2013-234}\\
    \texttt{DCPT-13/47}
  }

  \vspace{2\baselineskip}

  \begin{center}

    \textbf{\LARGE Bethe Ansatz for Yangian Invariants:\\[1ex]
      Towards Super Yang-Mills Scattering Amplitudes}

    \vspace{2\baselineskip}

    Rouven Frassek\textsuperscript{\textit{a,b,c}},
    Nils Kanning\textsuperscript{\textit{b,c}},
    Yumi Ko\textsuperscript{\textit{b,d}} and
    Matthias Staudacher\textsuperscript{\textit{b,c}}
 
    \vspace{2\baselineskip}

    \textit{
      \textsuperscript{a}
      Department of Mathematical Sciences, Durham University,\\
      South Road, Durham DH1 3LE, United Kingdom\\
      \vspace{0.5\baselineskip}
      \textsuperscript{b} 
      Institut für Mathematik und Institut für Physik, Humboldt-Universität zu Berlin,\\
      IRIS-Adlershof, Zum Großen Windkanal 6, 12489 Berlin, Germany\\
      \vspace{0.5\baselineskip}
      \textsuperscript{c}
      Max-Planck-Institut für Gravitationsphysik, Albert-Einstein-Institut,\\
      Am Mühlenberg 1, 14476 Potsdam, Germany\\
      \vspace{0.5\baselineskip}
      \textsuperscript{d}
      Asia Pacific Center for Theoretical Physics,\\ 
      Pohang, Gyeongbuk 790-784, Republic of Korea 
    }

    \vspace{2\baselineskip}

    \texttt{
      \href{mailto:rouven.frassek@durham.ac.uk}{rouven.frassek@durham.ac.uk},
      \href{mailto:kanning@mathematik.hu-berlin.de}{kanning@mathematik.hu-berlin.de},
      \href{mailto:koyumi@mathematik.hu-berlin.de}{koyumi@mathematik.hu-berlin.de},
      \href{mailto:matthias@aei.mpg.de}{matthias@aei.mpg.de}
    }

    \vspace{2\baselineskip}
    
    \textbf{Abstract}
    
  \end{center}

  \noindent
  We propose that Baxter's Z-invariant six-vertex model at
  the rational $\mathfrak{gl}(2)$ point on a planar but in general not
  rectangular lattice provides a way to study Yangian invariants.
  These are identified with eigenfunctions of certain monodromies of
  an auxiliary inhomogeneous spin chain.
  As a consequence they are special solutions to the eigenvalue problem
  of the associated transfer matrix.
  Excitingly, this allows to construct them using Bethe ansatz
  techniques.
  Conceptually, our construction generalizes to general (super) Lie
  algebras and general representations.
  Here we present the explicit form of sample invariants for totally
  symmetric, finite-dimensional representations of $\mathfrak{gl}(n)$
  in terms of oscillator algebras.
  In particular, we discuss invariants of three- and four-site
  monodromies that can be understood respectively as intertwiners of the
  bootstrap and Yang-Baxter equation.
  We state a set of functional relations significant for these
  representations of the Yangian and discuss their solutions in terms
  of Bethe roots. They arrange themselves into exact strings in the
  complex plane.
  In addition, it is shown that the sample invariants can be expressed
  analogously to Graßmannian integrals.
  This aspect is closely related to a recent on-shell formulation of
  scattering amplitudes in planar $\mathcal{N}=4$ super Yang-Mills
  theory.

\end{titlepage}

\setcounter{page}{2}

\setcounter{tocdepth}{2}
\tableofcontents{}

\newpage

\section{Introduction  and overview}
\label{sec:intro}

Some time ago, a remarkable observation has been made in the field of
scattering amplitudes of planar $\mathcal{N}=4$ super Yang-Mills
theory, namely their Yangian structure \cite{Drummond:2009fd}. It was
obtained by combining superconformal symmetry and a hidden dual
superconformal symmetry \cite{Drummond:2008vq}. It holds for planar
tree-level scattering amplitudes, and there are indications that it
also plays a role at loop-level. Originally the Yangian algebra,
commonly abbreviated to ``Yangian'', was defined by Drinfeld as the
algebraic consequence of the Yang-Baxter equation underlying
one-dimensional quantum integrable models in the so-called rational
case. It is an infinite generalization of finite-dimensional Lie
algebras.  Thus the Yangian structure appearing in the
four-dimensional scattering amplitudes naturally suggests the
existence of a hidden quantum integrability. Such a structure has
already been unearthed in the last 11 years for the spectral problem
of anomalous dimensions in $\mathcal{N}=4$ theory, where it has led to
spectacular progress. See \cite{Beisert:2010jr} for a recent, fairly
up-to-date multi-author review series, and
specifically~\cite{Beisert:2010jq} for a juxtaposition of the Yangian
symmetry in the scattering and spectral problems. However, while
integrability has been essential for the (conjectured) solution of the
spectral problem, in the scattering problem it has not yet directly
led to any practical advantages in computations, with the notable
exception of the recent, very promising conjectural approach of
\cite{Basso:2013vsa,Basso:2013aha,Papathanasiou:2013uoa}, see also
\cite{CaronHuot:2011kk,Bullimore:2011}. The reason is that the
associated large integrability toolbox, the quantum inverse scattering
method (QISM), is so far available only for the calculation of
anomalous dimensions. Its application usually leads to powerful Bethe
ansatz methods. In contradistinction, apparently no such methods exist
to-date for directly exploiting Yangian invariance.

Our question starts here. What is the nature of Yangian symmetry, as
it appears in the scattering amplitudes, from the view point of
integrability and the QISM? In order to answer this question, we focus
on \emph{Yangian invariants} $|\Psi\rangle$, which are defined in the
following way,
\begin{equation}
  \label{mon_ev1}
  \mon_{ab}(\spec)|\Psi\rangle = \delta_{ab}|\Psi\rangle\,, 
\end{equation}
as a key to connect the scattering amplitudes and the Bethe
ansatz. Here $\mon(\spec)$ is a monodromy matrix, given by a product
of suitable R-matrices, $\spec$ is a spectral parameter, and $a,b$ are
indices in an auxiliary space, taking values in the fundamental
representation of the underlying symmetry algebra. The generators of
the Yangian algebra are obtained as the coefficients $\mon_{ab}^{(r)}$
of an expansion of the monodromy matrix $\mon(\spec)$ in powers $r$ of
the inverse spectral parameter $\spec^{-1}$. From \eqref{mon_ev1} with
$M_{ab}^{(0)}=\delta_{ab}$ one then sees that $|\Psi\rangle$ is
annihilated by all Yangian generators. By definition, $|\Psi\rangle$
is thus a Yangian invariant. Furthermore, finding all solutions of
\eqref{mon_ev1} for all suitable $\mon(\spec)$ should then lead
to the complete set of such invariants.

The first main observation we would like to present in this paper is
the following. In the simplest case of $\mathfrak{gl}(2)$ equation
\eqref{mon_ev1} can be derived from the rational limit of a
two-dimensional integrable model, the so-called Z-invariant vertex
model introduced by Baxter. It has a description as an inhomogeneous
spin chain \cite{Baxter:1987}. Introducing a certain oscillator
formalism, and thereby considering more general representations, one
can then obtain $\mathfrak{gl}(2)$ Yangian generators. Excitingly,
they take forms analogous to the ones acting on the scattering
amplitudes in $\mathcal{N}=4$ super Yang-Mills theory
\cite{Drummond:2009fd}. Furthermore, the procedure readily generalizes
to higher rank cases. Supersymmetry should also pose no obstacles.

A further interesting aspect of \eqref{mon_ev1} is that it allows one
to consider the Bethe ansatz for the spin chain, which is the second
main point of this paper. Equation \eqref{mon_ev1} represents a system
of eigenvalue problems for the matrix elements $\mon_{ab}(\spec)$ of
the monodromy matrix $\mon(\spec)$, with rather trivial eigenvalues
$0$ or $1$ for a common eigenvector $|\Psi\rangle$. In addition, by
taking a trace on both sides, \eqref{mon_ev1} becomes an eigenvalue
problem for the transfer matrix $T(\spec)$ of the spin chain,
\begin{equation}
  \label{trans_ev1}
  T(\spec)=\tr\mon(\spec)\,,\quad 
  T(\spec)|\Psi\rangle= n|\Psi\rangle\,,
\end{equation}
where we already generalized from $\mathfrak{gl}(2)$ to
$\mathfrak{gl}(n)$, hence $\mon(\spec)$ is an $n\times n$ matrix in
the auxiliary space. We conclude that any such Yangian invariant
$|\Psi\rangle$ must then be a special eigenvector of the transfer
matrix $T(\spec)$ with prescribed eigenvalue $n$. It is important to
stress that the Yangian invariant $|\Psi\rangle$ in \eqref{trans_ev1}
and thus also in \eqref{mon_ev1} does \emph{not} depend on the
spectral parameter $\spec$. This is a key feature of the QISM: The
diagonalization of $T(\spec)$ involves an $\spec$-independent change
of basis.

In this study we will content ourselves with compact representations
of $\mathfrak{gl}(n)$. This should play the role of a toy model of the
$\mathcal{N}=4$ scattering amplitudes, where suitable non-compact
representations of $\mathfrak{gl}(4|4)$ are needed instead. The latter
are built from continuous generalizations of the oscillators mentioned
above, which are essentially the spinor-helicity variables and their
derivatives. The basic philosophy based on \eqref{mon_ev1} should
nevertheless remain applicable, at least in the case of the tree-level
amplitudes, where Yangian invariance is unequivocal. Each
$\sites$-particle tree-level amplitude should then be identical to an
invariant $|\Psi\rangle$ solving \eqref{mon_ev1} with a monodromy of
``length'' $\sites$, and thus amenable to analysis by the QISM. The
monodromy is built from $L$ suitable R-matrices, just as in the case
of integrable spin chains. Thus amplitudes should turn into
``special'' spin chain states, similar, as we shall see, to
$\mathfrak{gl}(n)$ symmetric antiferromagnetic ground states of the
chain. The spin chain monodromy is again inhomogeneous, and the
external scattering data is encoded in the representing
``oscillators'' = spinor-helicity variables. Alternatively, we can
think of the tree-level amplitudes as appropriately generalized Baxter
lattices, i.e.\ special vertex models.

Just like in the toy model, it is imperative that the Yangian
invariants and therefore the tree-level amplitudes do not depend on
the spectral parameter $\spec$. The latter merely serves as a suitable
device for applying the QISM and for employing (an adequate
generalization of) the Bethe ansatz to the problem. On the other hand,
spectral parameters were recently introduced as certain natural
``helicity'' deformations of $\mathcal{N}=4$ scattering amplitudes in
\cite{Ferro:2012xw,Ferro:2013dga}. This is not a contradiction. In the
present framework, these parameters simply correspond to a freedom in
the choice of the inhomogeneities of the monodromy in
\eqref{mon_ev1}. In the $\mathfrak{gl}(n)$ toy model, the
representation labels in general do not fix the inhomogeneities
completely. The same holds true in the $\mathcal{N}=4$ case. In fact,
R-matrices of rational models in arbitrary representations are also
Yangian invariants. They may therefore be found from special solutions
of \eqref{mon_ev1}. For instance, a standard four-legged
$\mathfrak{gl}(n)$ \emph{R-matrix} acting on the tensor product of two
arbitrary compact representations may be deduced from the
\emph{eigenvector} $|\Psi\rangle$ of a length-four monodromy
$\mon(\spec)$. Here a difference of \emph{inhomogeneities}, denoted by
$\inhdiff$, in the monodromy $\mon(\spec)$ is to be interpreted as a
\emph{spectral parameter} of the R-matrix $R(\inhdiff)$. But
$\inhdiff$ is not the spectral parameter $\spec$ used to solve the
spectral problem \eqref{mon_ev1} for Yangian invariants.

This paper is organized as follows. In section~\ref{sec:pba} we
review Baxter's Z-invariant six-vertex model in the $\mathfrak{gl}(2)$
limit, as well as its remarkable solution through the little-known
perimeter Bethe ansatz \cite{Baxter:1987}. Its key feature is, rather
unusually, that the Bethe equations may be explicitly solved with
comparative ease. In section~\ref{sec:yi} we show that Baxter's
approach may be generalized to an important class of compact
representations of $\mathfrak{gl}(n)$, and reinterpreted as a
systematic way to define and derive Yangian invariants. This opens the
way to derive a perimeter Bethe ansatz for the latter.  In
section~\ref{sec:osc} we illustrate the method for the case of compact
oscillator representations of $\mathfrak{gl}(n)$ by presenting
explicit Yangian invariants for three specific examples. Pictorially
they correspond to a line, a three-vertex and a four-vertex. The
invariants are expressed in oscillator notation. They look somewhat
different from the Yangian-invariant tree-level scattering amplitudes,
which is surely due to the different nature of the representations
under investigation. However, in section~\ref{sec:amp} we demonstrate
that our examples may be rewritten as Graßmannian contour
integrals. Interestingly, this manifestly turns them into close
analogues of the scattering amplitudes, see
\cite{ArkaniHamed:2009dn}. An added benefit of our approach is that
the (multi)-contours are precisely defined by the construction.  In
section~\ref{sec:bethe} we then discuss the perimeter Bethe ansatz for
the Yangian invariants of our toy model. We illustrate it for
$\mathfrak{gl}(2)$ for the sample invariants of section~\ref{sec:osc}
and section~\ref{sec:amp}. Remarkably, the Bethe roots assemble into
exact strings in the complex spectral parameter plane, and are thus
explicitly determined. We also sketch the generalization to
$\mathfrak{gl}(n)$, where a nested perimeter Bethe ansatz is
required. Finally section~\ref{sec:concl} provides conclusions and an
outlook on the application of our novel approach to the computation of
actual scattering amplitudes of $\mathcal{N}=4$ Yang-Mills
theory. Some facts useful in section~\ref{sec:amp} on the connections
between oscillators and the well-known Bargmann representations as
well as the less-known conjugate Bargmann representations are deferred
to appendix~\ref{sec:bargmann}.

\section{Perimeter Bethe ansatz}
\label{sec:pba}

We begin our discussion with the six-vertex model, an important
example of an exactly solvable lattice model in two-dimensional
statistical mechanics, see e.g.\ \cite{Baxter:2007}. This model is
usually studied on a regular square lattice with periodic boundary
conditions. Its exact solution for the partition function of finite
size lattices is well known
\cite{Lieb:1967zz,Lieb:1967bg,Sutherland:1967}, albeit in an implicit
form requiring the solution of Bethe ansatz equations.  The six-vertex
model has also been studied on more general planar lattices
\cite{Baxter:1978xr}, the so-called Baxter lattices, which are
typically non-rectangular. It is probably less known that the
partition function on such lattices for fixed boundary conditions was
also obtained using a \emph{perimeter Bethe ansatz} by Baxter
\cite{Baxter:1987}. In this construction a Bethe wave function is
identified with the partition function. Remarkably, in this case the
solutions of the Bethe equations are given explicitly, in difference
to most other applications of the Bethe ansatz.

Here we review the perimeter Bethe ansatz of the six-vertex model in
the rational limit, which has $\mathfrak{su}(2)$ symmetry in the
spin~$\frac{1}{2}$ representation. However, our notation differs
considerably from the original work \cite{Baxter:1987}. In
section~\ref{sec:pba-model} we introduce the Baxter lattice along with
the model defined on it. Its solution in terms of the perimeter Bethe
ansatz is discussed in section~\ref{sec:pba-solution}.  For brevity we
refrain from repeating Baxter's proof of this result here. Instead, we
will understand it later in section~\ref{sec:bethe-pba} as a special
case of a more general connection between partition functions of
vertex models and Bethe vectors.

\subsection{Rational six-vertex model on Baxter lattices}
\label{sec:pba-model}

\begin{figure}[!t]
  \begin{center}  
    \begin{align*}
      \begin{aligned}
        \begin{tikzpicture}
          \draw[thick,densely dotted]
          (0,0) +(-180:2.2cm) arc (-180:180:2.2cm) 
          node[draw,solid,fill=black,inner sep=0.75pt,shape=circle] {}
          node[left] {$B$};
          \draw[thick,
          decoration={
            markings, mark=at position 0.9 with {\arrow{latex reversed}}},
          postaction={decorate}
          ] 
          (-160:2.2cm) 
          node[left] {$\epe_1\!=\!1$}
           -- (50:2.2cm)
          node[above right] {$\epb_1\!=\!9$};
          \draw[thick,
          decoration={
            markings, mark=at position 0.9 with {\arrow{latex reversed}}},
          postaction={decorate}
          ] 
          (-135:2.2cm) 
          node[below=0.1cm,left] {$\epe_2\!=\!2$}
          -- (75:2.2cm)
          node[right,above=0.05cm] {$\epb_2\!=\!10$};
          \draw[thick,
          decoration={
            markings, mark=at position 0.95 with {\arrow{latex reversed}}},
          postaction={decorate}
          ] 
          (-110:2.2cm) 
          node[below left] {$\epe_3\!=\!3$}
          -- (28:2.2cm)
          node[right=0.2cm] {$\epb_3\!=\!8$};
          \draw[thick,
          decoration={
            markings, mark=at position 0.85 with {\arrow{latex reversed}}},
          postaction={decorate}
          ] 
          (-85:2.2cm) 
          node[left=0.2cm,below] {$\epe_4\!=\!4$}
          -- (150:2.2cm)
          node[left] {$\epb_4\!=\!12$};
          \draw[thick,
          decoration={
            markings, mark=at position 0.5 with {\arrow{latex reversed}}},
          postaction={decorate}
          ] 
          (-65:2.2cm) 
          node[right=0.4cm,below=0.1cm] {$\epe_5\!=\!5$}
          -- (-20:2.2cm)
          node[right] {$\epb_5\!=\!6$};
          \draw[thick,
          decoration={
            markings, mark=at position 0.85 with {\arrow{latex reversed}}},
          postaction={decorate}
          ] 
          (3:2.2cm) 
          node[right] {$\epe_6\!=\!7$}
          -- (120:2.2cm)
          node[left=0.3cm,above=0.1cm] {$\epb_6\!=\!11$};
        \end{tikzpicture}
      \end{aligned}
       \end{align*}
       \caption{Sample Baxter lattice with $\lines=6$ lines whose
         configuration is given by the endpoints
         $\graph=((1,9),(2,10),(3,8),(4,12),(5,6),(7,11))$. Each
         line $k$ has endpoints $(\epe_k,\epb_k)$, an orientation indicated
         by an arrow and carries a rapidity $\rap_k$, which is not
         shown in this figure.}
    \label{fig:pba-lattice}
  \end{center}  
\end{figure}
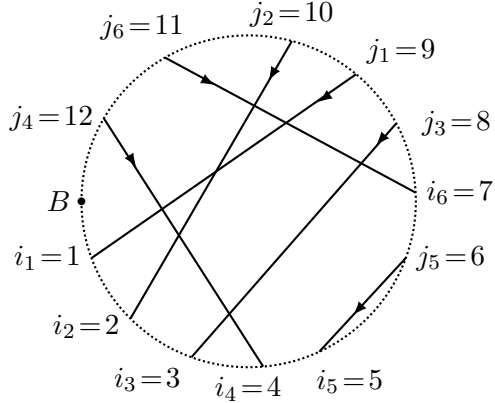
The lattice is defined by $\lines$ straight lines in the interior of a
circle which start and end at points on the perimeter. The lines can
be arranged in an arbitrary way. The intersection points of the lines
divide each line into a number of edges. However, only two lines are
allowed to intersect at a point. An example is shown in figure
\ref{fig:pba-lattice}, where the perimeter is represented by a dotted
circle. The $\lines$ lines and their $2\lines$ endpoints are labeled
counterclockwise starting at a reference point $B$ on the
perimeter. Each line has an orientation, which for the $k$-th line
with endpoints $(\epe_k, \epb_k)$ obeying $1\leq \epe_k <
\epb_k\leq2\lines$ is given by an arrow pointing from $\epb_k$ towards
$\epe_k$. In addition, we assign a rapidity $\rap_k$ to the $k$-th
line. We refer to this configuration of lines as a \emph{Baxter
  lattice}. It is specified by the ordered sets
\begin{align}
  \label{eq:pba-g-theta}
  \graph=((\epe_1,\epb_1),\ldots,(\epe_\lines, \epb_\lines))\,,
  \quad
  \rapset=(\rap_1,\ldots,\rap_\lines)\,.
\end{align}

To each vertex, i.e.\ intersection of two lines, we associate the
Boltzmann weights of the six-vertex model in the rational limit, which
may be conveniently expressed as elements of an R-matrix. They depend
on the rapidities of the lines and state labels $1$ or $2$ which are
assigned to the adjacent edges of the vertex:
\begin{align}
  \label{eq:pba-boltzmann}
  \begin{aligned}
    \langle \alpha,\gamma|
    R(\rap-\rap')
    |\beta,\delta\rangle
    =\,\,\,\\\phantom{}
  \end{aligned}
  \begin{aligned}
    \begin{tikzpicture}
      \draw[thick,
      decoration={
        markings, mark=at position 0.85 with {\arrow{latex reversed}}},
      postaction={decorate}
      ] 
      (0,0) 
      node[left=0.4cm] {$\rap$}
      node[left] {$\alpha$} -- 
      (1,0)
      node[right] {$\beta$};
      \draw[thick,
      decoration={
        markings, mark=at position 0.85 with {\arrow{latex reversed}}},
      postaction={decorate}
      ] 
      (0.5,-0.5) node[below=0.5cm] {$\rap'$}
      node[below] {$\gamma$} -- 
      (0.5,0.5)
      node[above] {$\delta$};
    \end{tikzpicture}
  \end{aligned}
  \begin{aligned}
    .\\\phantom{}
  \end{aligned}
\end{align}
These weights are defined as elements of an $\mathfrak{su}(2)$
spin~$\frac{1}{2}$ R-matrix
\begin{align}
  \label{eq:pba-r-matrix}
  R(\rap-\rap')=\frac{1}{\rap-\rap'+1}
    \begin{pmatrix}
      \rap-\rap'+1 & 0 & 0 & 0\\
      0&\rap-\rap'&1&0\\
      0&1&\rap-\rap'&0\\
      0&0&0&\rap-\rap'+1\\
    \end{pmatrix}.
\end{align}
The Greek indices $\alpha, \beta,\gamma,\delta$ assigned to the edges
take the values $1$ or $2$. These correspond to the states $|1\rangle
= \bigl(\begin{smallmatrix}1\\0\end{smallmatrix}\bigr)$ or $|2\rangle
= \bigl(\begin{smallmatrix}0\\1\end{smallmatrix}\bigr)$, or the
respective bras. The R-matrix acts on the tensor product of two states
$|\beta, \delta \rangle := |\beta\rangle \otimes |\delta\rangle$ and
the matrix element is built with $\langle \alpha, \gamma| := \langle
\alpha| \otimes \langle \gamma|$. The six non-zero elements of the
matrix denote the six configurations of a vertex for which the number
of incoming states $|1\rangle$, $|2\rangle$ and outgoing states
$\langle 1|$, $\langle 2|$ are equal, respectively. This
``conservation law'' is the so-called \emph{ice rule}.

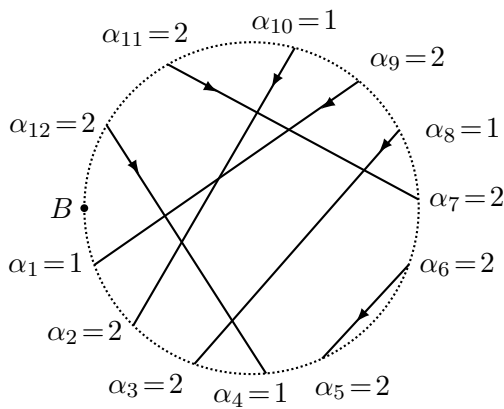
\begin{figure}[!t]
  \begin{center}  
    \begin{align*}
      \begin{aligned}
        \begin{tikzpicture}
          \draw[thick,densely dotted]
          (0,0) +(-180:2.2cm) arc (-180:180:2.2cm) 
          node[draw,solid,fill=black,inner sep=0.75pt,shape=circle] {}
          node[left] {$B$};
          \draw[thick,
          decoration={
            markings, mark=at position 0.9 with {\arrow{latex reversed}}},
          postaction={decorate}
          ] 
          (-160:2.2cm) 
          node[left] {$\alpha_1\!=\!1$}
           -- (50:2.2cm)
          node[above right] {$\alpha_9\!=\!2$};
          \draw[thick,
          decoration={
            markings, mark=at position 0.9 with {\arrow{latex reversed}}},
          postaction={decorate}
          ] 
          (-135:2.2cm) 
          node[below=0.1cm,left] {$\alpha_2\!=\!2$}
          -- (75:2.2cm)
          node[right,above=0.05cm] {$\alpha_{10}\!=\!1$};
          \draw[thick,
          decoration={
            markings, mark=at position 0.95 with {\arrow{latex reversed}}},
          postaction={decorate}
          ] 
          (-110:2.2cm) 
          node[below left] {$\alpha_3\!=\!2$}
          -- (28:2.2cm)
          node[right=0.2cm] {$\alpha_8\!=\!1$};
          \draw[thick,
          decoration={
            markings, mark=at position 0.85 with {\arrow{latex reversed}}},
          postaction={decorate}
          ] 
          (-85:2.2cm) 
          node[left=0.2cm,below] {$\alpha_4\!=\!1$}
          -- (150:2.2cm)
          node[left] {$\alpha_{12}\!=\!2$};
          \draw[thick,
          decoration={
            markings, mark=at position 0.5 with {\arrow{latex reversed}}},
          postaction={decorate}
          ] 
          (-65:2.2cm) 
          node[right=0.4cm,below=0.1cm] {$\alpha_5\!=\!2$}
          -- (-20:2.2cm)
          node[right] {$\alpha_6\!=\!2$};
          \draw[thick,
          decoration={
            markings, mark=at position 0.85 with {\arrow{latex reversed}}},
          postaction={decorate}
          ] 
          (3:2.2cm) 
          node[right] {$\alpha_7\!=\!2$}
          -- (120:2.2cm)
          node[left=0.3cm,above=0.1cm] {$\alpha_{11}\!=\!2$};
        \end{tikzpicture}
      \end{aligned}
       \end{align*}
       \caption{The sample Baxter lattice of
         figure~\ref{fig:pba-lattice} with a certain assignment of
         states, labeled by $\stateset$, to the endpoints. Note that
         the number of endpoints $\epe_k$ with a state labeled
         $\alpha_{\epe_k}=1$ and that of endpoints $\epb_k$ with
         $\alpha_{\epb_k}=1$ agrees, i.e.\ the ice rule
         \eqref{eq:pba-ice-rule-global} is satisfied. In this case
         $\stateset$ gives rise to $\magnset=(1,4,6,9,11,12)$, see
         \eqref{eq:pba-alpha-position}, which is used to express the
         partition function $\mathcal{Z}(\graph,\rapset,\stateset)$ in
         terms of a Bethe wave function
         $\Phi(\inhtset,\brtset,\magnset)$ in
         \eqref{eq:pba-partition-wave}.}
    \label{fig:pba-lattice-states}
  \end{center}  
\end{figure}
The boundary conditions of this vertex model are given by labels
$\alpha_{\epe_k}$ and $\alpha_{\epb_k}$ that can take the values $1$
or $2$ and are assigned to the endpoints $(\epe_k, \epb_k)$ of the
lines, see figure~\ref{fig:pba-lattice-states}. We denote them by
\begin{align}
  \label{eq:pba-alpha}
  \stateset=(\alpha_1, \ldots, \alpha_{2 \lines})\,.
\end{align}
These labels correspond to the states at the boundary edges of the
lattice.

The partition function of a Baxter lattice is defined as
\begin{align}
  \label{eq:pba-partition}
  \mathcal{Z}(\graph,\rapset,\stateset)
  =
  \sum_{\substack{\text{internal}\\\text{state}\\\text{config.}}}\,
  \prod_{\text{vertices}} \text{Boltzmann weight}\,.
\end{align} 
The sum runs over all possible state configurations of the internal
edges. We have to add an additional prescription for lines not
containing any vertex, see e.g.\ the line with endpoints $(5,6)$ in
figure~\ref{fig:pba-lattice-states}.  If such a line $k$ has equal
state labels $\alpha_{\epe_k}=\alpha_{\epb_k}$ at the endpoints, it
contributes a factor of unity to the partition function. In case of
differing labels $\alpha_{\epe_k} \neq \alpha_{\epb_k}$, the partition
function is set to zero.

As a consequence of the ice rule at each vertex, the partition
function can only be non-zero if the number of endpoints $\epe_k$ at
outward pointing boundary edges with $\alpha_{\epe_k}=1$ is equal to
that of endpoints $\epb_k$ at inward pointing edges with
$\alpha_{\epb_k}=1$,
\begin{align}
  \label{eq:pba-ice-rule-global}
  \big|\{\epe_k\,|\,\alpha_{\epe_k}\!=\!1\}\big|
  =
  \big|\{\epb_k\,|\,\alpha_{\epb_k}\!=\!1\}\big|\,.
\end{align}
The same condition then also holds for endpoints with state labels
$2$.

The R-matrix \eqref{eq:pba-r-matrix} and thus the Boltzmann weights at
the vertices satisfy a Yang-Baxter equation, see section~\ref{sec:yi}
below. This means that the partition function does not change if a
line of the lattice is moved through a vertex without changing the
order of the endpoints in $\graph$. This is usually referred to as
Z-invariance.

\subsection{Solution by Baxter}
\label{sec:pba-solution}

An exact expression for the partition function
\eqref{eq:pba-partition} was obtained in terms of a Bethe wave
function in \cite{Baxter:1987}. The wave function of the Heisenberg
spin chain with $\mathfrak{su}(2)$ spin~$\frac{1}{2}$ symmetry can be
derived from a coordinate Bethe ansatz \cite{Bethe:1931hc}, as nicely
explained e.g.\ in \cite{Karbach:1997,Sutherland:2004}. This was
generalized to a spin chain with inhomogeneities in
\cite{Yang:1967bm,Gaudin:1967}, which is the case needed here. For a
spin chain of length $\sites$ with $\brts$ excitations (``magnons'')
the wave function is parametrized by
\begin{align}
  \label{eq:pba-psi-param}
  \inhtset=(\inht_1,\ldots,\inht_{\sites})\,,
  \quad
  \brtset=(\brt_1,\ldots,\brt_\brts)\,,
  \quad
  \magnset=(\magn_1,\ldots,\magn_{\brts})\,,
\end{align}
denoting respectively the inhomogeneities, Bethe roots and positions
of the magnons with $1\leq \magn_1<\ldots<\magn_\brts\leq \sites$. The
wave function takes the form
\begin{align}
  \label{eq:pba-psi}
  \Phi(\inhtset,\brtset,\magnset)
  =
  \sum_\rho
  \mathrm{A}(\brt_{\rho(1)},\ldots,\brt_{\rho(P)})
  \prod_{k=1}^{\brts}\phi_{\magn_k}(\brt_{\rho(k)},\inhtset)\,,
\end{align} 
where the sum is over all permutations $\rho$ of $\brts$ elements. The
inhomogeneity-independent part is given by
\begin{align}
  \label{eq:pba-amp}
  \mathrm{A}(\brt_{\rho(1)},\ldots,\brt_{\rho(P)})=
  \prod_{1\leq k<l\leq \brts}
  \!\!\!
  \frac{\brt_{\rho(k)}-\brt_{\rho(l)}+1}{\brt_{\rho(k)}-\brt_{\rho(l)}}\,,
\end{align}
and the single particle wave function reads
\begin{align}
  \label{eq:pba-phi}
  \phi_{\magn}(\brt,\inhtset)
  =
  \prod_{j=1}^{\magn-1}(\brt-\inht_j+1)
  \prod_{j=\magn+1}^{\sites}(\brt-\inht_j)\,,
\end{align}
see also \cite{Essler:2010}.\footnote{In the homogeneous case, i.e\
  $\inht_j=0$, the wave function \eqref{eq:pba-psi} can be recast into
  a more familiar form containing the S-matrix by dividing
  \eqref{eq:pba-psi} by $\mathrm{A}(\brt_1,\ldots,\brt_P)$ and
  changing variables to $p_k=-i\log\tfrac{\brt_k+1}{\brt_k}$.}
Imposing periodicity of \eqref{eq:pba-psi} in the magnon positions,
one obtains the Bethe equations
\begin{align}
  \label{eq:pba-bethe}
  \prod_{i=1}^{\sites}\frac{\brt_k-\inht_i+1}{\brt_k-\inht_i}
  =
  -\prod_{l=1}^{\brts}\frac{\brt_k-\brt_l+1}{\brt_k-\brt_l-1}
\end{align}
with $1\leq k \leq \brts$. They guarantee that the wave functions
\eqref{eq:pba-psi} for different magnon configurations $\magnset$
build up a transfer matrix eigenvector of the closed inhomogeneous
Heisenberg spin chain. See also section~\ref{sec:bethe-gl2-review}
below for a recap of the Bethe ansatz in the algebraic
formulation. Often \eqref{eq:pba-psi} for generic Bethe roots
$\brtset$ is referred to as ``off-shell'' Bethe wave-function, while
it is ``on-shell'' in case the Bethe roots satisfy
\eqref{eq:pba-bethe}.

Now, we are ready to express the partition function
\eqref{eq:pba-partition} in terms of the Bethe wave function
\eqref{eq:pba-psi}. It is only non-trivial if the ice rule
\eqref{eq:pba-ice-rule-global} applies, hence we restrict to these
cases. We stress again that our notation differs from
\cite{Baxter:1987}. The relation is established by the following
procedure, where in particular the parameters $\inhtset$,
$\brtset$ and $\magnset$ of \eqref{eq:pba-psi} are related to the
variables $\graph$, $\rapset$ and
$\stateset$ of \eqref{eq:pba-partition}:
\begin{enumerate}
\item For a Baxter lattice with $\lines$ lines, we employ a wave
  function with length $\sites=2\lines$ and $\brts=\lines$
  excitations, a situation usually termed ``half-filling''.
\item The magnon coordinates $\magnset$ are related to
  $\stateset$ and $\graph$. They are given by the
  endpoint positions $\epe_k$ at outward pointing edges with
  $\alpha_{\epe_k}=1$ and $\epb_k$ at edges directed inwards with
  $\alpha_{\epb_k}=2$:
  \begin{align}
    \label{eq:pba-alpha-position}
    \{\magn_k\}
    =
    \{\epe_k|\alpha_{\epe_k}\!=\!1\}
    \cup
    \{\epb_k|\alpha_{\epb_k}\!=\!2\}\,.
  \end{align}
  These $\magn_k$ are then ordered as $1\leq
  \magn_1<\ldots<\magn_\lines\leq 2\lines$. See the example in
  figure~\ref{fig:pba-lattice-states}.
\item Most importantly, the inhomogeneities $\inhtset$ and the Bethe
  roots $\brtset$ are given in terms of the rapidities $\rapset$ and
  $\graph$. For each line $k$ with endpoints $(\epe_k,\epb_k)$ we set
  \begin{align}
    \label{eq:pba-inhomo-rap}
    \inht_{\epe_k}=\rap_k+1\,,
    \quad
    \inht_{\epb_k}=\rap_k+2\,,
    \quad
    \brt_k=\rap_k+1\,.
  \end{align}
  Remarkably, this is an explicit solution of the Bethe
  equations~\eqref{eq:pba-bethe}. It is easily seen after writing the
  Bethe equations in polynomial form in order to avoid divergencies,
  cf.\ section~\ref{sec:bethe-gl2-review}. Note also that the wave
  function \eqref{eq:pba-psi} is invariant under a permutation of the
  Bethe roots.
\end{enumerate}

Under these identifications, we finally obtain the desired expression
for the partition function \eqref{eq:pba-partition} in terms of the
Bethe wave function \eqref{eq:pba-psi}:
\begin{align}
  \label{eq:pba-partition-wave}
  \mathcal{Z}(\graph,\rapset,\stateset)
  =
  \mathcal{C}(\graph,\rapset)^{-1}
  (-1)^{\mathcal{K}(\graph,\stateset)}
  \Phi(\inhtset,\brtset,\magnset)\,.
\end{align}
The exponent $\mathcal{K}(\graph,\stateset)$ is the
number of endpoints $\epe_k$ with state label $\alpha_{\epe_k}=2$,
\begin{align}
  \label{eq:pba-partition-wave-exp}
  \mathcal{K}(\graph,\stateset)
  =
  \big|\{\epe_k\,|\,\alpha_{\epe_k}\!=\!2\}\big|\,.
\end{align}
The $\stateset$-independent normalization is given by
\begin{align}
  \label{eq:-partition-wave-norm}
  \mathcal{C}(\graph,\rapset)
  =
  \Phi(\inhtset,\brtset,\magnset_0)\,,
\end{align}
where $\magnset_0=(\epe_1,\ldots,\epe_\lines)$ is obtained from
\eqref{eq:pba-alpha-position} with $\stateset_0=(1,\ldots,1)$, which
means the state labels are $1$ at all $2\lines$ endpoints. Expression
\eqref{eq:pba-partition-wave} is the \emph{perimeter Bethe ansatz}
solution of the six-vertex model on a Baxter lattice in the rational
limit. A derivation of this solution, different from the original one
in \cite{Baxter:1987}, will be presented in
section~\ref{sec:bethe-pba} as a special case of a more general
result.

\section{From vertex models to Yangian invariance}
\label{sec:yi}

In the previous section the computation of the partition function of
vertex models on typically non-rectangular Baxter lattices using the
perimeter Bethe ansatz was reviewed. Here we vastly generalize the
class of vertex models, and we establish a new perspective on the
computation of the partition functions $\mathcal{Z}$. This is achieved
by connecting the problem to the powerful Quantum Inverse Scattering
Method (QISM) and relating $\mathcal{Z}$ to invariants $|\Psi\rangle$
of Yangian algebras. We are led to a characterization of invariants
$|\Psi\rangle$, which provides the conceptual basis of all further
studies. In particular, it will enable us later in section
\ref{sec:bethe} to construct Yangian invariants using a Bethe ansatz.

In section~\ref{sec:yi-baxter-lattice} we generalize the Baxter
lattice of section~\ref{sec:pba} in two respects. Firstly, we extend
the algebra from $\mathfrak{su}(2)\subset\mathfrak{gl}(2)$ to
$\mathfrak{gl}(n)$. Secondly, we replace the spin~$\frac{1}{2}$
representation of $\mathfrak{su}(2)$ carried by every line with a more
general representation $\Lambda$ of $\mathfrak{gl}(n)$, which in
addition may differ for each line. The resulting lattices will still
be referred to as ``Baxter lattices'', and we will define the
partition function of vertex models associated with them. In
section~\ref{sec:yi-monodromy} we derive identities satisfied by these
partition functions, which are then translated into a set of
eigenvalue equations within the context of the QISM. To this end, the
partition function $\mathcal{Z}$ is identified with a component of a
simultaneous eigenvector $|\Psi\rangle$ of all elements
$\mon_{ab}(\spec)$ of a spin chain monodromy with specific
representations and inhomogeneities.  More precisely, the eigenvalue
of this eigenvector $|\Psi\rangle$ is $1$ for all diagonal monodromy
elements $\mon_{aa}(\spec)$ and $0$ for the off-diagonal ones. In
section~\ref{sec:yi-yangian} we free ourselves from specific
representations and inhomogeneities, and consider the just derived set
of eigenvalue equations for general spin chain monodromies. Noting
that such monodromies provide realizations of the Yangian algebra
$\mathcal{Y}(\mathfrak{gl}(n))$, we observe that the set of eigenvalue
equations characterizes vectors $|\Psi\rangle$ that are Yangian
invariant. Later, in section~\ref{sec:osc-3vertices}, we will come
across examples of Yangian invariants that fall outside the framework
of Baxter lattices in the sense of
section~\ref{sec:yi-monodromy}. These require the generalizations of
section~\ref{sec:yi-yangian}.

\subsection{Vertex models on Baxter lattices}
\label{sec:yi-baxter-lattice}

\begin{figure}[!t]
  \begin{center}  
    \begin{align*}
      \begin{aligned}
        \begin{tikzpicture}
          \draw[thick,densely dotted]
          (0,0) +(-180:1.8cm) arc (-180:180:1.8cm) 
          node[draw,solid,fill=black,inner sep=0.75pt,shape=circle] {}
          node[left] {$B$};
          \draw[thick,
          decoration={
            markings, mark=at position 0.8 with {\arrow{latex reversed}}},
          postaction={decorate}
          ] 
          (-140:1.8cm) 
          node[left] {$\alpha_1$}
          node[left=0.75cm] {$\Lambda_{1}, \rap_{1}$}
          -- (-20:1.8cm)
          node[right] {$\alpha_3$};
          \draw[thick,
          decoration={
            markings, mark=at position 0.85 with {\arrow{latex reversed}}},
          postaction={decorate}
          ] 
          (-80:1.8cm) 
          node[below] {$\alpha_2$}
          node[below=0.5cm] {$\Lambda_{2}, \rap_{2}$}
          -- (95:1.8cm)
          node[above] {$\alpha_5$};
          \draw[thick,
          decoration={
            markings, mark=at position 0.8 with {\arrow{latex reversed}}},
          postaction={decorate}
          ] 
          (20:1.8cm) 
          node[right] {$\alpha_4$}
          node[right=0.75cm] {$\Lambda_{3}, \rap_{3}$}
          -- (155:1.8cm)
          node[left] {$\alpha_6$};
          \node at (0.3,0) {$\gamma$};
        \end{tikzpicture}
      \end{aligned}
      \quad
      \begin{aligned}
        &\mathcal{Z}
        (\graph,\repset,
        \rapset,\stateset)\\
        &=
        \smash[b]{\sum_\gamma}\;
        \langle\alpha_1,\alpha_2|
        R_{\Lambda_1\,\Lambda_2}(\rap_1-\rap_2)
        |\alpha_3,\gamma\rangle\\
        &\hspace{1cm}\cdot\langle\gamma,\alpha_4|
        R_{\Lambda_2\,\Lambda_3}(\rap_2-\rap_3)
        |\alpha_5,\alpha_6\rangle
      \end{aligned}
    \end{align*}
    \caption{Example of a (generalized) Baxter lattice with $\lines=3$
      lines and $\graph=((1,3),(2,5),(4,6))$, left side. Each line
      carries a $\mathfrak{gl}(n)$ representation $\Lambda_k$, a
      spectral parameter $\rap_k$, two state labels $\alpha_{\epe_k}$
      and $\alpha_{\epb_k}$ at the endpoints $\epe_k<\epb_k$ and an
      orientation indicated by an arrow. The dotted circle and the
      reference point $B$ are not part of the Baxter lattice. The
      associated partition function $\mathcal{Z} (\graph,\repset,
      \rapset,\stateset)$ is shown on the right, cf.\
      \eqref{eq:yi-partition-function}.}
    \label{fig:yi-baxter-lattice}
  \end{center}  
\end{figure}
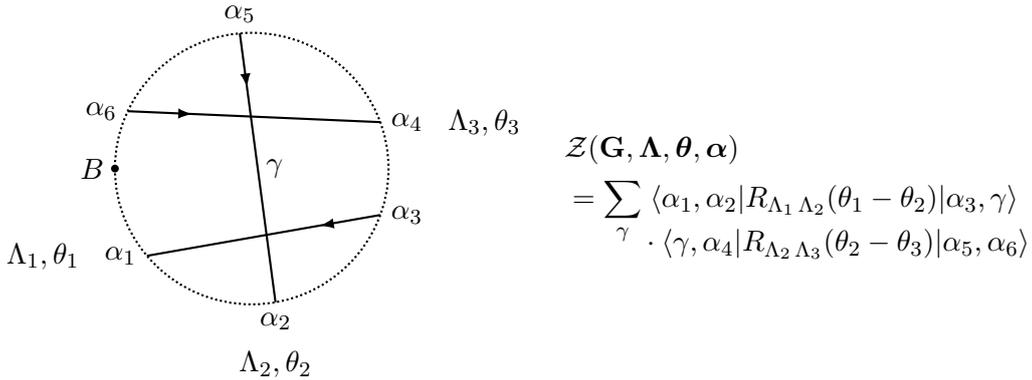

Let us repeat the definition of the Baxter lattice spelled out in
section~\ref{sec:pba}, and extend it by the generalizations just
mentioned. See also the simple example lattice in the left part of
figure~\ref{fig:yi-baxter-lattice}.
We start with a dotted circle on which we mark a reference point
$B$. Notice that the circle and the reference point are only used for
the construction and will not become part of the Baxter lattice
itself. $\lines$ straight lines, each connecting two points on the
dotted circle, are specified in such a way that in the interior of the
circle only two of the lines intersect at a single point. Starting at
the reference point, the $\lines$ lines and the $2\lines$ endpoints of
these lines are labeled counterclockwise. Each line has an
orientation, which for the $k$-th line with endpoints $\epe_k<\epb_k$
is given by an arrow pointing from $\epb_k$ towards $\epe_k$. The
choice of the reference point clearly affects the orientation of the
lines. In addition, we assign a $\mathfrak{gl}(n)$ representation
$\Lambda_k$ and a complex spectral parameter $\rap_k$ to the $k$-th
line. To the endpoints of this line we assign states of the
representation $\Lambda_k$ labeled by $\alpha_{\epe_k}$ and
$\alpha_{\epb_k}$. A (generalized) \emph{Baxter lattice} including
boundary conditions is defined with this data by the ordered sets
\begin{align}
  \label{eq:yi-baxter-data}
  \begin{gathered}
    \graph=((\epe_1,\epb_1),\ldots,(\epe_\lines,\epb_\lines))\,,\\
    \repset=(\Lambda_{1},\ldots,\Lambda_{\lines})\,,
    \quad
    \rapset=(\rap_{1},\ldots,\rap_{\lines})\,,
    \quad
    \stateset=(\alpha_1,\ldots,\alpha_{2\lines})\,.
  \end{gathered}
\end{align}

In order to introduce a vertex model on such a Baxter lattice we also
have to generalize the Boltzmann weights of section~\ref{sec:pba}. We
introduce them as
\begin{align}
  \label{eq:yi-boltzmann}
  \begin{aligned}
    \langle \alpha,\gamma|
    R_{\Lambda\,\Lambda'}(\rap-\rap')
    |\beta,\delta\rangle
    =
    \,\,\,\\\phantom{}
  \end{aligned}
  \begin{aligned}
    \begin{tikzpicture}
      \draw[thick,
      decoration={
        markings, mark=at position 0.85 with {\arrow{latex reversed}}},
      postaction={decorate}
      ] 
      (0,0) 
      node[left=0.4cm] {$\Lambda,\rap$}
      node[left] {$\alpha$} -- 
      (1,0)
      node[right] {$\beta$};
      \draw[thick,
      decoration={
        markings, mark=at position 0.85 with {\arrow{latex reversed}}},
      postaction={decorate}
      ] 
      (0.5,-0.5) node[below=0.5cm] {$\Lambda',\rap'$}
      node[below] {$\gamma$} -- 
      (0.5,0.5)
      node[above] {$\delta$};
    \end{tikzpicture}
  \end{aligned}
  \begin{aligned}
    ,\\\phantom{}
  \end{aligned}
\end{align}
which are matrix elements of the R-matrix
\begin{align}
  \label{eq:yi-r-matrix}
  \begin{aligned}
    R_{\Lambda\,\Lambda'}(\rap-\rap')
    =
    \,\,\,\\\phantom{}    
  \end{aligned}
  \begin{aligned}
    \begin{tikzpicture}
      \draw[thick,
      decoration={
        markings, mark=at position 0.8 with {\arrow{latex reversed}}},
      postaction={decorate}
      ] 
      (0,0) 
      node[left] {$\Lambda,\rap$} -- 
      (1,0);
      \draw[thick,
      decoration={
        markings, mark=at position 0.8 with {\arrow{latex reversed}}},
      postaction={decorate}
      ] 
      (0.5,-0.5) 
      node[below] {$\Lambda',\rap'$} -- 
      (0.5,0.5);
    \end{tikzpicture}
  \end{aligned}
  \begin{aligned}
    .\\\phantom{}
  \end{aligned}
\end{align}
This R-matrix is an operator acting on the tensor product
$V_\Lambda\otimes V_{\Lambda'}$ of the spaces of the two
$\mathfrak{gl}(n)$ representations $\Lambda$ and $\Lambda'$ with
spectral parameters $\rap$ and $\rap'$. The Boltzmann weights are
defined using orthonormal basis states of $V_\Lambda$ and
$V_{\Lambda'}$ labeled by Greek indices $\alpha$, $\beta$ and
$\gamma$, $\delta$, respectively.  Graphically each space is
associated with one line. The orientation of (i.e.\ arrow on) a line
specifies the order of multiple R-matrices acting on one
space. R-matrices ``earlier'' on the line are right of ``later'' ones
in the corresponding formula. In this sense the arrows in
\eqref{eq:yi-r-matrix} point from the ``inputs'' of the R-matrix
towards the ``outputs'' or, in component language
\eqref{eq:yi-boltzmann}, from the kets towards the bras. We will
switch between the operator language and the Boltzmann weights
whenever it is convenient. The R-matrix \eqref{eq:yi-r-matrix} is
required to be a solution of the Yang-Baxter equation
\begin{align}
  \label{eq:yi-ybe}
  \begin{aligned}
    &R_{\Lambda\,\Lambda'}(\rap-\rap')
    R_{\Lambda\,\Lambda''}(\rap-\rap'')
    R_{\Lambda'\,\Lambda''}(\rap'-\rap'')\\
    &=
    R_{\Lambda'\,\Lambda''}(\rap'-\rap'')
    R_{\Lambda\,\Lambda''}(\rap-\rap'')
    R_{\Lambda\,\Lambda'}(\rap-\rap')\,,
  \end{aligned}
\end{align}
which acts in the tensor product $V_\Lambda\otimes V_{\Lambda'}\otimes
V_{\Lambda''}$ and reads graphically
\begin{align}
  \label{eq:yi-ybe-graphical}
  \begin{aligned}
    \begin{tikzpicture}
      \draw[thick,
      decoration={
        markings, mark=at position 0.9 with {\arrow{latex reversed}}},
      postaction={decorate}
      ]
      (-0.25,0.5) 
      node[left] {$\Lambda,\rap$} -- 
      (1.75,0.5);
      \draw[thick,
      decoration={
        markings, mark=at position 0.85 with {\arrow{latex reversed}}},
      postaction={decorate}
      ]
      (0,0) 
      node[below] {$\Lambda',\rap'$} -- 
      (1.5,1.5);
      \draw[thick,
      decoration={
        markings, mark=at position 0.85 with {\arrow{latex reversed}}},
      postaction={decorate}
      ]
      (1.5,0)
      node[below] {$\Lambda'',\rap''$}  -- 
      (0,1.5);
    \end{tikzpicture}
  \end{aligned}
  \begin{aligned}
    \,\,\,=\,\,\,\\\phantom{}
  \end{aligned}
  \begin{aligned}
    \begin{tikzpicture}
      \draw[thick,
      decoration={
        markings, mark=at position 0.9 with {\arrow{latex reversed}}},
      postaction={decorate}
      ]
      (-0.25,1) 
      node[left] {$\Lambda,\rap$} -- 
      (1.75,1);
      \draw[thick,
      decoration={
        markings, mark=at position 0.85 with {\arrow{latex reversed}}},
      postaction={decorate}
      ]
      (0,0) 
      node[below] {$\Lambda',\rap'$} -- 
      (1.5,1.5);
      \draw[thick,
      decoration={
        markings, mark=at position 0.85 with {\arrow{latex reversed}}},
      postaction={decorate}
      ]
      (1.5,0) 
      node[below] {$\Lambda'',\rap''$} -- 
      (0,1.5);
    \end{tikzpicture}
  \end{aligned}
  \begin{aligned}
    .\\\phantom{}
  \end{aligned}
\end{align}

Now we can define the partition function of a vertex model on a
(generalized) Baxter lattice, see again the example in
figure~\ref{fig:yi-baxter-lattice}, employing the component language
of Boltzmann weights. To each internal edge of the lattice we assign a
state of the given representation $\Lambda$, while the states at the
boundary edges are naturally fixed by $\stateset$. Recall that
$\Lambda$ is associated to the entire line, and therefore all internal
and boundary edges which make up the line carry states in this
representation.  Each vertex of the lattice is then translated into a
Boltzmann weight as shown in \eqref{eq:yi-boltzmann}. The partition
function $\mathcal{Z}(\graph,\repset, \rapset,\stateset)$ is the sum,
ranging over all possible configurations of states at the internal
edges, of the product of all Boltzmann weights of the lattice. As
already in \eqref{eq:pba-partition-wave}, we write symbolically
\begin{align}
  \label{eq:yi-partition-function}  
  \mathcal{Z}
  (\graph,\repset,
  \rapset,\stateset)
  =
  \sum_{\substack{\text{internal}\\\text{state}\\\text{config.}}}\,
  \prod_{\text{vertices}}
  \text{Boltzmann weight}\,.
\end{align}
If there is a line consisting of a single edge with differing states
at the boundary, the partition function vanishes.

In the operator description this partition function is a matrix
element of a product of R-matrices. These R-matrices and their order
in the product are given by the form of the Baxter lattice. A line
with a single edge is translated into an identity operator on the
corresponding representation space. The matrix element is specified by
$\stateset$, and the sum and product in
\eqref{eq:yi-partition-function} translate into matrix multiplication.

\subsection{Partition function as eigenvalue problem}
\label{sec:yi-monodromy}

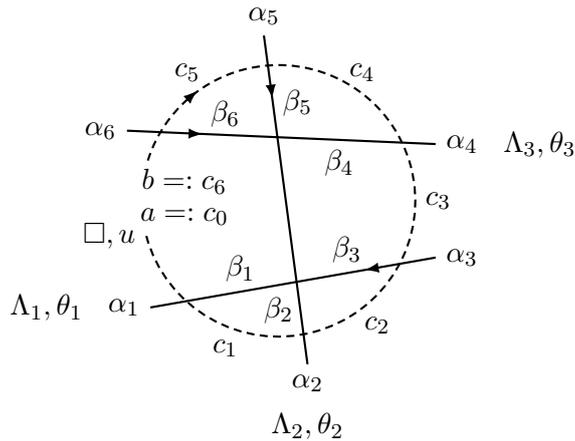
\begin{figure}[!t]
  \begin{center}
    \begin{tikzpicture}
      \draw[thick,densely dashed,
      decoration={
        markings, mark=at position 0.9 with {\arrow{latex reversed}}},
      postaction={decorate}
      ]
      (0,0) +(-165:1.8cm) 
      node[right=0.5cm,above] {$a=:c_0$}
      node[left] {$\square, \spec$}
      arc (-165:165:1.8cm)
      node[right=0.5cm,below=-0.1cm] {$b=:c_6$};
      \node at (-0.7,-1.95) {$c_1$};
      \node at (1.3,-1.65) {$c_2$};
      \node at (2.1,0) {$c_3$};
      \node at (1.1,1.7) {$c_4$};
      \node at (-1.2,1.7) {$c_5$};
      \node at (-0.5,-0.9) {$\beta_1$};
      \node at (-0.0,-1.45) {$\beta_2$};
      \node at (0.9,-0.65) {$\beta_3$};
      \node at (0.8,0.5) {$\beta_4$};
      \node at (0.25,1.3) {$\beta_5$};
      \node at (-0.7,1.15) {$\beta_6$};
      \draw[thick,
      decoration={
        markings, mark=at position 0.8 with {\arrow{latex reversed}}},
      postaction={decorate}
      ] 
      (-140:2.2cm) 
      node[left] {$\alpha_1$}
      node[left=0.75cm] {$\Lambda_{1}, \rap_{1}$}
      -- (-20:2.2cm)
      node[right] {$\alpha_3$};
      \draw[thick,
      decoration={
        markings, mark=at position 0.85 with {\arrow{latex reversed}}},
      postaction={decorate}
      ] 
      (-80:2.2cm) 
      node[below] {$\alpha_2$}
      node[below=0.5cm] {$\Lambda_{2}, \rap_{2}$}
      -- (95:2.2cm)
      node[above] {$\alpha_5$};
      \draw[thick,
      decoration={
        markings, mark=at position 0.8 with {\arrow{latex reversed}}},
      postaction={decorate}
      ] 
      (20:2.2cm) 
      node[right] {$\alpha_4$}
      node[right=0.75cm] {$\Lambda_{3}, \rap_{3}$}
      -- (155:2.2cm)
      node[left] {$\alpha_6$};
    \end{tikzpicture}
    \caption{The Baxter lattice introduced in the example of
      figure~\ref{fig:yi-baxter-lattice} after the dotted circle has
      been replaced by a dashed auxiliary space line in the
      fundamental representation $\square$ with spectral parameter
      $\spec$ and states labeled $a$, $b$ at the endpoints. The
      indices $c_i$ are assigned to the edges of this auxiliary
      space. The states at the edges connecting this space with the
      Baxter lattice are labeled $\beta_i$.}
    \label{fig:yi-baxter-lattice-aux}
  \end{center}
\end{figure}

As a first step to understand the partition function of a Baxter
lattice as an eigenvalue problem within the QISM, we derive an
identity satisfied by $\mathcal{Z} (\graph,\repset,
\rapset,\stateset)$. Recall that the construction of Baxter lattices
involves a dotted circle. Here we replace this circle by an arc, which
is opened at the reference point $B$ and is to represent an actual
\emph{space} called auxiliary space. In addition, the lines of the
Baxter lattice are slightly extended such that they intersect the
arc. This is depicted by the dashed line in
figure~\ref{fig:yi-baxter-lattice-aux}.
The auxiliary space $V_\square=\mathbb{C}^n$ carries the fundamental
representation $\square$ of $\mathfrak{gl}(n)$ as well as a spectral
parameter $\spec$. The orientation is chosen clockwise. The bra and
ket states at the endpoints of the dashed line are labeled by the
indices $a$ and $b$, respectively, which may take the values
$1,\ldots,n$.  An auxiliary space intersects all other lines twice, it
introduces a layer of additional vertices at the boundary of the
Baxter lattice. The Boltzmann weights at these vertices correspond to
elements of R-matrices of the type $R_{\square\,\Lambda}(\spec-\rap)$
or $R_{\Lambda\,\square}(\rap-\spec)$, which are referred to as Lax
operators. These Lax operators also satisfy a Yang-Baxter equation of
the form
\begin{align}
  \label{eq:yi-ybe-boundary}
  \begin{aligned}
    \begin{tikzpicture}
      \draw[thick,
      decoration={
        markings, mark=at position 0.85 with {\arrow{latex reversed}}},
      postaction={decorate}
      ]
      (0,0) 
      node[below] {$\Lambda,\rap$} -- 
      (1.5,1.5);
      \draw[thick,
      decoration={
        markings, mark=at position 0.85 with {\arrow{latex reversed}}},
      postaction={decorate}
      ]
      (1.5,0) 
      node[below] {$\Lambda',\rap'$} -- 
      (0,1.5);
      \draw[thick,densely dashed,
      decoration={
        markings, mark=at position 0.9 with {\arrow{latex reversed}}},
      postaction={decorate}
      ]
      (-0.25,0.5)
      node[left] {$\square,\spec$} -- 
      (1.75,0.5);
    \end{tikzpicture}
  \end{aligned}
  \begin{aligned}
    \,\,\,=\,\,\,\\\phantom{}
  \end{aligned}
  \begin{aligned}
    \begin{tikzpicture}
      \draw[thick,
      decoration={
        markings, mark=at position 0.85 with {\arrow{latex reversed}}},
      postaction={decorate}
      ]
      (0,0) 
      node[below] {$\Lambda,\rap$} -- 
      (1.5,1.5);
      \draw[thick,
      decoration={
        markings, mark=at position 0.85 with {\arrow{latex reversed}}},
      postaction={decorate}
      ]
      (1.5,0) 
      node[below] {$\Lambda',\rap'$} -- 
      (0,1.5);
      \draw[thick,densely dashed,
      decoration={
        markings, mark=at position 0.9 with {\arrow{latex reversed}}},
      postaction={decorate}
      ]
      (-0.25,1) 
      node[left] {$\square,\spec$} -- 
      (1.75,1);
    \end{tikzpicture}
  \end{aligned}
  \begin{aligned}
    ,\\\phantom{}
  \end{aligned}
\end{align}
which is a special case of \eqref{eq:yi-ybe-graphical}. In addition,
we demand the unitarity condition
\begin{align}
  \label{eq:yi-unitarity}
  \begin{aligned}
    R_{\square\,\Lambda}(\spec-\rap)
    R_{\Lambda\,\square}(\rap-\spec)
    =
    1\,,
  \end{aligned}
  \quad\text{i.e.}\quad
  \begin{aligned}
    \begin{aligned}
      \begin{tikzpicture}
        \draw[thick,densely dashed,
        decoration={
          markings, mark=at position 0.90 with {\arrow{latex reversed}}},
        postaction={decorate}
        ]
        (0,0) 
        node[left] {$\square,\spec$} -- 
        (2.0,0);
        \draw[thick,
        decoration={
          markings, mark=at position 0.95 with {\arrow{latex reversed}}},
        postaction={decorate}
        ]
        (0.5,-0.5)
        node[below] {$\Lambda,\rap$} -- 
        (0.5,0.25) .. controls (0.5,1) and (1.5,1) .. (1.5,0.25) -- 
        (1.5,-0.5);
        \path
        (0,1.25) node[left] {\phantom{$\square,\spec$}} -- 
        (2.0,1.25);
      \end{tikzpicture}
    \end{aligned}\;
    \,\,\,=
    \begin{aligned}
      \begin{tikzpicture}
        \draw[thick,densely dashed,
        decoration={
          markings, mark=at position 0.9 with {\arrow{latex reversed}}},
        postaction={decorate}
        ]
        (0,1.25) 
        node[left] {$\square,\spec$} -- 
        (2.0,1.25);
        \draw[thick,
        decoration={
          markings, mark=at position 0.95 with {\arrow{latex reversed}}},
        postaction={decorate}
        ]
        (0.5,-0.5) 
        node[below] {$\Lambda,\rap$} -- 
        (0.5,0.25) .. controls (0.5,1) and (1.5,1) .. (1.5,0.25) -- 
        (1.5,-0.5);
      \end{tikzpicture}
    \end{aligned}
  \end{aligned}
\end{align}
using the graphical notation.

Making use of the Yang-Baxter equation \eqref{eq:yi-ybe-boundary} and
the unitarity condition \eqref{eq:yi-unitarity} we can completely
disentangle the auxiliary space (dashed line) from the $\lines$ spaces
defining the Baxter lattice (solid lines). Graphically one easily sees
that this leads to a non-trivial identity for the partition function
$\mathcal{Z}(\graph,\repset,\rapset,\stateset)$ of a Baxter lattice,
see the example in figure~\ref{fig:yi-disentangle}.
\begin{figure}[!t]
  \begin{center}
    \begin{align*}
      \begin{aligned}
        \begin{tikzpicture}
          \draw[thick,densely dashed,
          decoration={
            markings, mark=at position 0.95 with {\arrow{latex reversed}}},
          postaction={decorate}
          ]
          (0,0) 
          node[left=0.4cm] {$\square,\spec$}
          node[left] {$a$} -- 
          (3.5,0)
          node[right] {$b$};
          \draw[thick,
          decoration={
            markings, mark=at position 0.96 with {\arrow{latex reversed}}},
          postaction={decorate}
          ] 
          (0.5,-0.5) 
          node[below] {$\alpha_1$}-- 
          (0.5,0.5)
          node[above=0.5cm,left=-0.1cm] {$\Lambda_1,\rap_1$} 
          .. controls (0.5,1.25) and (1.5,1.25) .. (1.5,0.5) -- 
          (1.5,-0.5)
          node[below] {$\alpha_3$};
          \draw[thick,
          decoration={
            markings, mark=at position 0.95 with {\arrow{latex reversed}}},
          postaction={decorate}
          ] 
          (2,-0.5)
          node[below] {$\alpha_4$} -- 
          (2,0.25) .. controls (2,1) and (3,1) .. 
          (3,0.25) 
          node[above=0.5cm,right=-0.1cm] {$\Lambda_3,\rap_3$} -- 
          (3,-0.5)
          node[below] {$\alpha_6$};
          \draw[thick,
          decoration={
            markings, mark=at position 0.965 with {\arrow{latex reversed}}},
          postaction={decorate}
          ] 
          (1,-0.5) 
          node[below] {$\alpha_2$} -- 
          (1,0.5) 
          node[above=1.1cm,right=0cm] {$\Lambda_2,\rap_2$} 
          .. controls (1,1.5) and (2.5,1.5) .. (2.5,0.5) -- 
          (2.5,-0.5)
          node[below] {$\alpha_5$};
          \path
          (0,2.25) node[left] {\phantom{$\square,\spec$}} -- 
          (2,2.25);
        \end{tikzpicture}
      \end{aligned}\;
      \,\,\,=
      \begin{aligned}
        \begin{tikzpicture}
          \draw[thick,densely dashed,
          decoration={
            markings, mark=at position 0.95 with {\arrow{latex reversed}}},
          postaction={decorate}
          ]
          (0,2.25) 
          node[left=0.4cm] {$\square,\spec$}
          node[left] {$a$} -- 
          (3.5,2.25)
          node[right] {$b$};
          \draw[thick,
          decoration={
            markings, mark=at position 0.96 with {\arrow{latex reversed}}},
          postaction={decorate}
          ] 
          (0.5,-0.5) 
          node[below] {$\alpha_1$}-- 
          (0.5,0.5)
          node[above=0.5cm,left=-0.1cm] {$\Lambda_1,\rap_1$} 
          .. controls (0.5,1.25) and (1.5,1.25) .. (1.5,0.5) -- 
          (1.5,-0.5)
          node[below] {$\alpha_3$};
          \draw[thick,
          decoration={
            markings, mark=at position 0.95 with {\arrow{latex reversed}}},
          postaction={decorate}
          ] 
          (2,-0.5)
          node[below] {$\alpha_4$} -- 
          (2,0.25) .. controls (2,1) and (3,1) .. 
          (3,0.25) 
          node[above=0.5cm,right=-0.1cm] {$\Lambda_3,\rap_3$} -- 
          (3,-0.5)
          node[below] {$\alpha_6$};
          \draw[thick,
          decoration={
            markings, mark=at position 0.965 with {\arrow{latex reversed}}},
          postaction={decorate}
          ] 
          (1,-0.5) 
          node[below] {$\alpha_2$} -- 
          (1,0.5) 
          node[above=1.1cm,right=0cm] {$\Lambda_2,\rap_2$} 
          .. controls (1,1.5) and (2.5,1.5) .. (2.5,0.5) -- 
          (2.5,-0.5)
          node[below] {$\alpha_5$};
        \end{tikzpicture}
      \end{aligned}
    \end{align*}
    \caption{An identity for $\mathcal{Z}(\graph,\repset,
      \rapset,\stateset)$ of the sample Baxter lattice in
      figure~\ref{fig:yi-baxter-lattice} is derived by disentangling
      the dashed auxiliary line from the solid lines using
      \eqref{eq:yi-ybe-boundary} and \eqref{eq:yi-unitarity}. The
      lattice has been deformed to emphasize that the row of vertices
      involving the auxiliary line will be written as a monodromy
      shortly, cf. figure~\ref{fig:yi-mono-element}.}
    \label{fig:yi-disentangle}
  \end{center}
\end{figure}
To obtain this identity for a general Baxter lattice, we start by
denoting the Boltzmann weights involving the auxiliary space by
\begin{align}
  \label{eq:yi-add-boltzmann}
    \begin{aligned}
      \mathcal{M}_{ab}(\spec,\graph,
      \repset,
      \rapset,
      \stateset,
      \boldsymbol{\beta})\\
      =
      \smash[b]{\sum_{c_1,\ldots,c_{2\lines-1}=1}^n}\;
      \Bigg(\;
      &
      \smash[b]{\prod_{k=1}^\lines}\;
      \langle c_{\epe_k-1},\alpha_{\epe_k}|
      R_{\square\,\Lambda_k}(\spec-\rap_k)
      |c_{\epe_k},\beta_{\epe_k}\rangle\\
      &\quad\cdot
      \langle \beta_{\epb_k},c_{\epb_k-1}|
      R_{\Lambda_k\,\square}(\rap_k-\spec)
      |\alpha_{\epb_k},c_{\epb_k}\rangle
      \smash[t]{
        \Bigg)_{
          \begin{subarray}{l}
            c_0:=a\\c_{2\lines}:=b
          \end{subarray}
        }
      }\,,
    \end{aligned}
\end{align}
where each of the $\lines$ lines of the lattice contributes two
weights.  In figure~\ref{fig:yi-baxter-lattice-aux} we see an example
for the assignment of the indices $c_i$ and $\beta_i$ to the
edges. Also, recall from \eqref{eq:yi-baxter-data} that the $k$-th
line of a Baxter lattice has the endpoints $\epe_k<\epb_k$. The states
labeled $c_i$ with $i=0,\ldots,2\lines$ are assigned to the edges of
the auxiliary space. The state labels
$\boldsymbol{\beta}=(\beta_1,\ldots,\beta_{2\lines})$ are placed at
the edges that connect the layer of vertices involving the auxiliary
space to the Baxter lattice on which the partition function is
defined.  Equating the Baxter lattice entangled with the auxiliary
space to the disentangled situation, we find
\begin{align}
  \label{eq:yi-condition-z-general}
  \sum_{\boldsymbol{\beta}}
  \mathcal{M}_{ab}(\spec,\graph,
  \repset,\rapset,
  \stateset,\boldsymbol{\beta})\,
  \mathcal{Z}
  (\graph,
  \repset,\rapset,\boldsymbol{\beta})
  =
  \delta_{ab}\,
  \mathcal{Z}
  (\graph,
  \repset,\rapset,\stateset)\,.
\end{align}
We see that the unraveled auxiliary line simply translates into
$\delta_{ab}$ on the r.h.s.\ of \eqref{eq:yi-condition-z-general}. The
entire equation is depicted for an example in
figure~\ref{fig:yi-disentangle}.

As will be shown next, the summed-over Boltzmann weights in
$\mathcal{M}_{ab}(\spec,\graph,\repset,\rapset,\stateset,\boldsymbol{\beta})$
can be rewritten as matrix elements of an inhomogeneous spin chain
monodromy $\mon(\spec)$ with $\sites$ sites. This allows us to link
with the QISM. The monodromy is introduced as
\begin{align}
  \label{eq:yi-mono}
  \begin{aligned}
    \mon(\spec)
    =
    R_{\square\,\Xi_1}(\spec-\inh_1)
    \cdots R_{\square\,\Xi_\sites}(\spec-\inh_\sites)
    =
    \,\,\,\\\phantom{}
  \end{aligned}
  \begin{aligned}
    \begin{tikzpicture}
      \draw[thick,densely dashed,
      decoration={
        markings, mark=at position 0.95 with {\arrow{latex reversed}}},
      postaction={decorate}
      ] 
      (0,0) 
      node[left] {$\square,\spec$} -- 
      (3,0);
      \draw[thick,
      decoration={
        markings, mark=at position 0.85 with {\arrow{latex reversed}}},
      postaction={decorate}
      ] 
      (0.5,-0.5) 
      node[below] {$\Xi_1,\inh_1$} -- 
      (0.5,0.5);
      \node at (1.5,-0.25) {$\ldots$};
      \draw[thick,
      decoration={
        markings, mark=at position 0.85 with {\arrow{latex reversed}}},
      postaction={decorate}
      ] 
      (2.5,-0.5) 
      node[below] {$\Xi_\sites,\inh_\sites$} -- 
      (2.5,0.5);
    \end{tikzpicture}
  \end{aligned}
  \begin{aligned}
    .\\\phantom{}
  \end{aligned}
\end{align}
The $j$-th site carries a $\mathfrak{gl}(n)$ representation $\Xi_j$
acting on the local quantum space $V_j$ and it has an inhomogeneity
$\inh_j\in\mathbb{C}$. The total quantum space of the spin chain is
the tensor product $V_{1}\otimes\cdots\otimes V_{\sites}$. The
auxiliary space carries the fundamental representation $\square$ and
the matrix elements of the monodromy with respect to this space are
denoted by
\begin{align}
  \label{eq:yi-mono-elements}
  \mon_{ab}(\spec):=\langle a|\mon(\spec)|b\rangle\,.
\end{align}
They are still operators in the total quantum space. In what follows,
we require the Lax operators associated with the Boltzmann weights in
\eqref{eq:yi-add-boltzmann} to satisfy the crossing relation
\begin{align}
  \label{eq:yi-crossing}
  R_{\square\,\bar\Lambda}(\spec-\rap+\kappa_\Lambda)
  =
  R_{\Lambda\,\square}(\rap-\spec)^\dagger\,,
\end{align}
where $\kappa_\Lambda$ is a representation-dependent crossing
parameter, and the conjugation only acts on the space $V_\Lambda$. For
a representation $\Lambda$ realized by $\mathfrak{gl}(n)$ generators
$J_{ab}$ on $V_\Lambda$, the conjugate representation $\bar\Lambda$
appearing in \eqref{eq:yi-crossing} is defined by the generators
\begin{align}
  \label{eq:yi-conj-rep}
  \bar J_{ab}=-J_{ab}^\dagger\,.
\end{align}
Assuming the matrix elements $\langle\alpha|J_{ab}|\beta\rangle$ of
the generators to be real, we obtain from \eqref{eq:yi-crossing}
\begin{align}
  \label{eq:yi-crossing-coord}
  \begin{aligned}
    &\langle c,\beta|
    R_{\square\,\bar\Lambda}(\spec-\rap+\kappa_\Lambda)
    |d,\alpha\rangle\\
    &=
    \langle \alpha,c|
    R_{\Lambda\,\square}(\rap-\spec)
    |\beta,d\rangle\,,
  \end{aligned}
  \begin{aligned}
    \quad\text{i.e.}\quad
  \end{aligned}
  \begin{aligned}
    \begin{aligned}
      \begin{tikzpicture}
        \draw[thick,densely dashed,
        decoration={
          markings, mark=at position 0.85 with {\arrow{latex reversed}}},
        postaction={decorate}
        ]
        (0,0)
        node[left=0.4cm] {$\square,\spec$}
        node[left] {$c$} -- 
        (1,0)
        node[right] {$d$};
        \draw[thick,
        decoration={
          markings, mark=at position 0.85 with {\arrow{latex reversed}}},
        postaction={decorate}
        ] 
        (0.5,-0.5)
        node[below=0.5cm] {$\bar\Lambda,\rap-\kappa_\Lambda$}
        node[below] {$\beta$} -- 
        (0.5,0.5)
        node[above] {$\alpha$};
      \end{tikzpicture}
    \end{aligned}
    \begin{aligned}
      \,\,\,=\,\,\,\\\phantom{}  
    \end{aligned}
    \begin{aligned}
      \begin{tikzpicture}
        \draw[thick,densely dashed,
        decoration={
          markings, mark=at position 0.85 with {\arrow{latex reversed}}},
        postaction={decorate}
        ] 
        (0,0) 
        node[left=0.4cm] {$\square,\spec$}
        node[left] {$c$} -- 
        (1,0)
        node[right] {$d$};
        \draw[thick,
        decoration={
          markings, mark=at position 0.3 with {\arrow{latex}}},
        postaction={decorate}
        ] 
        (0.5,-0.5) 
        node[below=0.5cm] {$\Lambda,\rap$}
        node[below] {$\beta$} -- 
        (0.5,0.5)
        node[above] {$\alpha$};
      \end{tikzpicture}
    \end{aligned}
  \end{aligned}
  \begin{aligned}
    ,
  \end{aligned}
\end{align}
where we have given both the equation and its graphic
representation. Applying \eqref{eq:yi-crossing-coord} to the weights
in the second line of \eqref{eq:yi-add-boltzmann} yields
\begin{align}
  \label{eq:yi-add-boltzmann-cross}
  \begin{aligned}
    \mathcal{M}_{ab}(\spec,\graph,
    \repset,\rapset,
    \stateset,\boldsymbol{\beta})\\
    =
    \smash[b]{\sum_{c_1,\ldots,c_{2\lines-1}=1}^n}\;
    \Bigg(\;
    &
    \smash[b]{\prod_{k=1}^\lines}\;
    \langle c_{\epe_k-1},\alpha_{\epe_k}|
    R_{\square\,\Lambda_k}(\spec-\rap_k)
    |c_{\epe_k},\beta_{\epe_k}\rangle\\
    &\quad\cdot
    \langle c_{\epb_k-1},\alpha_{\epb_k}|
    R_{\square\,\bar\Lambda_k}(\spec-\rap_k+\kappa_{\Lambda_k})
    |c_{\epb_k},\beta_{\epb_k}\rangle
    \smash[t]{
      \Bigg)_{
        \begin{subarray}{l}
          c_0:=a\\c_{2\lines}:=b
        \end{subarray}
      }
    }\,.
  \end{aligned}
\end{align}
In this form the index structure is such that all weights combine into
matrix elements of the monodromy \eqref{eq:yi-mono} with
$\sites=2\lines$ sites,
\begin{align}
  \label{eq:yi-mono-element}
  \begin{aligned}
    \mathcal{M}_{ab}(\spec,\graph,
    \repset,\rapset,
    \stateset,\boldsymbol{\beta})
    =
    \langle\stateset|
    \mon_{ab}(\spec)
    |\boldsymbol{\beta}\rangle\,.
  \end{aligned}
\end{align}
As is usual for a monodromy, the labels of the total quantum space are
hidden. Thus there is no analogue of the labels $\graph$, $\repset$,
$\rapset$ on the r.h.s.\ of \eqref{eq:yi-mono-element}.
\begin{figure}[!t]
  \begin{center}
    \begin{align*}
      \begin{aligned}
        \begin{tikzpicture}
          \draw[thick,densely dashed,
          decoration={
            markings, mark=at position 0.97 with {\arrow{latex reversed}}},
          postaction={decorate}
          ] 
          (0,0) 
          node[left=0.4cm] {$\square,\spec$}
          node[left] {$a$} -- 
          (6.5,0)
          node[right] {$b$};
          \draw[thick,
          decoration={
            markings, mark=at position 0.85 with {\arrow{latex reversed}}},
          postaction={decorate}
          ] 
          (0.5,-0.5) 
          node[below] {$\alpha_1$}
          node[below=0.5cm] {$\Lambda_1,\rap_1$} -- 
          (0.5,0.5)
          node[above] {$\beta_1$};
          \draw[thick,
          decoration={
            markings, mark=at position 0.85 with {\arrow{latex reversed}}},
          postaction={decorate}
          ] 
          (1.6,-0.5) 
          node[below] {$\alpha_2$}
          node[below=0.5cm] {$\Lambda_2,\rap_2$} -- 
          (1.6,0.5)
          node[above] {$\beta_2$};
          \draw[thick,
          decoration={
            markings, mark=at position 0.3 with {\arrow{latex}}},
          postaction={decorate}
          ] 
          (2.7,-0.5) 
          node[below] {$\alpha_3$}
          node[below=0.5cm] {$\Lambda_1,\rap_1$} -- 
          (2.7,0.5)
          node[above] {$\beta_3$};
          \draw[thick,
          decoration={
            markings, mark=at position 0.85 with {\arrow{latex reversed}}},
          postaction={decorate}
          ] 
          (3.8,-0.5) 
          node[below] {$\alpha_4$}
          node[below=0.5cm] {$\Lambda_3,\rap_3$} -- 
          (3.8,0.5)
          node[above] {$\beta_4$};
          \draw[thick,
          decoration={
            markings, mark=at position 0.3 with {\arrow{latex}}},
          postaction={decorate}
          ] 
          (4.9,-0.5) 
          node[below] {$\alpha_5$}
          node[below=0.5cm] {$\Lambda_2,\rap_2$} -- 
          (4.9,0.5)
          node[above] {$\beta_5$};
          \draw[thick,
          decoration={
            markings, mark=at position 0.3 with {\arrow{latex}}},
          postaction={decorate}
          ] 
          (6.0,-0.5) 
          node[below] {$\alpha_6$}
          node[below=0.5cm] {$\Lambda_3,\rap_3$} -- 
          (6.0,0.5)
          node[above] {$\beta_6$};
        \end{tikzpicture}
      \end{aligned}
      \begin{aligned}
        \,\,\,=\,\,\,\\\phantom{}
      \end{aligned}
      \begin{aligned}
        \begin{tikzpicture}
          \draw[thick,densely dashed,
          decoration={
            markings, mark=at position 0.95 with {\arrow{latex reversed}}},
          postaction={decorate}
          ] 
          (0,0) 
          node[left=0.4cm] {$\square,\spec$}
          node[left] {$a$} -- 
          (2.5,0)
          node[right] {$b$};
          \draw[thick,
          decoration={
            markings, mark=at position 0.85 with {\arrow{latex reversed}}},
          postaction={decorate}
          ] 
          (0.5,-0.5) 
          node[below] {$\alpha_1$}
          node[below=0.5cm] {$\Xi_1,\inh_1$} -- 
          (0.5,0.5)
          node[above] {$\beta_1$};
          \node at (1.25,-0.25) {$\ldots$};
          \draw[thick,
          decoration={
            markings, mark=at position 0.85 with {\arrow{latex reversed}}},
          postaction={decorate}
          ] 
          (2.0,-0.5) 
          node[below] {$\alpha_6$}
          node[below=0.5cm] {$\Xi_6,\inh_6$} 
          -- 
          (2.0,0.5)
          node[above] {$\beta_6$};
        \end{tikzpicture}
      \end{aligned}
    \end{align*}
    \caption{Rewriting of the summed Boltzmann weights in
      $\mathcal{M}_{ab}(\spec,\graph,\repset,
      \rapset,\stateset,\boldsymbol{\beta})$ on the l.h.s.\ as a
      matrix element
      $\langle\stateset|\mon_{ba}(\spec)|\boldsymbol{\beta}\rangle$ of
      a monodromy on the r.h.s.\ for the example discussed in
      figure~\ref{fig:yi-disentangle}. After applying
      \eqref{eq:yi-crossing-coord} to the l.h.s.\ all vertical lines
      have the same orientation. $\Xi_i$ and $\inh_i$ of the resulting
      monodromy are given by \eqref{eq:yi-ident-rep-inhomo} with
      $\graph$ specified in the caption of
      figure~\ref{fig:yi-baxter-lattice}.}
    \label{fig:yi-mono-element}
  \end{center}
\end{figure}
Here we use the notation
$|\boldsymbol{\beta}\rangle:=|\beta_1\rangle\otimes\cdots\otimes|\beta_{2\lines}\rangle\in
V_1\otimes\cdots\otimes V_{2\lines}$. For each line $k$ of the Baxter
lattice with endpoints $\epe_k<\epb_k$ specified in $\graph$, see
\eqref{eq:yi-baxter-data}, we obtain two spin chain sites with
representations and inhomogeneities
\begin{align}
  \label{eq:yi-ident-rep-inhomo}
  \Xi_{\epe_k}=\Lambda_k\,,
  \quad
  \inh_{\epe_k}=\rap_k
  \quad
  \text{and}
  \quad
  \Xi_{\epb_k}=\bar\Lambda_k\,,
  \quad
  \inh_{\epb_k}=\rap_k-\kappa_{\Lambda_k}\,.
\end{align}
See also the example in figure~\ref{fig:yi-mono-element}. In addition,
the partition function of the vertex model defines a vector
$|\Psi\rangle$ in the total quantum space of the spin chain via
\begin{align}
  \label{eq:yi-partition-psi}
  \langle\stateset|\Psi\rangle
  :=
  \mathcal{Z}(\graph,
  \repset,\rapset,\stateset)\,.
\end{align}
With \eqref{eq:yi-mono-elements}, \eqref{eq:yi-partition-psi} and the
orthonormality of the states $|\boldsymbol{\beta}\rangle$ the identity
\eqref{eq:yi-condition-z-general} for the partition function
translates into
\begin{align}
  \label{eq:yi-eigenvalue-bra}
  \langle\stateset|\mon_{ab}(\spec)|\Psi\rangle
  =
  \delta_{ab}\langle\stateset|\Psi\rangle\,.
\end{align}
Dropping the bra $\langle\stateset|$,
\eqref{eq:yi-eigenvalue-bra} is the sought-for set of eigenvalue
equations. These equations characterize the vector $|\Psi\rangle$,
which, according to \eqref{eq:yi-partition-psi}, is built out of the
partition functions of a Baxter lattice for all possible boundary
configurations $\stateset$. Equation
\eqref{eq:yi-eigenvalue-bra} tells us that $|\Psi\rangle$ is a
\emph{special} simultaneous eigenvector of all matrix elements of the
\emph{specific} monodromy defined by \eqref{eq:yi-mono} with
\eqref{eq:yi-ident-rep-inhomo}. The eigenvector $|\Psi\rangle$ is
special because its eigenvalues are fixed to be $1$ for diagonal
monodromy elements and $0$ for off-diagonal ones. Remarkably,
\eqref{eq:yi-eigenvalue-bra} is an eigenvalue problem within the realm
of the QISM.

\subsection{Yangian algebra and invariants}
\label{sec:yi-yangian}

In this section we will analyze \eqref{eq:yi-eigenvalue-bra} in the
context of Yangians. We will continue to employ the monodromy
\eqref{eq:yi-mono}. However, we shall allow for general
representations $\Xi_i$ and inhomogeneities $\inh_i$, which in general do
not have to obey the restrictions
\eqref{eq:yi-ident-rep-inhomo}. Furthermore, an odd number of sites
$\sites$ is now also permitted. This was not meaningful in the context of
section~\ref{sec:yi-monodromy}, where each line of the Baxter lattice
gave rise to exactly two sites.

Let us use the well-known explicit expression for the Lax operators at
the sites of the monodromy,
\begin{align}
  \label{eq:yi-lax-fund-xi}
  \begin{aligned}
    R_{\square\,\Xi}(\spec-\inh)
    &=
    f_{\Xi}(\spec-\inh)
    \Bigg(1+(\spec-\inh)^{-1}\sum_{a,b=1}^ne_{ab}J_{ba}\Bigg)
    =
    \,\,\,\\\phantom{}
  \end{aligned}
  \begin{aligned}
    \begin{tikzpicture}
      \draw[thick,densely dashed,
      decoration={
        markings, mark=at position 0.85 with {\arrow{latex reversed}}},
      postaction={decorate}
      ] 
      (0,0) 
      node[left] {$\square,\spec$} -- 
      (1,0);
      \draw[thick,
      decoration={
        markings, mark=at position 0.85 with {\arrow{latex reversed}}},
      postaction={decorate}
      ] 
      (0.5,-0.5) 
      node[below] {$\Xi,\inh$} -- 
      (0.5,0.5);
    \end{tikzpicture}
  \end{aligned}
  \begin{aligned}
    ,\\\phantom{}
  \end{aligned}
\end{align}
where the generators $J_{ab}$ of the representation $\Xi$ satisfy the
$\mathfrak{gl}(n)$ algebra
\begin{align}
  \label{eq:yi-gln}
  [J_{ab},J_{cd}]=\delta_{cb}J_{ad}-\delta_{ad}J_{cb}\,.
\end{align}
The $n\times n$ matrices $e_{ab}$ are generators of the fundamental
representation $\square$ of $\mathfrak{gl}(n)$. Their components are
$\langle c|e_{ab}|d\rangle=\delta_{ac}\delta_{bd}$, where $|a\rangle$
with $a=1,\ldots,n$ are the standard basis vectors of
$V_\square=\mathbb{C}^n$. Hence,
$e_{ab}e_{cd}=\delta_{bc}e_{ad}$. Moreover, $f_\Xi(\spec-\inh)$ is a
scalar normalization factor and $1$ stands for the appropriate
identity operator. Here it is that in $\mathbb{C}^n\otimes V_\Xi$. The
monodromy \eqref{eq:yi-mono} built out of these Lax operators
satisfies the RTT-relation\footnote{The name stems from the frequent
  use of the symbol ``$T(\spec)$'' for the monodromy $\mon(\spec)$ in
  the literature.}
\begin{align}
  \label{eq:yi-ybe-rmm}
  R_{\square\,\square'}(\spec-\specp)\mon(\spec)\mon'(\specp)
  =
  \mon'(\specp)\mon(\spec)R_{\square\,\square'}(\spec-\specp)\,.
\end{align}
This is proven with the so-called ``train argument'', see e.g.\
\cite{Faddeev:1994nk}, making use of the Yang-Baxter equations for the
individual Lax operators. The RTT-relation is an equation in the
tensor product of two fundamental auxiliary spaces
$V_\square=\mathbb{C}^n$ and $V_{\square'}=\mathbb{C}^n$, and the
total quantum space of the spin chain $V_1\otimes\cdots\otimes
V_\sites$. The monodromies $\mon(\spec)$ and $\mon'(\specp)$ act
respectively on the auxiliary spaces $V_{\square}$ and $V_{\square'}$,
and on the same total quantum space. The remaining R-matrix in
\eqref{eq:yi-ybe-rmm} is
\begin{align}
  \label{eq:yi-r-fund-fund}
  R_{\square\,\square'}(\spec-\specp)
  =
  1+(\spec-\specp)^{-1}\sum_{a,b=1}^ne_{ab}\,e'_{ba}\,,
\end{align}
where the sum in the last term is the permutation operator on
$\mathbb{C}^n\otimes\mathbb{C}^n$. Written in terms of the monodromy
elements \eqref{eq:yi-mono-elements}, the RTT-relation
\eqref{eq:yi-ybe-rmm} becomes
\begin{align}
  \label{eq:yi-rmm-comp}
  (\specp-\spec)[\mon_{ab}(\spec),\mon_{cd}(\specp)]
  =
  \mon_{cb}(\spec)\mon_{ad}(\specp)-\mon_{cb}(\specp)\mon_{ad}(\spec)\,.
\end{align}

Importantly, the RTT-relation \eqref{eq:yi-ybe-rmm} is the defining
relation of the \emph{Yangian algebra} $\mathcal{Y}(\mathfrak{gl}(n))$
in the QISM language, see e.g. \cite{Molev:2007}. A formal Laurent
expansion of the monodromy elements \eqref{eq:yi-mono-elements} in
inverse powers of the spectral parameter $\spec$,
\begin{align}
  \label{eq:yi-mono-expanded}
  \mon_{ab}(\spec)
  =
  \mon_{ab}^{(0)}+\mon_{ab}^{(1)}\spec^{-1}+\mon_{ab}^{(2)}\spec^{-2}
  +\ldots\,,
\end{align}
yields the generators $\mon_{ab}^{(r)}$ of the Yangian, where one demands
\begin{align}
  \label{eq:yi-yangian-gen-norm}
  \mon_{ab}^{(0)}=\delta_{ab}\,.
\end{align}
Inserting the expansion \eqref{eq:yi-mono-expanded} into
\eqref{eq:yi-rmm-comp}, one obtains the commutation relations for
these generators
\begin{align}
  \label{eq:yi-rmm-expanded}
  [\mon^{(r)}_{ab},\mon^{(s)}_{cd}]
  =
  \sum_{q=1}^{\Min(r,s)}
  \!\!
  \left(\mon_{cb}^{(r+s-q)}\mon_{ad}^{(q-1)}-\mon_{cb}^{(q-1)}\mon_{ad}^{(r+s-q)}\right)\,.
\end{align}
The formulation of the algebra in terms of $\mon_{ab}^{(r)}$
satisfying \eqref{eq:yi-rmm-expanded} is closely related to Drinfeld's
first realization of the Yangian
\cite{Drinfeld:1985rx,Drinfeld:1986in}. See also
\cite{Bernard:1992ya,MacKay:2004tc} for further reviews. Setting
$r=s=1$ in \eqref{eq:yi-rmm-expanded} shows that the
$-\mon_{ab}^{(1)}$ generate the $\mathfrak{gl}(n)$ symmetry, and the
generators with $r\geq 2$ correspond to its Yangian extension. We also
note that the monodromy elements transform in the adjoint
representation of this $\mathfrak{gl}(n)$ symmetry,
\begin{align}
  \label{eq:yi-rmm-gln-comp}
  [\mon^{(1)}_{ab},\mon_{cd}(\spec)]
  =
  \mon_{cb}(\spec)\delta_{ad}-\mon_{ad}(\spec)\delta_{cb}\,,
\end{align}
as may be seen by expanding \eqref{eq:yi-rmm-comp} in only one of the
two spectral parameters. Condition \eqref{eq:yi-yangian-gen-norm} is
satisfied up to a scalar factor by the monodromy \eqref{eq:yi-mono}
built out of the Lax operators \eqref{eq:yi-lax-fund-xi}. We use a
normalization of the Lax operators for which
\eqref{eq:yi-yangian-gen-norm} holds.

Finally, we come to the primary objective of this section, the role of
the main equation \eqref{eq:yi-eigenvalue-bra} of
section~\ref{sec:yi-monodromy} from a Yangian perspective. Let us
recall \eqref{eq:yi-eigenvalue-bra} in the more general context of the
current section.  Omitting the bra $\langle\stateset|$, the set of
eigenvalue equations\footnote{In \cite{deVega:1984wk,Destri:1993qh}
  such a set of equations is shown to be satisfied by the physical
  vacuum state of integrable two-dimensional quantum field theories.}
\begin{align}
  \label{eq:yi-inv-rmm}
  \mon_{ab}(\spec)|\Psi\rangle
  =
  \delta_{ab}|\Psi\rangle
\end{align}
may be graphically represented with the help of \eqref{eq:yi-mono} as
\begin{align}
  \label{eq:yi-inv-rmm-pic}
  \begin{aligned}
    \begin{tikzpicture}
      \draw[thick,densely dashed,
      decoration={
        markings, mark=at position 0.95 with {\arrow{latex reversed}}},
      postaction={decorate}
      ] 
      (0,0) 
      node[left=0.4cm] {$\square,\spec$}
      node[left] {$a$} -- 
      (3,0)
      node[right] {$b$};
      \draw[thick,
      decoration={
        markings, mark=at position 0.85 with {\arrow{latex reversed}}},
      postaction={decorate}
      ] 
      (0.5,-0.5) 
      node[below] {$\Xi_1,\inh_1$} -- 
      (0.5,0.5)
      node[above] {\phantom{$\Xi_1,\inh_1$}};
      \node at (1.5,-0.25) {$\ldots$};
      \draw[thick,
      decoration={
        markings, mark=at position 0.85 with {\arrow{latex reversed}}},
      postaction={decorate}
      ] 
      (2.5,-0.5) 
      node[below] {$\Xi_\sites,\inh_\sites$} -- 
      (2.5,0.5);
      \draw (1.5,1) 
      node[minimum height=1cm,minimum width=2.5cm,draw,
      thick,rounded corners=8pt,densely dotted] 
      {$|\Psi\rangle$};
      \path
      (0,2) 
      node[left] {\phantom{$\square,\spec$}} -- (3,2);
    \end{tikzpicture}
  \end{aligned}
  \,\,\,=
  \begin{aligned}
    \begin{tikzpicture}
      \draw[thick,densely dashed,
      decoration={
        markings, mark=at position 0.95 with {\arrow{latex reversed}}},
      postaction={decorate}
      ] 
      (0,2) 
      node[left=0.4cm] {$\square,\spec$}
      node[left] {$a$} -- 
      (3,2)
      node[right] {$b$};
      \draw[thick,
      decoration={
        markings, mark=at position 0.85 with {\arrow{latex reversed}}},
      postaction={decorate}
      ] 
      (0.5,-0.5) 
      node[below] {$\Xi_1,\inh_1$} -- 
      (0.5,0.5)
      node[above] {\phantom{$\Xi_1,\inh_1$}};
      \node at (1.5,-0.25) {\ldots};
      \draw[thick,
      decoration={
        markings, mark=at position 0.85 with {\arrow{latex reversed}}},
      postaction={decorate}
      ] 
      (2.5,-0.5) 
      node[below] {$\Xi_\sites,\inh_\sites$} -- 
      (2.5,0.5);
      \draw (1.5,1) 
      node[minimum height=1cm,minimum width=2.5cm,draw,
      thick,rounded corners=8pt,densely dotted] 
      {$|\Psi\rangle$};
    \end{tikzpicture}
  \end{aligned}\,.
\end{align}
Expanding in $\spec^{-1}$, we see that \eqref{eq:yi-inv-rmm} is equivalent
to
\begin{align}
  \label{eq:yi-inv-expanded}
  \mon^{(r)}_{ab}|\Psi\rangle=0
\end{align}
for $r\geq 1$. This means that $|\Psi\rangle$ forms a one-dimensional
representation of the Yangian as it is annihilated by all generators
$\mon^{(r)}_{ab}$. Hence we call $|\Psi\rangle$ \emph{Yangian
  invariant}. The observation that \eqref{eq:yi-inv-rmm} characterizes
Yangian invariants is the main result of this section. Compared to the
expanded version \eqref{eq:yi-inv-expanded}, which is on the level of
Drinfeld's first realization, equation \eqref{eq:yi-inv-rmm} has the
advantage that it may be understood within the QISM. As a result,
powerful mathematical tools become applicable. In particular, the
formulation \eqref{eq:yi-inv-rmm} will be exploited in
section~\ref{sec:bethe}, where this equation is solved using an
algebraic Bethe ansatz.

With this interpretation of \eqref{eq:yi-inv-rmm} the partition
function $\mathcal{Z}$ of a vertex model on a Baxter lattice discussed
in section~\ref{sec:yi-monodromy} is a component of a Yangian
invariant vector $|\Psi\rangle$, cf.\
\eqref{eq:yi-partition-psi}. However, a generic solution
$|\Psi\rangle$ of \eqref{eq:yi-inv-rmm} with a more general monodromy
$\mon(\spec)$ built from representations $\Xi_i$ and inhomogeneities
$\inh_i$ not obeying \eqref{eq:yi-ident-rep-inhomo} and possibly
containing an odd number of sites $\sites$ does \emph{not} correspond to a
partition function in the sense of
section~\ref{sec:yi-monodromy}. Hence $|\Psi\rangle$ is symbolized in
\eqref{eq:yi-inv-rmm-pic} by a dotted ``black box'' without specifying
the interior. In section~\ref{sec:osc-3vertices} we will indeed find
solutions of the Yangian invariance condition \eqref{eq:yi-inv-rmm}
that go beyond the Baxter lattices of
section~\ref{sec:yi-monodromy}. The graphical representation of these
solutions not only contains lines and four-valent vertices, i.e.\
R-matrices, but also trivalent-vertices, which are associated with
solutions of bootstrap equations, see \cite{Zamolodchikov:1989zs} and
e.g.\ \cite{Dorey:1997}, \cite{Samaj:2013}.

We end this section with a remark on a reformulation of Yangian
invariance. The condition in the form \eqref{eq:yi-inv-rmm} can
naturally be understood as an intertwining relation of the tensor
product of the first $\dsites$ with the remaining $\sites-\dsites$
spaces of the total quantum space. For this purpose we split the
monodromy \eqref{eq:yi-mono} as
\begin{align}
  \label{eq:yi-mono-split}
  \begin{aligned}
    \mon(\spec)
    =
    &R_{\square\,\Xi_1}(\spec-\inh_1)\cdots R_{\square\,\Xi_\dsites}(\spec-\inh_\dsites)\\
    &\cdot R_{\square\,\Xi_{\dsites+1}}(\spec-\inh_{\dsites+1})\cdots R_{\square\,\Xi_\sites}(\spec-\inh_\sites)\,.
  \end{aligned}
\end{align}
Conjugating \eqref{eq:yi-inv-rmm} in the first $\dsites$ spaces and using
\eqref{eq:yi-unitarity} and \eqref{eq:yi-crossing} for these spaces
yields the intertwining relation
\begin{align}
  \label{eq:yi-intertwiner}
  \begin{aligned}
    &R_{\square\,\Xi_{\dsites+1}}(\spec-\inh_{\dsites+1})\cdots R_{\square\,\Xi_\sites}(\spec-\inh_\sites)
    \mathcal{O}_\Psi\\
    &=
    \mathcal{O}_\Psi
    R_{\square\,\bar\Xi_\dsites}(\spec-\inh_\dsites+\kappa_{\Xi_\dsites})\cdots
    R_{\square\,\bar\Xi_1}(\spec-\inh_1+\kappa_{\Xi_1})\,,
  \end{aligned}
\end{align}
where $\mathcal{O}_\Psi:=|\Psi\rangle^{\dagger_1\cdots\dagger_\dsites}$.
This is depicted graphically as
\begin{align}
  \label{eq:yi-intertwiner-pic}
  \begin{aligned}
    \begin{tikzpicture}
      \draw[thick,densely dashed,
      decoration={
        markings, mark=at position 0.95 with {\arrow{latex reversed}}},
      postaction={decorate}
      ] 
      (0,0) 
      node[left] {$\square,\spec$} -- 
      (2.5,0);
      \node at (1.25,-0.25) {$\ldots$};
      \node at (1.25,2.25) {$\ldots$};
      \draw[thick,
      decoration={
        markings, mark=at position 0.85 with {\arrow{latex reversed}}},
      postaction={decorate}
      ] 
      (0.5,-0.5) 
      node[left=0.25cm,below] {$\Xi_{\dsites+1},\inh_{\dsites+1}$} -- 
      (0.5,0.5);
      \draw[thick,
      decoration={
        markings, mark=at position 0.85 with {\arrow{latex reversed}}},
      postaction={decorate}
      ] 
      (2,-0.5) 
      node[right=0.25cm,below] {$\Xi_\sites,\inh_\sites$} -- 
      (2,0.5);
      \draw[thick,
      decoration={
        markings, mark=at position 0.85 with {\arrow{latex reversed}}},
      postaction={decorate}
      ] 
      (2,1.5) -- 
      (2,2.5)
      node[right=0.5cm,above] {$\bar\Xi_1,\inh_1-\kappa_{\Xi_1}$};
      \draw[thick,
      decoration={
        markings, mark=at position 0.85 with {\arrow{latex reversed}}},
      postaction={decorate}
      ] 
      (0.5,1.5) -- 
      (0.5,2.5)
      node[left=0.5cm,above] {$\bar\Xi_\dsites,\inh_\dsites-\kappa_{\Xi_\dsites}$};
      \draw (1.25,1) 
      node[minimum height=1cm,minimum width=2cm,draw,
      thick,rounded corners=8pt,densely dotted] 
      {$\mathcal{O}_\Psi$};
    \end{tikzpicture}
  \end{aligned}
  =
  \begin{aligned}
    \begin{tikzpicture}
      \draw[thick,densely dashed,
      decoration={
        markings, mark=at position 0.95 with {\arrow{latex reversed}}},
      postaction={decorate}
      ] 
      (0,2) 
      node[left] {$\square,\spec$} -- 
      (2.5,2);
      \node at (1.25,-0.25) {$\ldots$};
      \node at (1.25,2.25) {$\ldots$};
      \draw[thick,
      decoration={
        markings, mark=at position 0.85 with {\arrow{latex reversed}}},
      postaction={decorate}
      ] 
      (0.5,-0.5) 
      node[left=0.25cm,below] {$\Xi_{\dsites+1},\inh_{\dsites+1}$} -- 
      (0.5,0.5);
      \draw[thick,
      decoration={
        markings, mark=at position 0.85 with {\arrow{latex reversed}}},
      postaction={decorate}
      ] 
      (2,-0.5) 
      node[right=0.25cm,below] {$\Xi_\sites,\inh_\sites$} -- 
      (2,0.5);
      \draw[thick,
      decoration={
        markings, mark=at position 0.85 with {\arrow{latex reversed}}},
      postaction={decorate}
      ] 
      (2,1.5) -- 
      (2,2.5)
      node[right=0.5cm,above] {$\bar\Xi_1,\inh_1-\kappa_{\Xi_1}$};
      \draw[thick,
      decoration={
        markings, mark=at position 0.85 with {\arrow{latex reversed}}},
      postaction={decorate}
      ] 
      (0.5,1.5) -- 
      (0.5,2.5)
      node[left=0.5cm,above] {$\bar\Xi_\dsites,\inh_\dsites-\kappa_{\Xi_\dsites}$};
      \draw (1.25,1) 
      node[minimum height=1cm,minimum width=2cm,draw,
      thick,rounded corners=8pt,densely dotted] 
      {$\mathcal{O}_\Psi$};
    \end{tikzpicture}
  \end{aligned}.
\end{align}
In case $\mathcal{O}_\Psi$ corresponds to the partition function
$\mathcal{Z}$ of a vertex model, this equation is nothing but a
consequence of Z-invariance, cf.\ \cite{Baxter:1978xr} and also
section~\ref{sec:pba-model}: The (dashed) fundamental auxiliary line
is moved through the entire Baxter lattice. An equation of the type
\eqref{eq:yi-intertwiner} also appeared recently in
\cite{Ferro:2013dga} in the context of a spectral parameter
deformation of planar $\mathcal{N}=4$ super Yang-Mills scattering
amplitudes. There it was referred to as ``generalized Yang-Baxter
equation''. In the scattering problem, Yangian invariance of
undeformed tree-level amplitudes is usually formulated in the sense of
\eqref{eq:yi-inv-expanded}, see \cite{Drummond:2009fd} and
e.g. \cite{Beisert:2010jq}. Bearing in mind our ambitions to construct
Yangian invariants using a Bethe ansatz in section~\ref{sec:bethe}, we
focus in this paper on \eqref{eq:yi-inv-rmm} instead of
\eqref{eq:yi-inv-expanded} or \eqref{eq:yi-intertwiner}.

\section{Yangian invariants in oscillator formalism}
\label{sec:osc}

In section~\ref{sec:yi-yangian} we characterized Yangian invariants by
the (system of) eigenvalue equations \eqref{eq:yi-inv-rmm} for matrix
elements of a monodromy and equivalently by the associated
intertwining relation \eqref{eq:yi-intertwiner}. Here we will begin
our study of \eqref{eq:yi-inv-rmm} by working out explicit solutions
$|\Psi\rangle$ in a number of concrete examples. We restrict our
analysis to monodromies $\mon(\spec)$, where the total quantum space
is built by tensoring finite-dimensional totally symmetric
representations $\s$ and their conjugates $\bs$, i.e.\ $\Xi_i=\s_i$ or
$\Xi_i=\bs_i$ for all $i=1, \ldots \sites$ in \eqref{eq:yi-mono}. We
need these conjugate representations to make sure that the total
quantum space contains a $\mathfrak{gl}(n)$ singlet, which is a
necessary criterion for Yangian invariants, see
\eqref{eq:yi-inv-expanded} for the case $r=1$.

The representations $\s$ and $\bs$ are realized in terms of oscillator
algebras, see section~\ref{sec:osc-rep-lax}. Since the non-zero
eigenvalues appearing in \eqref{eq:yi-inv-rmm} are identical to $1$,
the normalization of the Lax operators used in the construction is
clearly important and will be discussed in some detail. After that we
are in place to construct the sought solutions in
section~\ref{sec:osc-examples}. Our first and simplest examples are
the two-site monodromies of section~\ref{sec:osc-line}, where the
representations of the two sites are necessarily conjugate to each
other. The inhomogeneities are then fixed by demanding Yangian
invariance, i.e.\ \eqref{eq:yi-inv-rmm}. This solution $|\Psi\rangle$
is graphically represented by a Baxter lattice consisting of a single
line. In section~\ref{sec:osc-3vertices} we construct three-site
invariants. The corresponding intertwiner $\mathcal{O}_\Psi$
satisfying \eqref{eq:yi-intertwiner} is to be interpreted as a
solution of a bootstrap equation in analogy with
\cite{Zamolodchikov:1989zs}. Although these invariants leave the
framework of section~\ref{sec:yi-monodromy}, they are naturally
included in our definition of Yangian invariants. Finally, in
section~\ref{sec:osc-4vertex} we study the Yangian invariant related
to the first non-trivial Baxter lattice consisting of two intersecting
lines. The associated intertwiner $\mathcal{O}_\Psi$ contains a free
parameter $\inhdiff$ and actually turns out to be the
$\mathfrak{gl}(n)$ symmetric R-matrix $R_{\s\,\s'}(\inhdiff)$ for
arbitrary totally symmetric representations $\s$, $\s'$. We obtain a
compact expression for this R-matrix in a certain oscillator
basis. The spectral parameter $\inhdiff$ of the R-matrix should not be
confused with that of the auxiliary space in section \ref{sec:yi}
denoted by $\spec$.

\subsection{Oscillators, Lax operators and monodromies}
\label{sec:osc-rep-lax}

We start by specifying the two types of oscillator representations
$\s$ and $\bs$ of the $\mathfrak{gl}(n)$ algebra \eqref{eq:yi-gln},
which will be used for the local quantum spaces of the monodromy
\eqref{eq:yi-mono}. These representations are labeled by their highest
weight. Consider the totally symmetric representation of
$\mathfrak{gl}(n)$ with highest weight $\s=(s,0,\ldots,0)$, where $s$
is a non-negative integer. We build these representations from a
single family of oscillators $\osca_a$ with $a=1,\ldots,n$.
Furthermore, consider the $\mathfrak{gl}(n)$ representation with
highest weight $\bs=(0,\ldots,0,-s)$, and construct it using a second
family of $n$ oscillators $\oscb_a$. The $n^2$ generators $J_{ab}$ of
the representation $\s$ and the second set of $n^2$ generators $\bar
J_{ab}$ of $\bs$ are given by
\begin{align}
  \label{eq:osc-gen-s-bs}
  \begin{aligned}
    J_{ab}&=+\bar\osca_a\osca_b&
    \quad\text{with}\quad&&
    [\osca_a,\bar\osca_b]=\delta_{ab}\,,&&
    \osca_a|0\rangle=0\,,&&
    \bar\osca_a=\osca_a^\dagger\,,&\\
    \bar J_{ab}&=-\bar\oscb_b\oscb_a&
    \quad\text{with}\quad&&
    [\oscb_a,\bar\oscb_b]=\delta_{ab}\,,&&
    \oscb_a|0\rangle=0\,,&&
    \bar\oscb_a=\oscb_a^\dagger\,.&
  \end{aligned}
\end{align}
Commutators of oscillators that are not specified by these relations
vanish. See e.g.\ \cite{Biedenharn:1981} for a review of such
Jordan-Schwinger-type realizations of the $\mathfrak{gl}(n)$
algebra. The generators \eqref{eq:osc-gen-s-bs} act on the
representation spaces $V_{\s}$ and $V_{\bs}$. These spaces consist of
homogeneous polynomials of degree $s$ in, respectively, the creation
operators $\bar\osca_a$ and $\bar\oscb_a$ acting on the Fock vacuum
$|0\rangle$. Therefore the number operators
$\sum_{a=1}^n\bar\osca_a\osca_a$ and $\sum_{a=1}^n\bar\oscb_a \oscb_a$
both take the value $s$. The highest weight states in $V_\s$ and
$V_{\bs}$ are, respectively,
\begin{align}
  \label{eq:osc-hws}
  \begin{aligned}
    |\sigma\rangle&=(\bar\osca_1)^s|0\rangle
    &\quad\text{with}\quad&&
    J_{aa}|\sigma\rangle&=s\,\delta_{1\,a}|\sigma\rangle\,,\quad&
    J_{ab}|\sigma\rangle&=0\quad\text{for}\quad a<b\,,\\
    |\bar\sigma\rangle&=(\bar\oscb_n)^s|0\rangle
    &\quad\text{with}\quad&&
    \bar J_{aa}|\bar\sigma\rangle&=-s\,\delta_{n\,a}|\bar\sigma\rangle\,,\quad&
    \bar J_{ab}|\bar\sigma\rangle&=0\quad\text{for}\quad a<b\,.
  \end{aligned}
\end{align}
The representation $\bs$ is conjugate to
$\s$ in the sense of \eqref{eq:yi-conj-rep},
\begin{align}
  \label{eq:osc-gen-conj}
  \bar J_{ab}\big|_{\oscb_a\mapsto\osca_a}
  =
  -J_{ab}^\dagger\,.
\end{align}

The Lax operators \eqref{eq:yi-lax-fund-xi} for the two
representations defined in \eqref{eq:osc-gen-s-bs} read
\begin{align}
  \label{eq:osc-lax-fund-s}
  \begin{aligned}
    R_{\square\,\s}(\spec-\inh)
    &=
    f_{\s}(\spec-\inh)
    \Bigg(1+(\spec-\inh)^{-1}\sum_{a,b=1}^ne_{ab}\bar\osca_b\osca_a\Bigg)
    =
    \,\,\,\\\phantom{}
  \end{aligned}
  \begin{aligned}
    \begin{tikzpicture}
      \draw[thick,densely dashed,
      decoration={
        markings, mark=at position 0.85 with {\arrow{latex reversed}}},
      postaction={decorate}
      ] 
      (0,0) 
      node[left] {$\square,\spec$} -- 
      (1,0);
      \draw[thick,
      decoration={
        markings, mark=at position 0.85 with {\arrow{latex reversed}}},
      postaction={decorate}
      ] 
      (0.5,-0.5) 
      node[below] {$\s,\inh$} -- 
      (0.5,0.5);
    \end{tikzpicture}
  \end{aligned}
  \begin{aligned}
    ,\\\phantom{}
  \end{aligned}
  \\
  \label{eq:osc-lax-fund-bs}
  \begin{aligned}
    R_{\square\,\bs}(\spec-\inh)
    &=
    f_{\bs}(\spec-\inh)
    \Bigg(1-(\spec-\inh)^{-1}\sum_{a,b=1}^ne_{ab}\bar\oscb_a\oscb_b\Bigg)
    =
    \,\,\,\\\phantom{}
  \end{aligned}
  \begin{aligned}
    \begin{tikzpicture}
      \draw[thick,densely dashed,
      decoration={
        markings, mark=at position 0.85 with {\arrow{latex reversed}}},
      postaction={decorate}
      ] 
      (0,0) 
      node[left] {$\square,\spec$} -- 
      (1,0);
      \draw[thick,
      decoration={
        markings, mark=at position 0.85 with {\arrow{latex reversed}}},
      postaction={decorate}
      ] 
      (0.5,-0.5) 
      node[below] {$\bs,\inh$} -- 
      (0.5,0.5);
    \end{tikzpicture}
  \end{aligned}
  \begin{aligned}
    .\\\phantom{}
  \end{aligned}
\end{align}
As discussed in section~\ref{sec:yi-monodromy}, we require these Lax
operators to possess the properties of unitarity
\eqref{eq:yi-unitarity} and crossing \eqref{eq:yi-crossing}, which
will impose constraints on the normalizations $f_{\s}(\spec)$ and
$f_{\bs}(\spec)$. The first property \eqref{eq:yi-unitarity} yields
\begin{align}
  \label{eq:osc-unitarity}
  R_{\square\,\s}(\spec-\inh)R_{\s\,\square}(\inh-\spec)=1\,,
  \quad
  R_{\square\,\bs}(\spec-\inh)R_{\bs\,\square}(\inh-\spec)=1\,.
\end{align}
These equations contain the two additional Lax operators
\begin{align}
  \label{eq:osc-lax-sbs-fund}
  \begin{aligned}
    R_{\s\,\square}(\inh-\spec)
    =
    \,\,\,\\\phantom{}
  \end{aligned}
  \begin{aligned}
    \begin{tikzpicture}
      \draw[thick,
      decoration={
        markings, mark=at position 0.85 with {\arrow{latex reversed}}},
      postaction={decorate}
      ] 
      (0,0) 
      node[left] {$\s,\inh$} -- 
      (1,0);
      \draw[thick,densely dashed,
      decoration={
        markings, mark=at position 0.85 with {\arrow{latex reversed}}},
      postaction={decorate}
      ] 
      (0.5,-0.5) 
      node[below] {$\square,\spec$} -- 
      (0.5,0.5);
    \end{tikzpicture}
  \end{aligned}
  \begin{aligned}
    ,\\\phantom{}
  \end{aligned}
  \quad
  \begin{aligned}
    R_{\bs\,\square}(\inh-\spec)
    =
    \,\,\,\\\phantom{}
  \end{aligned}
  \begin{aligned}
    \begin{tikzpicture}
      \draw[thick,
      decoration={
        markings, mark=at position 0.85 with {\arrow{latex reversed}}},
      postaction={decorate}
      ] 
      (0,0) 
      node[left] {$\bs,\inh$} -- 
      (1,0);
      \draw[thick,densely dashed,
      decoration={
        markings, mark=at position 0.85 with {\arrow{latex reversed}}},
      postaction={decorate}
      ] 
      (0.5,-0.5) 
      node[below] {$\square,\spec$} -- 
      (0.5,0.5);
    \end{tikzpicture}
  \end{aligned}
\end{align}
with exchanged order of auxiliary and quantum space. These are
obtained as solutions of the Yang-Baxter equation in
$V_{\square}\otimes V_{\s}\otimes V_\square$ and $V_{\square}\otimes
V_{\bs}\otimes V_\square$, respectively, where they are the only
unknowns, cf.\ \eqref{eq:yi-ybe}. The Lax operators
\eqref{eq:osc-lax-sbs-fund} are symmetric in the sense that, up to a
shift of the spectral parameter, they can be expressed in terms of
\eqref{eq:osc-lax-fund-s} and \eqref{eq:osc-lax-fund-bs},
\begin{align}
  \label{eq:osc-lax-symm}
  R_{\s\,\square}(\inh-\spec)
  =
  R_{\square\,\s}(\inh-\spec-s+1)\,,\quad
  R_{\bs\,\square}(\inh-\spec)
  =
  R_{\square\,\bs}(\inh-\spec+n+s-1)\,.
\end{align}
Then the unitarity conditions in \eqref{eq:osc-unitarity} turn into
constraints on the normalization of the Lax operators,
\begin{gather}
  \label{eq:osc-norm-unitarity}
  f_{\s}(\spec)f_{\s}(-\spec-s+1)
  =
  \frac{\spec(\spec+s-1)}{\spec(\spec+s-1)-s}\,,\quad
  f_{\bs}(\spec)f_{\bs}(-\spec+s-1+n)=1\,.
\end{gather}
The other condition, the crossing relation \eqref{eq:yi-crossing},
reads for the representations $\s$ and $\bs$, respectively,
\begin{align}
  \label{eq:osc-crossing}
  R_{\square\,\bs}(\spec+\kappa_{\s})\big|_{\oscb_a\mapsto\osca_a}
  =
  R_{\s\,\square}(-\spec)^\dagger\,,\quad
  R_{\square\,\s}(\spec+\kappa_{\bs})
  =
  R_{\bs\,\square}(-\spec)^\dagger\big|_{\oscb_a\mapsto\osca_a}\,,
\end{align}
where $\kappa_{\s}$ and $\kappa_{\bs}$ are the crossing parameters.
These conditions imply
\begin{align}
  \label{eq:osc-crossing-norm}
  \kappa_{\s}=s-1\,,
  \quad
  \kappa_{\bs}=-s+1-n\,,
  \quad
  f_{\bs}(\spec)=f_{\s}(-\spec)\,.
\end{align}
Notice that the two equations in \eqref{eq:osc-crossing} lead to only
one constraint on the normalizations. Relations
\eqref{eq:osc-norm-unitarity} and \eqref{eq:osc-crossing-norm} are
solved by the well-known normalization
\begin{align}
  \label{eq:osc-norm-solution}
  f_{\s}(\spec)
  =
  \frac{
    \Gamma\big(\frac{1-\spec}{n}\big)
    \Gamma\big(\frac{n+\spec}{n}\big)
  }
  {
    \Gamma\big(\frac{1-s-\spec}{n}\big)
    \Gamma\big(\frac{n+s+\spec}{n}\big)
  }\,.
\end{align}
For $s=1$ this solution was obtained in \cite{Berg:1977dp}. The
solution for higher integer values of $s$ can be constructed using the
additional recursion relation
\begin{align}
  \label{eq:osc-norm-add-eq}
  f_\s(\spec) f_{\s'}(\spec+s)=f_{\s+\s'}(\spec)\,,
\end{align}
where $\s+\s'=(s+s',0,\ldots,0)$ denotes the addition of weights. Note
that the solution \eqref{eq:osc-norm-solution} is not unique.

Now, we concentrate on monodromies $\mon(\spec)$ of the form
\eqref{eq:yi-mono}, which are built entirely out of the two types of
Lax operators \eqref{eq:osc-lax-fund-s} and \eqref{eq:osc-lax-fund-bs}
with the proper normalization
\eqref{eq:osc-norm-solution}. Consequently, at the $i$-th site of the
monodromy the representation of the local quantum space is
$\Xi_i=\s_i$ or $\Xi_i=\bs_i$ and the oscillator families building
these representations are labeled $\osca_a^i$ or $\oscb_a^i$,
respectively. Further restricting to monodromies that allow for
solutions $|\Psi\rangle$ of the Yangian invariance condition
\eqref{eq:yi-inv-rmm}, one finds severe constraints on the
representation labels $s_i$ and inhomogeneities $\inh_i$.

One large class of such monodromies is obtained by considering Baxter
lattices in the sense of section~\ref{sec:yi-monodromy}, where each
line carries either a symmetric representation or a conjugate one.  If
the $k$-th line of the Baxter lattice with endpoints $\epe_k<\epb_k$
and spectral parameter $\rap_k$ carries a symmetric representation
labeled by $\Lambda_k=\s_{\epe_k}$, then according to
\eqref{eq:yi-ident-rep-inhomo} and using \eqref{eq:osc-crossing-norm}
the monodromy $\mon(\spec)$ contains two sites
\begin{align}
  \label{eq:osc-line-s}
  \begin{gathered}
  \Xi_{\epe_k}=\s_{\epe_k}\,,
  \quad
  \inh_{\epe_k}=\rap_k
  \quad
  \text{and}
  \quad
  \Xi_{\epb_k}=\bs_{\epb_k}\,,
  \quad
  \inh_{\epb_k}=\rap_k-s_{\epe_k}+1\\
  \text{with}
  \quad
  s_{\epe_k}=s_{\epb_k}\,.
  \end{gathered}
\end{align}
As a consequence, in $\mon(\spec)$ the Lax operator
$R_{\square\,\s_{\epe_k}}(\spec-\inh_{\epe_k})$ with the symmetric
representation is placed left of
$R_{\square\,\bs_{\epb_k}}(\spec-\inh_{\epb_k})$ with the conjugate
representation. If instead the $k$-th line carries the conjugate
representation $\Lambda_k=\bs_{\epe_k}$, we obtain from
\eqref{eq:yi-ident-rep-inhomo} with \eqref{eq:osc-crossing-norm}
\begin{align}
  \label{eq:osc-line-bs}
  \begin{gathered}
    \Xi_{\epe_k}=\bs_{\epe_k}\,,
    \quad
    \inh_{\epe_k}=\rap_k
    \quad
    \text{and}
    \quad
    \Xi_{\epb_k}=\s_{\epb_k}\,,
    \quad
    \inh_{\epb_k}=\rap_k+s_{\epe_k}-1+n\\
    \text{with}
    \quad
    s_{\epe_k}=s_{\epb_k}\,.
  \end{gathered}
\end{align}
In this case the Lax operator with the conjugate representation is to
the left of the one with the symmetric representation. In the
following, we will also study solutions $|\Psi\rangle$ of
\eqref{eq:yi-inv-rmm} where the representation labels and
inhomogeneities do not obey \eqref{eq:osc-line-s} or
\eqref{eq:osc-line-bs}. These do not correspond to a Baxter lattice in
the sense of section~\ref{sec:yi-monodromy}.

Let us comment on the normalization of the monodromies considered in
the remainder of section~\ref{sec:osc}. The constraints on their
representation labels and inhomogeneities guarantee that the gamma
functions in the normalizations of the different Lax operators cancel
and only a rational function in $\spec$ remains.

\subsection{Sample invariants}
\label{sec:osc-examples}

After these preparations, we are in a position to actually solve
\eqref{eq:yi-inv-rmm} in a number of simple examples. From now on, we
label the monodromies $\mon_{\sites,\dsites}(\spec)$ and the Yangian
invariants $|\Psi_{\sites,\dsites}\rangle$ by the total number of
sites $\sites$ and the number $\dsites$ of sites carrying a conjugate
representation of type $\bs$. This is motivated by
section~\ref{sec:amp}, where the invariant
$|\Psi_{\sites,\dsites}\rangle$ is compared with the $\sites$-particle
$\text{N}^{\dsites-2}\text{MHV}$ tree-level scattering amplitude of
planar $\mathcal{N}=4$ super Yang-Mills theory. In addition, we focus
on monodromies $\mon_{\sites,\dsites}(\spec)$ whose sites with
conjugate representations of type $\bs$ are all to the left of the
sites with $\s$. This order corresponds to the gauge fixing used in
the Graßmannian integral formulation in section \ref{sec:amp}.

\subsubsection{Line and identity operator}
\label{sec:osc-line}

The simplest Yangian invariant $|\Psi_{2,1}\rangle$ solving
\eqref{eq:yi-inv-rmm} corresponds to a Baxter lattice consisting of a
single line. In order to obtain the associated monodromy
$\mon_{2,1}(\spec)$ where the site with the conjugate representation
is situated to the left of the symmetric one, we choose the line in
the Baxter lattice to carry a conjugate representation, cf.\
\eqref{eq:osc-line-bs} and see figure~\ref{fig:osc-psi21}.
\begin{figure}[!t]
  \begin{center}  
    \begin{align*}
      \begin{aligned}
        \begin{tikzpicture}
          \draw[densely dashed,thick,
          decoration={
            markings, mark=at position 0.95 with {\arrow{latex reversed}}},
          postaction={decorate}]
          (0,0)
          node[left] {$a$}
          node[left=0.4cm] {$\square,\spec$} -- 
          (2.5,0)
          node[right] {$b$};
          \draw[thick,
          decoration={
            markings, mark=at position 0.5 with {\arrow{latex reversed}}},
          postaction={decorate}] 
          (0.5,-0.5)
          node[below] {$\alpha_1$}
          -- 
          (0.5,0.5) 
          node[above=1.2cm,right=0.2cm] {$\bs_1,\inh_1$} 
          .. controls (0.5,1.5) and (2.0,1.5) .. (2.0,0.5) -- 
          (2.0,-0.5)
          node[below] {$\alpha_2$};
        \end{tikzpicture}
      \end{aligned}
      \quad
      \begin{aligned}
        \mon_{2,1}(\spec)=\,\,\,\\\vphantom{}
      \end{aligned}
      \begin{aligned}
        \begin{tikzpicture}
          \draw[densely dashed,thick,
          decoration={
            markings, mark=at position 0.95 with {\arrow{latex reversed}}},
          postaction={decorate}]
          (0,0)
          node[left] {$\square,\spec$} -- 
          (2.5,0);
          \draw[thick,
          decoration={
            markings, mark=at position 0.85 with {\arrow{latex reversed}}},
          postaction={decorate}] 
          (0.5,-0.5)
          node[below] {$\bs_1,\inh_1$} -- 
          (0.5,0.5); 
          \draw[thick,
          decoration={
            markings, mark=at position 0.85 with {\arrow{latex reversed}}},
          postaction={decorate}] 
          (2.0,-0.5) 
          node[below] {$\vphantom{\bs_2}\s_2,\inh_2$} -- 
          (2.0,0.5); 
        \end{tikzpicture}
      \end{aligned}
    \end{align*}
    \caption{A Baxter lattice with one line specified by
      $\graph=((1,2))$, $\repset=(\bs_1)$, $\rapset=(\inh_1)$,
      $\stateset=(\alpha_1,\alpha_2)$, cf.\ \eqref{eq:yi-baxter-data},
      and intersected by a dashed auxiliary space, left part. This
      arrangement of Boltzmann weights corresponds to the l.h.s.\ of
      the invariance condition \eqref{eq:yi-inv-rmm} for
      $|\Psi_{2,1}\rangle$, i.e.\ to
      $\mon_{2,1}(\spec)|\Psi_{2,1}\rangle$. The elements of the
      monodromy $\mon_{2,1}(\spec)$ in the right part of the figure
      are obtained from the Boltzmann weights on the left side using
      the crossing relation \eqref{eq:yi-crossing-coord}. The
      representation labels and inhomogeneities of this monodromy obey
      \eqref{eq:osc-m21-vs}.}
    \label{fig:osc-psi21}
  \end{center} 
\end{figure} 

This leads to the length-two monodromy
\begin{align}
  \label{eq:osc-m21}
  \mon_{2,1}(\spec)
  =
  R_{\square\,\bs_1}(\spec-\inh_1)R_{\square\,\s_2}(\spec-\inh_2)
\end{align}
with the following constraints on the representation labels and
inhomogeneities:
\begin{align}
  \label{eq:osc-m21-vs}
    \inh_1=\inh_2-n-s_2+1\,,
    \quad
    s_1=s_2\,.
\end{align}
Recalling the Baxter lattice associated with this particular monodromy,
we happily notice that \eqref{eq:osc-m21-vs} agrees with
\eqref{eq:osc-line-bs}. The overall normalization of the monodromy
\eqref{eq:osc-m21} originating from those of the Lax operators
\eqref{eq:osc-lax-fund-s} and \eqref{eq:osc-lax-fund-bs} trivializes,
\begin{align}
  \label{eq:osc-m21-norm}
  f_{\bs_1}(\spec-\inh_1)f_{\s_2}(\spec-\inh_2)=1\,,
\end{align}
where we used \eqref{eq:osc-m21-vs} and subsequently the unitarity
condition for $f_{\bs}(\spec)$ in \eqref{eq:osc-norm-unitarity} and
the relation between the two normalizations $f_{\s}(\spec)$ and
$f_{\bs}(\spec)$ in \eqref{eq:osc-crossing-norm}. We can now easily
solve \eqref{eq:yi-inv-rmm} to obtain the explicit form of the
invariant,
\begin{align}
  \label{eq:osc-psi21}
  |\Psi_{2,1}\rangle
  =
  (\bar\oscb^1\cdot \bar\osca^2)^{s_2}|0\rangle
  \quad
  \text{with}
  \quad
  \bar\oscb^i\cdot\bar\osca^j:=\sum_{a=1}^n\bar\oscb^i_a\bar\osca^j_a\,,
\end{align} 
where we recall that the upper indices on the oscillators refer to the
sites of the monodromy. This solution is unique up to a scalar factor,
which clearly drops out of \eqref{eq:yi-inv-rmm}. To obtain the
intertwiner associated to the invariant $|\Psi_{2,1}\rangle$ we employ
\eqref{eq:yi-intertwiner} with $\dsites=1$ and use the value of the
crossing parameter $\kappa_{\bs_1}$ given in
\eqref{eq:osc-crossing-norm}. This leads to
\begin{align}
  \label{eq:osc-int-psi21}  
  R_{\square\,\s_2}(\spec-\inh_2)\mathcal{O}_{\Psi_{2,1}}
  =
  \mathcal{O}_{\Psi_{2,1}}R_{\square\,\s_1}(\spec-\inh_2)
\end{align}
with
\begin{align}
  \label{eq:osc-opsi21}
  \mathcal{O}_{\Psi_{2,1}}:=|\Psi_{2,1}\rangle^{\dagger_1}
  =
  \sum_{a_1,\ldots,a_{s_2}=1}^n
  \!\!\!
  \bar\osca^2_{a_1}\cdots\bar\osca^2_{a_{s_2}}
  |0\rangle\langle 0|
  \oscb^1_{a_1}\cdots\oscb^1_{a_{s_2}}\,.
\end{align} 
After identifying the representation spaces $V_{\s_1}$ and $V_{\s_2}$,
which is possible because of $s_1=s_2$ in \eqref{eq:osc-m21-vs}, we
see that $\mathcal{O}_{\Psi_{2,1}}$ reduces to $s_2!$ times the
identity operator.

\subsubsection{Three-vertices and bootstrap equations}
\label{sec:osc-3vertices}

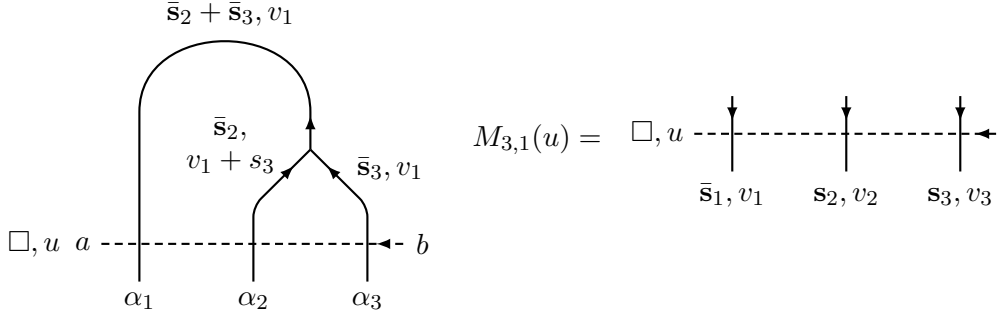
\begin{figure}[!t]
  \begin{center}
    \begin{align*}
      \begin{aligned}
        \begin{tikzpicture}
          \draw[densely dashed,thick,
          decoration={
            markings, mark=at position 0.95 with {\arrow{latex reversed}}},
          postaction={decorate}]
          (0,0)
          node[left] {$a$}
          node[left=0.4cm] {$\square,\spec$} --
          (4,0)
          node[right] {$b$};
          \draw[thick,rounded corners,
          decoration={markings, 
            mark=at position 0.85 with {\arrow{latex}}},postaction={decorate}]
          (2,-0.5)
          node[below] {$\alpha_2$} -- 
          (2,0.5) 
          node[left=0.3cm,above=0.3cm,align=center] 
          {$\bs_2,$\\[-4pt]$\inh_1+s_3$}
          -- (2.75,1.25);
          \draw[thick,rounded corners,
          decoration={markings, 
            mark=at position 0.85 with {\arrow{latex}}},postaction={decorate}]
          (3.5,-0.5) 
          node[below] {$\alpha_3$} -- 
          (3.5,0.5) 
          node[right=0.3cm,above=0.2cm] {$\bs_3,\inh_1$} 
          -- (2.75,1.25);
          \draw[thick,
          decoration={markings,
            mark=at position 0.075 with {\arrow{latex}}},postaction={decorate}] 
          (2.75,1.25) -- 
          (2.75,1.75) .. controls (2.75,3) and (0.5,3) .. (0.5,1.75) 
          node[right=1.2cm,above=1.0cm] {$\bs_2+\bs_3,\inh_1$} 
          -- 
          (0.5,-0.5)
          node[below] {$\alpha_1$};
        \end{tikzpicture}
      \end{aligned}
      \quad
      \begin{aligned}
        \mon_{3,1}(\spec)=\,\,\,\\\vphantom{}
      \end{aligned}
      \begin{aligned}
        \begin{tikzpicture}
          \draw[densely dashed,thick,
          decoration={
            markings, mark=at position 0.97 with {\arrow{latex reversed}}},
          postaction={decorate}]
          (0,0)
          node[left] {$\square,\spec$} --
          (4,0);
          \draw[thick,
          decoration={markings,
            mark=at position 0.85 with {\arrow{latex reversed}}},
          postaction={decorate}] 
          (0.5,-0.5)
          node[below] {$\vphantom{\bs_i}\bs_1,\inh_1$} -- 
          (0.5,0.5);
          \draw[thick,
          decoration={markings,
            mark=at position 0.85 with {\arrow{latex reversed}}},
          postaction={decorate}] 
          (2.0,-0.5)
          node[below] {$\vphantom{\bs_i}\s_2,\inh_2$} -- 
          (2.0,0.5);
          \draw[thick,
          decoration={markings,
            mark=at position 0.85 with {\arrow{latex reversed}}},
          postaction={decorate}] 
          (3.5,-0.5) 
          node[below] {$\vphantom{\bs_i}\s_3,\inh_3$} -- 
          (3.5,0.5);
        \end{tikzpicture}
      \end{aligned}
    \end{align*}
    \caption{The left part corresponds to the l.h.s.\ of
      \eqref{eq:yi-inv-rmm} for $|\Psi_{3,1}\rangle$, i.e.\
      $\mon_{3,1}(\spec)|\Psi_{3,1}\rangle$. It contains a (solid)
      trivalent vertex, which is an extension of the usual Baxter
      lattice, and a dashed auxiliary line. Using the crossing
      relation \eqref{eq:yi-crossing-coord} and the crossing
      parameters in \eqref{eq:osc-crossing-norm}, the Boltzmann
      weights involving the auxiliary line can be reformulated as
      elements of the monodromy $\mon_{3,1}(\spec)$, as is shown on the
      right side. The necessary constraints on the representation
      labels and inhomogeneities of the monodromy may be found in
      \eqref{eq:osc-m31-vs}.}
    \label{fig:osc-psi31}
  \end{center} 
\end{figure}

\begin{figure}[!t]
  \begin{center}
    \begin{align*}
      \begin{aligned}
        \begin{tikzpicture}
          \draw[densely dashed,thick,
          decoration={
            markings, mark=at position 0.95 with {\arrow{latex reversed}}},
          postaction={decorate}]
          (0,0)
          node[left] {$a$}
          node[left=0.4cm] {$\square,\spec$} -- 
          (4,0)
          node[right] {$b$};
          \draw[thick,rounded corners,
          decoration={markings,
            mark=at position 0.85 with {\arrow{latex reversed}}},
          postaction={decorate}]
          (0.5,-0.5) 
          node[below] {$\alpha_1$} -- 
          (0.5,0.5) 
          node[left=0.7cm,above=0.2cm] {$\bs_1,\inh_3'+s_2$} 
          -- (1.25,1.25);
          \draw[thick,rounded corners,
          decoration={markings, 
            mark=at position 0.85 with {\arrow{latex reversed}}},
          postaction={decorate}]
          (2,-0.5) 
          node[below] {$\alpha_2$} --
          (2,0.5)
          node[right=0.3cm,above=0.2cm] {$\bs_2,\inh_3'$} 
          -- (1.25,1.25);
          \draw[thick,
          decoration={markings, 
            mark=at position 0.06 with {\arrow{latex reversed}}},
          postaction={decorate}] 
          (1.25,1.25) -- 
          (1.25,1.75) .. controls (1.25,3) and (3.5,3) .. (3.5,1.75) 
          node[left=1.0cm,above=1.0cm,align=center] 
          {$\bs_1+\bs_2,$\\[-4pt]$\inh_3':=\inh_3-s_1-s_2+1-n$}
          -- 
          (3.5,-0.5)
          node[below] {$\alpha_3$};
        \end{tikzpicture}
      \end{aligned}
      \quad
      \begin{aligned}
        \mon_{3,2}(\spec)=\,\,\,\\\vphantom{}
      \end{aligned}
      \begin{aligned}
        \begin{tikzpicture}
          \draw[densely dashed,thick,
          decoration={
            markings, mark=at position 0.97 with {\arrow{latex reversed}}},
          postaction={decorate}]
          (0,0)
          node[left] {$\square,\spec$} --
          (4,0);
          \draw[thick,
          decoration={markings,
            mark=at position 0.85 with {\arrow{latex reversed}}},
          postaction={decorate}] 
          (0.5,-0.5)
          node[below] {$\vphantom{\bs_i}\bs_1,\inh_1$} -- 
          (0.5,0.5);
          \draw[thick,
          decoration={markings,
            mark=at position 0.85 with {\arrow{latex reversed}}},
          postaction={decorate}] 
          (2.0,-0.5)
          node[below] {$\vphantom{\bs_i}\bs_2,\inh_2$} -- 
          (2.0,0.5);
          \draw[thick,
          decoration={markings,
            mark=at position 0.85 with {\arrow{latex reversed}}},
          postaction={decorate}] 
          (3.5,-0.5) 
          node[below] {$\vphantom{\bs_i}\s_3,\inh_3$} -- 
          (3.5,0.5);
        \end{tikzpicture}
      \end{aligned}
    \end{align*}
    \caption{ The l.h.s.\ $\mon_{3,2}(\spec)|\Psi_{3,2}\rangle$ of
      \eqref{eq:yi-inv-rmm} for $|\Psi_{3,2}\rangle$ corresponds to
      the lattice in the left part. It consists of an extended Baxter
      lattice in form of a trivalent vertex and a dashed auxiliary
      space. The Boltzmann weights containing the auxiliary space can
      be formulated as elements of a monodromy $\mon_{3,2}(\spec)$ using
      the crossing relation \eqref{eq:yi-crossing-coord} with
      \eqref{eq:osc-crossing-norm}. This monodromy is shown in the
      right part and the parameters of the monodromy obey the
      constraints \eqref{eq:osc-m32-vs}.}
    \label{fig:osc-psi32}
  \end{center} 
\end{figure}
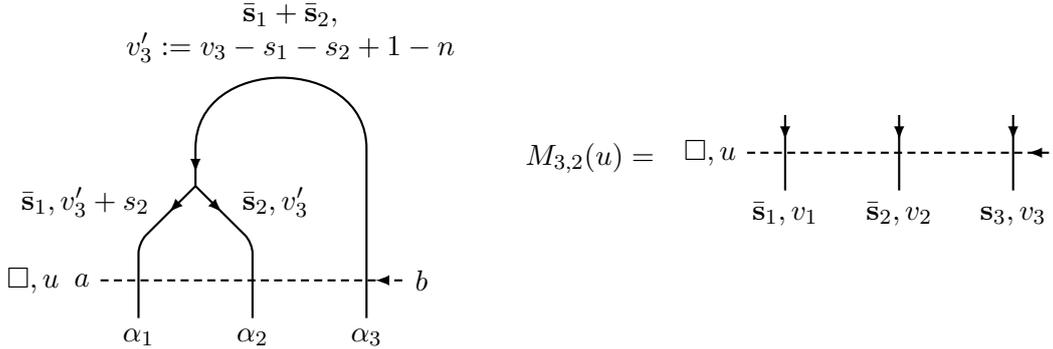

The next simplest Yangian invariants are characterized by monodromies
with three sites and are of the type $|\Psi_{3,1}\rangle$ or
$|\Psi_{3,2}\rangle$. We restrict once more to the case where the
sites with conjugate representations are to the left of those with
symmetric ones. These three-site invariants clearly leave the
framework of section~\ref{sec:yi-monodromy}. We represent them
graphically by an extension of the Baxter lattice, which in this case
consists of a trivalent vertex. See figures \ref{fig:osc-psi31} and
\ref{fig:osc-psi32} for the invariants $|\Psi_{3,1}\rangle$ and
$|\Psi_{3,2}\rangle$, respectively.

We start with a monodromy containing one conjugate site,
\begin{align}
  \label{eq:osc-m31}
  \mon_{3,1}(\spec)
  =
  R_{\square\,\bs_1}(\spec-\inh_1)
  R_{\square\,\s_2}(\spec-\inh_2)
  R_{\square\,\s_3}(\spec-\inh_3)\,,
\end{align}
see also the right part of figure~\ref{fig:osc-psi31}. Now the Yangian
invariance condition \eqref{eq:yi-inv-rmm} can be easily solved if the
parameters obey
\begin{equation}
  \label{eq:osc-m31-vs}
  \inh_2=\inh_1+n+s_2+s_3-1\,,
  \quad
  \inh_3=\inh_1+n+s_3-1\,,
  \quad
  s_1=s_2+s_3\,.
\end{equation}
In this case the normalizations of the Lax operators of types
\eqref{eq:osc-lax-fund-s} and \eqref{eq:osc-lax-fund-bs} appearing in
\eqref{eq:osc-m31} trivializes using the relation
\eqref{eq:osc-norm-add-eq} for $f_{\s}(\spec)$, the unitarity
condition for $f_{\bs}(\spec)$ and finally expressing $f_{\bs}(\spec)$
in terms of $f_{s}(\spec)$ with the help of
\eqref{eq:osc-crossing-norm}:
\begin{align}
  \label{eq:osc-m31-norm}
  f_{\bs_1}(\spec-\inh_1)f_{\s_2}(\spec-\inh_2)f_{\s_3}(\spec-\inh_3)=1\,.
\end{align}
Then one immediately checks that the solution of \eqref{eq:yi-inv-rmm}
is given by
\begin{equation}
  \label{eq:osc-psi31}
  |\Psi_{3,1}\rangle
  =
  (\bar\oscb^1\cdot\bar\osca^2)^{s_2}
  (\bar\oscb^1\cdot\bar\osca^3)^{s_3}
  |0\rangle\,,
\end{equation}
where we fixed a possible scalar prefactor. We once again proceed to
the corresponding intertwining relation. From its general form in
\eqref{eq:yi-intertwiner} we obtain for $\dsites=1$ and $\kappa_{\bs_1}$
given by \eqref{eq:osc-crossing-norm} the relation
\begin{equation}
  \label{eq:osc-int-psi31}
  R_{\square\,\s_2}(\spec-\inh_2)
  R_{\square\,\s_3}(\spec-\inh_2+s_2)
  \mathcal{O}_{\Psi_{3,1}}
  =
  \mathcal{O}_{\Psi_{3,1}}
  R_{\square\,\s_1}(\spec-\inh_2)
\end{equation} 
with
\begin{equation}
  \label{eq:osc-opsi31}
  \mathcal{O}_{\Psi_{3,1}}
  :=
  |\Psi_{3,1}\rangle^{\dagger_1}
  =
  \sum_{\substack{a_1,\ldots,a_{s_2}\\b_1,\ldots,b_{s_3}}}
  \!\!\!
  \bar\osca_{a_1}^2\cdots\bar\osca_{a_{s_2}}^2
  \bar\osca_{b_1}^3\cdots\bar\osca_{b_{s_3}}^3
  |0\rangle
  \langle 0|
  \oscb_{a_1}^1\cdots\oscb_{a_{s_2}}^1
  \oscb_{b_1}^1\cdots\oscb_{b_{s_3}}^1\,.
\end{equation}
The intertwining relation \eqref{eq:osc-int-psi31} is known as
bootstrap equation.

We move on to a monodromy with two conjugate sites on the left,
\begin{equation}
  \label{eq:osc-m32}  
  \mon_{3,2}(\spec)
  =
  R_{\square\,\bs_1}(\spec-\inh_1)
  R_{\square\,\bs_2}(\spec-\inh_2)
  R_{\square\,\s_3}(\spec-\inh_3)\,,
\end{equation}
see also the right part of figure~\ref{fig:osc-psi32}. Looking for
solutions $|\Psi_{3,2}\rangle$ of \eqref{eq:yi-inv-rmm} with this
monodromy again leads to constraints on the representation labels and
inhomogeneities,
\begin{equation}
  \label{eq:osc-m32-vs}
  \inh_1=\inh_3-n-s_1+1\,,
  \quad
  \inh_2=\inh_3-n-s_3+1\,,
  \quad
  s_3=s_1+s_2\,.
\end{equation}
Analogously to the discussion of the other three-site invariant, the
normalization of the monodromy \eqref{eq:osc-m32} trivializes using
\eqref{eq:osc-m32-vs}:
\begin{align}
  \label{eq:osc-m32-norm}
  f_{\bs_1}(\spec-\inh_1)f_{\bs_2}(\spec-\inh_2)f_{\s_3}(\spec-\inh_3)=1\,.
\end{align}
The explicit expression for the solution of \eqref{eq:yi-inv-rmm}
turns out to be
\begin{align}
  \label{eq:osc-psi32}
  |\Psi_{3,2}\rangle
  =
  (\bar\oscb^1\cdot\bar\osca^3)^{s_1}
  (\bar\oscb^2\cdot\bar\osca^3)^{s_2}
  |0\rangle\,.
\end{align}
Again we fixed a scalar prefactor. We employ the intertwining relation
\eqref{eq:yi-intertwiner} in this case with $\dsites=2$ and
$\kappa_{\bs_1}$, $\kappa_{\bs_2}$ specified in
\eqref{eq:osc-crossing-norm} to derive the bootstrap equation
\begin{equation}
  \label{eq:osc-int-psi32}
  R_{\square\,\s_3}(\spec-\inh_3)
  \mathcal{O}_{\Psi_{3,2}}
  =
  \mathcal{O}_{\Psi_{3,2}}
  R_{\square\,\s_2}(\spec-\inh_3+s_1)
  R_{\square\,\s_1}(\spec-\inh_3)
\end{equation}
with the solution
\begin{equation}
  \label{eq:osc-opsi32}
  \mathcal{O}_{\Psi_{3,2}}
  :=
  |\Psi_{3,2}\rangle^{\dagger_1\dagger_2}
  =
  \sum_{\substack{a_1,\ldots,a_{s_1}\\b_1,\ldots,b_{s_2}}}
  \!\!\!
  \bar\osca_{a_1}^3\cdots\bar\osca_{a_{s_1}}^3
  \bar\osca_{b_1}^3\cdots\bar\osca_{b_{s_2}}^3
  |0\rangle \langle 0|
  \oscb_{a_1}^1\cdots\oscb_{a_{s_1}}^1
  \oscb_{b_1}^2\cdots\oscb_{b_{s_2}}^2\,.
\end{equation}

\subsubsection{Four-vertex and Yang-Baxter equation}
\label{sec:osc-4vertex}

Let us proceed to Yangian invariants associated with four-site
monodromies. As an important check of our formalism we will rederive
the well-known $\mathfrak{gl}(n)$ invariant
R-matrix~\cite{Kulish:1981gi}. We therefore leave aside the rather
trivial cases where the Baxter lattice consists of two
non-intersecting lines, and focus on the invariants of type
$|\Psi_{4,2}\rangle$, where the Baxter lattice is a four-vertex. Once
again, we may a priori vary the positions of the conjugate sites
within the monodromy. We picked a particular assignment, where all
sites with conjugate representations are left of those with symmetric
representations, see figure \ref{fig:osc-psi42}.

We use the four-site monodromy
\begin{equation}
  \label{eq:osc-m42}
  \mon_{4,2}(\spec)
  =
  R_{\square\,\bs_1}(\spec-\inh_1)R_{\square\,\bs_2}(\spec-\inh_2)
  R_{\square\,\s_3}(\spec-\inh_3)R_{\square\,\s_4}(\spec-\inh_4)
\end{equation}
with
\begin{equation}
  \label{eq:osc-m42-vs}
  \inh_1=\inh_3-n-s_1+1\,,
  \quad
  \inh_2=\inh_4-n-s_2+1\,,
  \quad
  s_1=s_3\,,
  \quad
  s_2=s_4\,.
\end{equation}
This identification of the inhomogeneities and representation labels
corresponds to a Baxter lattice with two lines of type
\eqref{eq:osc-line-bs}. In order to simplify the normalizations of the
Lax operators in \eqref{eq:osc-m42}, we note that the relations in
\eqref{eq:osc-m42-vs} are two sets of conditions of the form appearing
for the two site-invariant in \eqref{eq:osc-m21-vs}. Hence the
normalization factors are simplified analogously to the discussion in
section~\ref{sec:bethe-gl2-sol-line}, which leads to
\begin{align}
  \label{eq:osc-m42-norm}
  f_{\bs_1}(\spec-\inh_1)f_{\bs_2}(\spec-\inh_2)
  f_{\s_3}(\spec-\inh_3)f_{\s_4}(\spec-\inh_4)=1\,.
\end{align}
For the solution of the eigenvalue equation \eqref{eq:yi-inv-rmm} with
this monodromy we make the $\mathfrak{gl}(n)$ invariant ansatz
\begin{equation}
  \label{eq:osc-psi42}
  |\Psi_{4,2}(\inh_3-\inh_4)\rangle
  :=
  |\Psi_{4,2}\rangle
  =
  \sum_{k=0}^{\Min(s_3,s_4)}
  \!\!\!
  d_k(\inh_3-\inh_4)
  |\Upsilon_k\rangle
\end{equation}
with
\begin{align}
  \label{eq:osc-phi}
  |\Upsilon_k\rangle
  =
  (\bar\oscb^1\cdot\bar\osca^3)^{s_3-k}
  (\bar\oscb^2\cdot \bar\osca^4)^{s_4-k}
  (\bar\oscb^2\cdot \bar\osca^3)^{k}
  (\bar\oscb^1\cdot \bar\osca^4)^{k}
  |0\rangle\,.
\end{align}
In our formalism, the spectral parameter dependence of this four-site
invariant emerges in a natural fashion as the difference of two
inhomogeneities which from now on is denoted by
\begin{align}
  \label{eq:osc-diffinh}
  \inhdiff:=\inh_3-\inh_4.
\end{align}
We have made this manifest by using the notation
$|\Psi_{4,2}(\inhdiff)\rangle$. Substituting \eqref{eq:osc-psi42} into
\eqref{eq:yi-inv-rmm} yields a recursion relation for the coefficients
$d_k$,
\begin{equation}
  \label{eq:osc-recrel}
  \frac{d_k(\inhdiff)}{d_{k+1}(\inhdiff)}
  =
  \frac{(k+1)(\inhdiff-s_3+k+1)}{(s_3-k)(s_4-k)}\,.
\end{equation}
It is solved, up to a function periodic in the index $k$ with period
$1$, by
\begin{equation}
  \label{eq:osc-coeff}
 d_k(\inhdiff)
 =
 \frac{1}{(s_3-k)!(s_4-k)!k!^2}\,
 \frac{k!}{\Gamma(\inhdiff-s_3+k+1)}\,.
\end{equation}

\begin{figure}[!t]
  \begin{center}
    \begin{align*}
      \begin{aligned}
        \begin{tikzpicture}
          \draw[densely dashed,thick,
          decoration={
            markings, mark=at position 0.97 with {\arrow{latex reversed}}},
          postaction={decorate}]
          (0,0)
          node[left=0.4cm] {$\square,\spec$}
          node[left] {$a$} -- 
          (5.5,0)
          node[right] {$b$};
          \draw[thick,decoration={markings, 
            mark=at position 0.5 with {\arrow{latex reversed}}},
          postaction={decorate}] 
          (0.5,-0.5) 
          node[below] {$\alpha_1$} -- 
          (0.5,0.5) 
          node[right=1.4cm,above=1.4cm] {$\bs_1,\inh_1$}
          .. controls (0.5,2.25) and (3.5,2.25) .. (3.5,0.5) -- 
          (3.5,-0.5) 
          node[below] {$\alpha_3$};
          \draw[thick,
          decoration={markings, 
            mark=at position 0.5 with {\arrow{latex reversed}}},
          postaction={decorate}] 
          (2,-0.5) 
          node[below] {$\alpha_2$} -- 
          (2,0.5)
          node[right=1.4cm,above=1.4cm] {$\bs_2,\inh_2$}
          .. controls (2,2.25) and (5,2.25) .. (5,0.5) -- 
          (5,-0.5) 
          node[below] {$\alpha_4$};
        \end{tikzpicture}
      \end{aligned}
      \quad
      \begin{aligned}
        &
        \begin{aligned}
          \mon_{4,2}(\spec)=\\[-8pt]\phantom{}
        \end{aligned}\\
        &
        \begin{aligned}
          \begin{tikzpicture}
            \draw[densely dashed,thick,
            decoration={
              markings, mark=at position 0.97 with {\arrow{latex reversed}}},
            postaction={decorate}]
            (0,0)
            node[left] {$\square,\spec$} --
            (5.5,0);
            \draw[thick,
            decoration={markings,
              mark=at position 0.85 with {\arrow{latex reversed}}},
            postaction={decorate}] 
            (0.5,-0.5)
            node[below] {$\vphantom{\bs_i}\bs_1,\inh_1$} -- 
            (0.5,0.5);
            \draw[thick,
            decoration={markings,
              mark=at position 0.85 with {\arrow{latex reversed}}},
            postaction={decorate}] 
            (2.0,-0.5)
            node[below] {$\vphantom{\bs_i}\bs_2,\inh_2$} -- 
            (2.0,0.5);
            \draw[thick,
            decoration={markings,
              mark=at position 0.85 with {\arrow{latex reversed}}},
            postaction={decorate}] 
            (3.5,-0.5) 
            node[below] {$\vphantom{\bs_i}\s_3,\inh_3$} -- 
            (3.5,0.5);
            \draw[thick,
            decoration={markings,
              mark=at position 0.85 with {\arrow{latex reversed}}},
            postaction={decorate}] 
            (5.0,-0.5) 
            node[below] {$\vphantom{\bs_i}\s_4,\inh_4$} -- 
            (5.0,0.5);
          \end{tikzpicture}
        \end{aligned}
      \end{aligned}
    \end{align*}
  \end{center}
  \caption{The left part shows a Baxter lattice with two lines
    specified by $\graph=((1,3),(2,4))$, $\repset=(\bs_1,\bs_2)$,
    $\rapset=(\inh_1,\inh_2)$,
    $\stateset=(\alpha_1,\alpha_2,\alpha_3,\alpha_4)$ and a dashed
    auxiliary space. It corresponds to
    $\mon_{4,2}(\spec)|\Psi_{4,2}\rangle$ as the l.h.s.\ of
    \eqref{eq:yi-inv-rmm}. The right part contains the monodromy
    $\mon_{4,2}(\spec)$, which is associated with this Baxter
    lattice. The necessary identifications of the representation
    labels and the inhomogeneities are written in
    \eqref{eq:osc-m42-vs}.}
  \label{fig:osc-psi42}
\end{figure}

Following the same logic as before we obtain the equation, which
determines the intertwiner corresponding to
$|\Psi_{4,2}(\inhdiff)\rangle$, from \eqref{eq:yi-intertwiner} with
$\dsites=2$ and $\kappa_{\bs_1}$, $\kappa_{\bs_2}$ found in
\eqref{eq:osc-crossing-norm}. This yields the Yang-Baxter equation in
the form
\begin{equation}
  \label{eq:osc-int-psi42}
 R_{\square\,\s_3}(\spec-\inh_3)
 R_{\square\,\s_4}(\spec-\inh_4)
 \mathcal{O}_{\Psi_{4,2}(\inhdiff)}
 =
 \mathcal{O}_{\Psi_{4,2}(\inhdiff)}
 R_{\square\,\s_2}(\spec-\inh_4)
 R_{\square\,\s_1}(\spec-\inh_3)\,,
\end{equation} 
where
\begin{equation}
  \label{eq:osc-opsi42}
  \mathcal{O}_{\Psi_{4,2}(\inhdiff)}
  :=
  |\Psi_{4,2}(\inhdiff)\rangle^{\dagger_1\dagger_2}
  =
  \sum_{k=0}^{\Min(s_3,s_4)}d_k(\inhdiff)
  \mathcal{O}_{\Upsilon_k}\,,
\end{equation}
with
\begin{align}
  \label{eq:osc-ophi}
  \begin{aligned}
    \mathcal{O}_{\Upsilon_k}
    :=
    |\Upsilon_k\rangle^{\dagger_1\dagger_2}
    =
    \smash{\sum_{\substack{a_1,\ldots,a_{s_3}\\b_1,\ldots,b_{s_4}}}}
    &\bar\osca^3_{a_1}\cdots\bar\osca^3_{a_{s_3}}
    \bar\osca^4_{b_1}\cdots\bar\osca^4_{b_{s_4}}
    |0\rangle\\
    &\quad\cdot
    \langle 0| \oscb^1_{a_1}\cdots\oscb^1_{a_{s_3-k}}
    \oscb^1_{b_{s_4-k+1}}\cdots\oscb^1_{b_{s_4}}\\
    &\quad\quad\cdot
    \oscb^2_{b_1}\cdots\oscb^2_{b_{s_4-k}}
    \oscb^2_{a_{s_3-k+1}}\cdots\oscb^2_{a_{s_3}}\,.
  \end{aligned}
\end{align}
In order to rewrite this form of the Yang-Baxter equation in a more
standard way, we identify space $V_{\s_1}$ with $V_{\s_3}$ and
$V_{\s_2}$ with $V_{\s_4}$, and simultaneously
$\mathcal{O}_{\Psi_{4,2}(\inhdiff)}$ with
$R_{\s_3\,\s_4}(\inhdiff)$. This then yields
\begin{equation}
  \label{eq:osc-psi42-ybe}
  \begin{aligned}
    R_{\square\,\s_3}(\spec-\inh_3)
    R_{\square\,\s_4}(\spec-\inh_4)
    R_{\s_3\,\s_4}(\inhdiff)
    =
    R_{\s_3\,\s_4}(\inhdiff)
    R_{\square\,\s_4}(\spec-\inh_4)
    R_{\square\,\s_3}(\spec-\inh_3)\,.
  \end{aligned}
\end{equation} 
Indeed, \eqref{eq:osc-psi42-ybe} establishes that
$R_{\s_3\,\s_4}(\inhdiff)$ must be the $\mathfrak{gl}(n)$ invariant
R-matrix \cite{Kulish:1981gi} for symmetric representations.

In our approach $R_{\s_3\,\s_4}(\inhdiff)$ is expressed in an oscillator
basis. To be as explicit as possible, it is convenient to introduce
the hopping operators\footnote{This formalism has been developed in
  joint discussions with Tomek {\L}ukowski. See \cite{Ferro:2013dga},
  where these hopping operators are also employed.}
\begin{equation}
  \label{eq:osc-hop}
  \text{Hop}_k
  =
  \frac{1}{k!^2}
  \sum_{\substack{a_1,\ldots,a_k\\b_1,\ldots,b_k}}
  \!
  \bar\osca^3_{a_1}\cdots\bar\osca^3_{a_k}
  \bar\osca^4_{b_1}\cdots\bar\osca^4_{b_k}
  \osca^3_{b_1}\cdots\osca^3_{b_k}
  \osca^4_{a_1}\cdots\osca^4_{a_k}\,.
\end{equation}
On $V_{\s_3}\otimes V_{\s_4}$ the operator $\text{Hop}_k$ agrees with
$\mathcal{O}_{\Upsilon_k}$, after the said identification of spaces,
up to a trivial combinatorial factor.  This hopping basis allows us to
express the R-matrix in the form
\begin{equation}
  \label{eq:osc-hoppingr}
  R_{\s_3\,\s_4}(\inhdiff)
  =
  \sum_{k=0}^{\Min(s_3,s_4)}
  \!\!\!
  \frac{k!}{\Gamma(\inhdiff-s_3+k+1)}\,\text{Hop}_k\,.
\end{equation} 
The operator $\text{Hop}_k$ produces a sum of states containing all
possibilities to exchange $k$ of the oscillators in space $V_{\s_3}$
with $k$ of the oscillators in space $V_{\s_4}$, i.e.\ it ``hops'' $k$
oscillators between the two spaces. See also \cite{Ferro:2013dga} for
its supersymmetric and non-compact version. Note that we can extend
the summation range in \eqref{eq:osc-hoppingr} to infinity as
$\text{Hop}_k$ with $k>\Min(s_3,s_4)$ will annihilate any state.
Note also that in \eqref{eq:osc-hoppingr} the dependence on the
representation labels of the coefficients can be absorbed by a shift
of the spectral parameter. Taken in conjunction, these two
observations allow to interpret the expression \eqref{eq:osc-hoppingr}
in a way that does not depend on a specific symmetric representation
$\s$.

Apart from the invariant \eqref{eq:osc-psi42} discussed so far, which
corresponds to the R-matrix, there exists another class of invariants
based on the monodromy \eqref{eq:osc-m42}. Relaxing the conditions in
\eqref{eq:osc-m42-vs} one finds further solutions with
$s_1+s_2=s_3+s_4$. However, in the general case with $s_1\neq s_3$
these invariants do not depend on a free complex spectral parameter.

\section{Toy model for super Yang-Mills scattering amplitudes}
\label{sec:amp}

The main result of section~\ref{sec:osc} is summarized by explicit
formulas for the sample invariants \eqref{eq:osc-psi21},
\eqref{eq:osc-psi31}, \eqref{eq:osc-psi32} and \eqref{eq:osc-psi42} of
the Yangian $\mathcal{Y}(\mathfrak{gl}(n))$. The aim of this section
is to establish a relation between these expressions and tree-level
scattering amplitudes of planar $\mathcal{N}=4$ super Yang-Mills
theory, which will often simply be referred to as ``scattering
amplitudes''. See e.g.\ \cite{Drummond:2011ic} for a recent review of
the latter.

The essential connection between the expressions of
section~\ref{sec:osc}, which are formulated using oscillator algebras,
and these amplitudes is Yangian invariance. For the amplitudes this
was shown in \cite{Drummond:2009fd} employing spinor-helicity
variables.\footnote{Special diligence is required if the particle
  momenta are not in generic position, but there are collinear
  particles \cite{Bargheer:2009qu}, see also \cite{Bargheer:2011mm}.}
A formal relation between these variables and certain oscillators was
indicated in \cite{Beisert:2010jq}.  Nevertheless, the Yangian is
different in both cases. In the present paper we are focusing on
finite-dimensional representations of $\mathcal{Y}(\mathfrak{gl}(n))$
and not on the infinite-dimensional representation of the Yangian of
$\mathfrak{psu}(2,2|4)\subset\mathfrak{gl}(4|4)$, which is the one
relevant for amplitudes. Furthermore, at first sight the said formulas
of section~\ref{sec:osc} look somewhat different from known
expressions for super Yang-Mills scattering amplitudes.

In order to compare and relate these two different types of Yangian
invariants, it turns out to be most appropriate to formulate the
scattering amplitudes as Graßmannian integrals in terms of super
twistors \cite{ArkaniHamed:2012nw}. In these variables the generators
of the superconformal algebra, i.e.\ the lowest level Yangian
generators, are realized as first order differential operators
\cite{Witten:2003nn}. The Yangian invariance of these integrals was
proven in \cite{Drummond:2010qh}, see also
\cite{Drummond:2010uq,Ferro:2011ph,Korchemsky:2010ut}. Furthermore,
the super twistor variables together with the associated differential
operators obey the commutation relations of the oscillator algebra. In
the way the invariants of section~\ref{sec:osc} are formulated within
the framework of the QISM, they naturally contain spectral parameters
in the form of inhomogeneities. Hence, we are led to compare these
invariants to a recent spectral parameter deformation
\cite{Ferro:2012xw,Ferro:2013dga} of these amplitudes.

Those aspects of the Graßmannian integral for undeformed and deformed
scattering amplitudes which are important for our discussion are
briefly summarized in section~\ref{sec:amp-grass}. In
section~\ref{sec:amp-osc} we reformulate the invariants obtained in
section~\ref{sec:osc} with the aim of comparing them to the deformed
amplitudes. As a first step, the invariants are expressed as
multi-dimensional contour integrals over exponential functions of
creation operators. Next, the oscillator algebras are realized in
terms of multiplication and differentiation operators, see
appendix~\ref{sec:bargmann} for details. This turns the exponential
functions into certain delta functions, which are characteristic of
Graßmannian integrals.

Rewritten in this way, the Yangian invariants of section~\ref{sec:osc}
are essentially $\mathfrak{gl}(n)$ analogues of the deformed
tree-level scattering amplitudes of planar $\mathcal{N}=4$ super
Yang-Mills theory. Hence, we may think of them as a ``toy model'' for
scattering amplitudes. Note that we will be able to explicitly specify
the multi-dimensional integration contour for the sample invariants at
hand.

\subsection{Graßmannian integral for (deformed) scattering
  amplitudes}
\label{sec:amp-grass}

All tree-level scattering amplitudes of planar $\mathcal{N}=4$ super
Yang-Mills theory can be packaged into a single compact Graßmannian
integral formula using super twistor variables, see
\cite{ArkaniHamed:2012nw} for a recent formulation, and
\cite{ArkaniHamed:2009dn} for the original proposal. In this formalism
the undeformed $\sites$-point $\text{N}^{\dsites-2}\text{MHV}$
amplitude is given by
\begin{align}
  \label{eq:amp-grassint}
  \mathcal{A}_{\sites,\dsites}
  =
  \int
  \frac{\prod_{k=1}^\dsites\prod_{i=\dsites+1}^\sites\D\!c_{ki}}
  {(1\hdots \dsites)
    (2\hdots \dsites+1)
    \cdots 
    (\sites\hdots \sites+\dsites-1)}
  \prod_{k=1}^\dsites
  \delta^{4|4}
  \Bigg(
  \mathcal{W}^k+\sum_{i=\dsites+1}^\sites\!\!\! c_{ki}\mathcal{W}^i
  \Bigg)\,.
\end{align}
These amplitudes are organized by the deviation $\dsites-2$ from the
maximally helicity violating (MHV) configuration. The minor $(i\ldots
i+\dsites)$ , i.e.\ the $\dsites\times \dsites$ subdeterminant, is
built from the columns $i,\ldots,i+\dsites$ of the $\dsites\times
\sites$ matrix
\begin{align}
  \label{eq:amp-gaugefix}
  \begin{pmatrix}
    1&&0&c_{1\,\dsites+1}&\hdots&c_{1\sites}\\
    &\ddots&&\vdots&&\vdots\\
    0&&1&c_{\dsites\,\dsites+1}&\hdots&c_{\dsites\sites}\\
  \end{pmatrix}.
\end{align}
A $\mathfrak{gl}(\dsites)$ gauge symmetry of the Graßmannian integral
\eqref{eq:amp-grassint} has already been fixed by the choice of the
first $\dsites$ columns in \eqref{eq:amp-gaugefix}. The delta functions
$\delta^{4|4}$ in \eqref{eq:amp-grassint} are given by the product of
four bosonic and four fermionic delta functions depending on the super
twistor variables $\mathcal{W}_a^i$ with a point index $i$ and a
fundamental $\mathfrak{gl}(4|4)$ index $a$,
\begin{align}
  \label{eq:amp-delta44}
  \delta^{4|4}
  \Bigg(
  \mathcal{W}^k+\sum_{i=\dsites+1}^\sites\!\!\! c_{ki}\mathcal{W}^i
  \Bigg)
  :=
  \prod_{a=1}^{4+4}
  \delta\Bigg(
  \mathcal{W}_a^k+\sum_{i=\dsites+1}^\sites\!\!\! c_{ki}\mathcal{W}_a^i
  \Bigg)\,.
\end{align}
The Graßmannian integral \eqref{eq:amp-grassint} is often treated in a
formal sense, neither explicitly specifying the domain of integration
nor the meaning of the delta functions of \emph{complex}
variables. See, however, e.g.\ \cite{Mason:2009qx} for a
mathematically more rigorous approach.

Recently, a spectral parameter deformation of the Graßmannian integral
for scattering amplitudes has been introduced
\cite{Ferro:2012xw,Ferro:2013dga} in order to establish connections
with the common language of quantum integrable systems. Here we
consider the deformations of the $3$-point $\overline{\text{MHV}}$
amplitude $\mathcal{A}_{3,1}$, the $3$-point $\text{MHV}$ amplitude
$\mathcal{A}_{3,2}$, and the $4$-point $\text{MHV}$ amplitude
$\mathcal{A}_{4,2}$. These will shortly be compared with the Yangian
invariants constructed in section~\ref{sec:osc}. The two $3$-point
amplitudes are of special importance as they provide the building
blocks for general $\sites$-point amplitudes. The $4$-point $\text{MHV}$
amplitude is the first non-trivial example that can be constructed
using these building block. The deformations of these amplitudes read
\cite{Ferro:2013dga}
\begin{align}
  \label{eq:amp-def-a31}
  \tilde{\mathcal{A}}_{3,1}
  &=
  \int
  \frac{\D\!c_{12}\D\!c_{13}}{c_{12}^{s_2+1}c_{13}^{s_3+1}}
  \delta^{n|m}(\mathcal{W}^1+c_{12}\mathcal{W}^2+c_{13}\mathcal{W}^3)\,,\\
  \label{eq:amp-def-a32}
  \tilde{\mathcal{A}}_{3,2}
  &=
  \int
  \frac{\D\!c_{13}\D\!c_{23}}{c_{13}^{s_1+1}c_{23}^{s_2+1}}
  \delta^{n|m}(\mathcal{W}^1+c_{13}\mathcal{W}^3)
  \delta^{n|m}(\mathcal{W}^2+c_{23}\mathcal{W}^3)\,,\\
  \label{eq:amp-def-a42}
  \begin{split}
    \tilde{\mathcal{A}}_{4,2}(\inhdiff)
    &=
    \int
    \frac{\D\!c_{13}\D\!c_{14}\D\!c_{23}\D\!c_{24}}
    {c_{13}c_{24}(c_{13}c_{24}-c_{23}c_{14})}
    \left(-\frac{c_{13}c_{24}}{c_{13}c_{24}-c_{23}c_{14}}\right)^\inhdiff
    \frac{c_{24}^{s_3-s_4}}{(-c_{13}c_{24}+c_{23}c_{14})^{s_3}}\\
    &\quad\quad\quad\quad\quad\quad\quad
    \cdot\,
    \delta^{n|m}(\mathcal{W}^1+c_{13}\mathcal{W}^3+c_{14}\mathcal{W}^4)
    \delta^{n|m}(\mathcal{W}^2+c_{23}\mathcal{W}^3+c_{24}\mathcal{W}^4)\,,
  \end{split}
\end{align}
where the deformation parameters $s_i\in\mathbb{C}$ can be understood
as representation labels. For these low values of $\sites$ and
$\dsites$ a spectral parameter $\inhdiff$ appears only in the last
expression \eqref{eq:amp-def-a42}. In addition, in these deformations
the super twistors are generalized to variables $\mathcal{W}_a^i$ with
a fundamental $\mathfrak{gl}(n|m)$ index $a$ and the delta functions
are to be understood as the corresponding extension of
\eqref{eq:amp-delta44}. In case of $n|m=4|4$, $s_i=0$ and $\inhdiff=0$
the deformations $\tilde{\mathcal{A}}_{\sites,\dsites}$ reduce to the
undeformed scattering amplitudes $\mathcal{A}_{\sites,\dsites}$
obtained from the Graßmannian integral \eqref{eq:amp-grassint}. For
our comparison in the next section we will need the case $n|0$ with
positive integer values of $s_i$, because we will be dealing with
finite-dimensional, purely bosonic representations, and generic
complex $\inhdiff$.

\subsection{Sample invariants as Graßmannian-like
  integrals}
\label{sec:amp-osc}

Let us collect the invariants \eqref{eq:osc-psi21},
\eqref{eq:osc-psi31}, \eqref{eq:osc-psi32}, \eqref{eq:osc-psi42} of
the Yangian $\mathcal{Y}(\mathfrak{gl}(n))$ constructed in
section~\ref{sec:osc} in terms of oscillators:
\begin{align}
  \label{eq:amp-osc21}
  |\Psi_{2,1}\rangle
  &=
  (\bar\oscb^1\cdot \bar\osca^2)^{s_2}
  |0\rangle\,,\\
  \label{eq:amp-osc31}
  |\Psi_{3,1}\rangle
  &=
  (\bar\oscb^1\cdot\bar\osca^2)^{s_2}
  (\bar\oscb^1\cdot\bar\osca^3)^{s_3}
  |0\rangle\,,\\
  \label{eq:amp-osc32}
  |\Psi_{3,2}\rangle
  &=
  (\bar\oscb^1\cdot\bar\osca^3)^{s_1}
  (\bar\oscb^2\cdot\bar\osca^3)^{s_2}
  |0\rangle\,,\\
  \label{eq:amp-osc42}
  \begin{split}
    |\Psi_{4,2}(\inhdiff)\rangle
    &=
    \sum_{k=0}^{\Min(s_3,s_4)}
    \!\!\!
    \frac{1}{(s_3-k)!(s_4-k)!k!^2}\,
    \frac{k!}{\Gamma(\inhdiff-s_3+k+1)}\\
    &\quad\quad\quad\quad
    \cdot\,
    (\bar\oscb^1\cdot\bar\osca^3)^{s_3-k}
    (\bar\oscb^2\cdot \bar\osca^4)^{s_4-k}
    (\bar\oscb^2\cdot \bar\osca^3)^{k}
    (\bar\oscb^1\cdot \bar\osca^4)^{k}
    |0\rangle\,.
  \end{split}
\end{align}
At first sight there seems to be little resemblance between these
formulas and the deformed amplitudes \eqref{eq:amp-def-a31},
\eqref{eq:amp-def-a32} and \eqref{eq:amp-def-a42}. Nevertheless, in
this section we will reformulate these sample
$\mathcal{Y}(\mathfrak{gl}(n))$ invariants
$|\Psi_{\sites,\dsites}\rangle$ and compare to the $\mathfrak{gl}(n)$
version of the deformed amplitudes
$\tilde{\mathcal{A}}_{\sites,\dsites}$.

We start by introducing complex contour integrals in some auxiliary
variables $c_{ki}$. In case of the simplest two-site invariant
\eqref{eq:amp-osc21} this is particularly easy and we write
\begin{align}
  \label{eq:amp-int21}
  |\Psi_{2,1}\rangle
  &=
  (\bar\oscb^1\cdot \bar\osca^2)^{s_2}|0\rangle
  =
  \frac{s_2!(-1)^{s_2}}{2\pi i}
  \oint
  \frac{\D\!c_{12}}{c_{12}^{s_2+1}}
  e^{-c_{12}\,\bar\oscb^1\cdot \bar\osca^2}
  |0\rangle\,,
\end{align}
where the closed contour encircles the pole at the origin of the
complex $c_{12}$-plane counterclockwise. In the same way each product
$\bar\oscb^k\cdot\bar\osca^i$ of oscillators appearing in the further
invariants \eqref{eq:amp-osc31}, \eqref{eq:amp-osc32} and
\eqref{eq:amp-osc42} is translated into one complex contour integral
in the variable $c_{ki}$,
\begin{align}
  \label{eq:amp-int31}
  |\Psi_{3,1}\rangle
  &=
  \frac{s_2!s_3!(-1)^{s_2+s_3}}{(2\pi i)^2}
  \oint
  \frac{\D\!c_{12}\D\!c_{13}}{c_{12}^{s_2+1}c_{13}^{s_3+1}}
  e^{-c_{12}\bar\oscb^1\cdot\bar\osca^2-c_{13}\bar\oscb^1\cdot\bar\osca^3}
  |0\rangle\,,\\
  \label{eq:amp-int32}
  |\Psi_{3,2}\rangle
  &=
  \frac{s_1!s_2!(-1)^{s_1+s_2}}{(2\pi i)^2}
  \oint
  \frac{\D\!c_{13}\D\!c_{23}}{c_{13}^{s_1+1}c_{23}^{s_2+1}}
  e^{-c_{13}\bar\oscb^1\cdot\bar\osca^3-c_{23}\bar\oscb^2\cdot\bar\osca^3}
  |0\rangle\,,\\
  \label{eq:amp-int42}
  \begin{split}
    |\Psi_{4,2}(\inhdiff)\rangle
    &=
    \frac{(-1)^{s_3+s_4}}{(2\pi i)^4}
    \oint
    \frac{\D\!c_{13}\D\!c_{14}\D\!c_{23}\D\!c_{24}}
    {c_{13}^{s_3+1}c_{24}^{s_4+1}c_{14}c_{23}}
    \sum_{k=0}^{\Min(s_3,s_4)}
    \!\!\!
    \frac{k!}{\Gamma(\inhdiff-s_3+k+1)}
    \left(\frac{c_{13}c_{24}}{c_{14}c_{23}}\right)^k\\
    &\quad\quad\quad\quad\quad\quad\quad\quad\quad\quad\quad\quad\quad\quad
    \cdot\,
    e^{-c_{13}\bar\oscb^1\cdot\bar\osca^3-c_{14}\bar\oscb^1\cdot\bar\osca^4
      -c_{23}\bar\oscb^2\cdot\bar\osca^3-c_{24}\bar\oscb^2\cdot\bar\osca^4}
    |0\rangle\,,
  \end{split}
\end{align}
where the contour in each of the variables $c_{ki}$ is again a closed
counterclockwise circle around the origin. The four-site invariant
\eqref{eq:amp-int42} can also be expressed in a slightly more compact
form. We notice that the range of the summation in
\eqref{eq:amp-int42} can be extended to infinity without changing the
value of the integral because the additional terms have a vanishing
residue. Furthermore, choosing a contour that satisfies
$|c_{13}c_{24}|<|c_{14}c_{23}|$, the infinite sum is a series
expansion of a hypergeometric function leading to\footnote{Naively
  this expression does not seem to be valid at the special points
  $s_3-\inhdiff=1,2,3,\ldots$ because in this case the series
  expansion of the hypergeometric function is not defined. However,
  the divergence of the expansion is regularized by the gamma
  function, see e.g. \cite{AbramowitzStegun:1964}, and
  \eqref{eq:amp-int42-2f1} is also valid at these points.}
\begin{align}
  \label{eq:amp-int42-2f1}
  \begin{split}
    |\Psi_{4,2}(\inhdiff)\rangle
    &=
    \frac{(-1)^{s_3+s_4}}{(2\pi i)^4}
    \oint
    \frac{\D\!c_{13}\D\!c_{14}\D\!c_{23}\D\!c_{24}}
    {c_{13}^{s_3+1}c_{24}^{s_4+1}c_{14}c_{23}}\,
    \frac{{}_2F_1\Big(1,1;\inhdiff-s_3+1;\frac{c_{13}c_{24}}{c_{14}c_{23}}\Big)}
    {\Gamma(\inhdiff-s_3+1)}\\
    &\quad\quad\quad\quad\quad\quad\quad\quad\quad\quad
    \cdot\,
    e^{-c_{13}\bar\oscb^1\cdot\bar\osca^3-c_{14}\bar\oscb^1\cdot\bar\osca^4
      -c_{23}\bar\oscb^2\cdot\bar\osca^3-c_{24}\bar\oscb^2\cdot\bar\osca^4}
    |0\rangle\,.
  \end{split}
\end{align}
After these reformulations the integral structure of the invariants
$|\Psi_{\sites,\dsites}\rangle$ already matches the one of the
deformed amplitudes $\tilde{\mathcal{A}}_{\sites,\dsites}$, in the
sense that in both cases there are $\sites\cdot \dsites$ integration
variables. The exponential functions of creation operators in the
integrands of the sample invariants $|\Psi_{\sites,\dsites}\rangle$
are reminiscent of the link representation of scattering amplitudes
\cite{ArkaniHamed:2009dn,ArkaniHamed:2009si}.

Next, we turn to the form of the integrand with the aim to express the
exponential functions of creation operators as appropriate delta
functions like those in \eqref{eq:amp-def-a31}, \eqref{eq:amp-def-a32}
and \eqref{eq:amp-def-a42}. For this purpose we employ different
representations of the oscillator algebras at sites carrying symmetric
representations of type $\s$ and at sites with conjugate
representations of type $\bs$, respectively:
\begin{align}
  \label{eq:amp-osc-delta}
  \begin{aligned}
    \bar\osca&
    \mathrel{\widehat{=}}\mathcal{W}\,,&\quad
    \osca&
    \mathrel{\widehat{=}}\partial_{\mathcal{W}}\,,&\quad 
    |0\rangle&
    \mathrel{\widehat{=}}1&\text{for sites with}&\quad
    \s\,,\\
    \bar\oscb&
    \mathrel{\widehat{=}}-\partial_{\mathcal{W}}\,,&\quad
    \oscb&
    \mathrel{\widehat{=}}\mathcal{W}\,,&\quad 
    |0\rangle&
    \mathrel{\widehat{=}}\delta(\mathcal{W})&
    \text{for sites with}&\quad
    \bs\,.
  \end{aligned}
\end{align}
The oscillators are realized as multiplication and differentiation
operators in a complex variable $\mathcal{W}$. Consequently, as we
already stressed above, $\delta(\mathcal{W})$ is a delta function of a
complex variable. These representations of the oscillator algebra are
discussed in detail in appendix~\ref{sec:bargmann}.

Before we apply \eqref{eq:amp-osc-delta} to the integral expressions
of the invariants $|\Psi_{\sites,\dsites}\rangle$ given in
\eqref{eq:amp-int21}, \eqref{eq:amp-int31}, \eqref{eq:amp-int32} and
\eqref{eq:amp-int42-2f1}, it is instructive to first look at the form
of the Yangian generators annihilating these invariants, recall
\eqref{eq:yi-inv-expanded}. The corresponding monodromies all have a
trivial overall normalization factor, cf.\ \eqref{eq:osc-m21-norm},
\eqref{eq:osc-m31-norm}, \eqref{eq:osc-m32-norm} and
\eqref{eq:osc-m42-norm}. Hence, their expansion
\eqref{eq:yi-mono-expanded} leads to the common Yangian generators
\begin{align}
  \label{eq:amp-gen-yangian}
  \mon_{ab}^{(1)}
  =
  \sum_{i=1}^\sites J_{ba}^i\,,
  \quad
  \mon_{ab}^{(2)}
  =
  \sum_{\substack{i,j=1\\i<j}}^\sites\sum_{c=1}^nJ_{ca}^i J_{bc}^j
  +\sum_{i=1}^\sites\inh_iJ_{ba}^i\,,
\end{align}
where the $\mathfrak{gl}(n)$ generators at the sites are
\begin{align}
  \label{eq:amp-gen-gln}
  J^i_{ab}=
  \left\{
  \begin{aligned}
    \bar\osca_a^i\osca_b^i&
    \mathrel{\widehat{=}}\mathcal{W}_a^i\partial_{\mathcal{W}_b^i}&
    \text{for sites with}&
    \quad\s_i\,,\\
    -\bar\oscb_b^i\oscb_a^i&
    \mathrel{\widehat{=}}\mathcal{W}_a^i\partial_{\mathcal{W}_b^i}+\delta_{ab}&
    \text{for sites with}&
    \quad\bs_i\,.
  \end{aligned}
  \right.\quad
\end{align}
The inhomogeneities $\inh_i$ depend on the chosen invariant and are
specified in section~\ref{sec:osc}. In this formulation, the variables
$\mathcal{W}_a^i$ can be thought of as analogous to the super twistors
used in scattering amplitudes, where in case of the latter $a$ is a
fundamental $\mathfrak{gl}(4|4)$ index. While the oscillator form of
the $\mathfrak{gl}(n)$ generators in \eqref{eq:amp-gen-gln} has a
different structure at the two distinct types of sites, the generators
are, up to the shift $\delta_{ab}$, identical when written in terms of
$\mathcal{W}_a^i$. The two distinct types of representations, $\s_i$
and $\bs_i$, nevertheless manifest themselves in the structure of the
states: The invariants are polynomials in $\mathcal{W}_a^i$ if the
$i$-th site carries a representation $\s_i$, and they contain delta
functions with argument $\mathcal{W}_a^i$ and derivatives thereof for
a site with $\bs_i$. In discussions of the Yangian invariance of
scattering amplitudes the $\mathfrak{gl}(4|4)$ generators also take an
identical form for all points of the amplitude, see e.g.\
\cite{Drummond:2010uq}.

Let us return to our main goal of applying \eqref{eq:amp-osc-delta} to
the sample invariants $|\Psi_{\sites,\dsites}\rangle$ in the form
\eqref{eq:amp-int21}, \eqref{eq:amp-int31}, \eqref{eq:amp-int32} and
\eqref{eq:amp-int42-2f1}. Note that with \eqref{eq:amp-osc-delta} an
exponential of creation operators becomes
\begin{align}
  \label{eq:amp-shift}
  e^{-c_{ki}\bar\osca_a^i\bar\oscb_b^k}|0\rangle
  \mathrel{\widehat{=}}
  e^{c_{ki}\mathcal{W}_a^i\partial_{\mathcal{W}_b^k}}\delta(\mathcal{W}_b^k)
  =
  \delta(\mathcal{W}_b^k+c_{ki}\mathcal{W}_a^i)\,.
\end{align}
Here $|0\rangle$ denotes the tensor product of the Fock vacua of the
two oscillator algebras. The vacuum of the oscillators $\osca_a^i$ is
realized as $1$ and that of $\oscb_b^k$ as a delta function. For the
invariants $|\Psi_{\sites,\dsites}\rangle$, the symbol $|0\rangle$
stands more generally for the tensor product of the Fock vacua of all
involved oscillator algebras. This means, using
\eqref{eq:amp-osc-delta}, that
\begin{align}
  \label{eq:amp-fock-vac}
  |0\rangle
  \mathrel{\widehat{=}}
  \prod_{k\in\{\text{sites with $\bs$}\}}
  \!\!\!\!\!\!
  \delta^n(\mathcal{W}^k)
  \quad
  \text{with}
  \quad
  \delta^n(\mathcal{W}^k)
  :=
  \prod_{a=1}^n\delta(\mathcal{W}_a^k)\,,
\end{align}
where the range of the first product extends over all sites carrying a
conjugate representation of type $\bs$. Using
\eqref{eq:amp-osc-delta}, \eqref{eq:amp-shift} and
\eqref{eq:amp-fock-vac} the two-site invariant \eqref{eq:amp-int21} is
expressed as
\begin{align}
  \label{eq:amp-delta21}
  |\Psi_{2,1}\rangle
  \mathrel{\widehat{=}}
  \bigg(-\sum_{a=1}^n \mathcal{W}^2_a\partial_{\mathcal{W}^1_a}\bigg)^{s_2}
  \delta^n(\mathcal{W}^1)
  =
  \frac{s_2!(-1)^{s_2}}{2\pi i}
  \oint
  \frac{\D\!c_{12}}{c_{12}^{s_2+1}}\,
  \delta^n(\mathcal{W}^1+c_{12}\mathcal{W}^2)\,.
\end{align}
To show the equality of the middle and the right expression in this
formula explicitly, we have to evaluate a contour integral where the
integrand contains a delta function. This is done by first acting on a
holomorphic test function depending only on the variables
$\mathcal{W}^1_a$ of the conjugate site and subsequently evaluating a
standard contour integral. Such test functions are discussed in more
detail in appendix~\ref{sec:bargmann}. Note that such test functions
do not introduce new poles in the $c_{12}$-plane. Proceeding
analogously in the cases of the invariants \eqref{eq:amp-int31},
\eqref{eq:amp-int32}, \eqref{eq:amp-int42-2f1} we
obtain\footnote{Similar formulas for invariants of the Yangian of
  $\mathfrak{gl}(n)$ were also obtained recently in
  \cite{Chicherin:2013sqa}. This was extended in
  \cite{Chicherin:2013ora} to $\mathfrak{gl}(n|m)$, which includes the
  $\mathfrak{gl}(4|4)$ case relevant to scattering amplitudes.}
\begin{align}
  \label{eq:amp-delta31}
  |\Psi_{3,1}\rangle
  &\mathrel{\widehat{=}}
  \frac{s_2!s_3!(-1)^{s_2+s_3}}{(2\pi i)^2}
  \oint
  \frac{\D\!c_{12}\D\!c_{13}}{c_{12}^{s_2+1}c_{13}^{s_3+1}}\,
  \delta^n(\mathcal{W}^1+c_{12}\mathcal{W}^2+c_{13}\mathcal{W}^3)\,,\\
  \label{eq:amp-delta32}
  |\Psi_{3,2}\rangle
  &\mathrel{\widehat{=}}
  \frac{s_1!s_2!(-1)^{s_1+s_2}}{(2\pi i)^2}
  \oint
  \frac{\D\!c_{13}\D\!c_{23}}{c_{13}^{s_1+1}c_{23}^{s_2+1}}\,
  \delta^n(\mathcal{W}^1+c_{13}\mathcal{W}^3)
  \delta^n(\mathcal{W}^2+c_{23}\mathcal{W}^3)\,,\\
  \label{eq:amp-delta42}
  \begin{split}
    |\Psi_{4,2}(\inhdiff)\rangle
    &\mathrel{\widehat{=}}
    \frac{(-1)^{s_3+s_4}}{(2\pi i)^4}
    \oint
    \frac{\D\!c_{13}\D\!c_{14}\D\!c_{23}\D\!c_{24}}
    {c_{13}^{s_3+1}c_{24}^{s_4+1}c_{14}c_{23}}\,
    \frac{{}_2F_1\Big(1,1;\inhdiff-s_3+1;\frac{c_{13}c_{24}}{c_{14}c_{23}}\Big)}
    {\Gamma(\inhdiff-s_3+1)}\\
    &\quad\quad\quad\quad\quad\quad\quad
    \cdot\,
    \delta^n(\mathcal{W}^1+c_{13}\mathcal{W}^3+c_{14}\mathcal{W}^4)
    \delta^n(\mathcal{W}^2+c_{23}\mathcal{W}^3+c_{24}\mathcal{W}^4)\,.
  \end{split}
\end{align}
Recall that for these invariants the integrations in all variables
$c_{ki}$ are closed counterclockwise contours encircling the origin
and for \eqref{eq:amp-delta42} we have to assume in addition
$|c_{13}c_{24}|<|c_{14}c_{23}|$.

Finally, we want to compare this version of the invariants to the
deformed amplitudes summarized in section~\ref{sec:amp-grass}. The
integrations and the delta functions appearing in these deformed
amplitudes are normally only understood in a formal sense, cf.\
\cite{Ferro:2013dga}. To be able to make the comparison, we chose
closed counterclockwise circles around the coordinate origins for the
integration contours in \eqref{eq:amp-def-a31}, \eqref{eq:amp-def-a32}
and \eqref{eq:amp-def-a42}. Furthermore, we interpret the delta
functions in these expressions in the sense of
appendix~\ref{sec:bargmann} as for our invariants. 

First of all, no deformed amplitude $\tilde{\mathcal{A}}_{2,1}$ is
presented in \cite{Ferro:2013dga}. However, at least for $s_2=0$ the
two-site invariant \eqref{eq:amp-delta21} is contained up to a
normalization factor in the general formula \eqref{eq:amp-grassint}
for $\mathcal{A}_{\sites,\dsites}$ after replacing the delta function
$\delta^{4|4}$ by $\delta^n$. Both three-site invariants
\eqref{eq:amp-delta31} and \eqref{eq:amp-delta32} agree (again up to a
constant normalization) with the $\mathfrak{gl}(n|0)$ version of the
deformed amplitudes provided in \eqref{eq:amp-def-a31} and
\eqref{eq:amp-def-a32},
\begin{align}
  \label{eq:amp-compare-3sites}
  |\Psi_{3,1}\rangle
  \propto
  \tilde{\mathcal{A}}_{3,1}\Big|_{n|0}\,,
  \quad
  |\Psi_{3,2}\rangle
  \propto
  \tilde{\mathcal{A}}_{3,2}\Big|_{n|0}\,.
\end{align}
As already mentioned, the $3$-point amplitudes can be understood as
the basic building blocks for more general amplitudes. Hence,
\eqref{eq:amp-compare-3sites} is an important check of our formalism.
Interestingly, however, the integrand of the deformed amplitude
$\tilde{\mathcal{A}}_{4,2}(\inhdiff)$ given in \eqref{eq:amp-def-a42}
does not fully agree with that of $|\Psi_{4,2}(\inhdiff)\rangle$ found
in \eqref{eq:amp-delta42}. To relate these two expressions we note
that at the special points $s_3-\inhdiff=1,2,3,\ldots$ of the spectral
parameter the series expansion of the hypergeometric function in
\eqref{eq:amp-delta42} simplifies to
\begin{align}
  \label{eq:amp-delta42-special}
  \begin{split}
    |\Psi_{4,2}(\inhdiff)\rangle
    &\mathrel{\widehat{=}}
    \frac{(-1)^{s_3+s_4}}{(2\pi i)^4}
    \oint
    \frac{\D\!c_{13}\D\!c_{14}\D\!c_{23}\D\!c_{24}}
    {c_{13}c_{24}(c_{13}c_{24}-c_{23}c_{14})}\,
    \frac{1}{c_{13}^{s_3}c_{24}^{s_4}}
    \left(-\frac{c_{13}c_{24}}{c_{13}c_{24}-c_{23}c_{14}}\right)^{\inhdiff-s_3}\\
    &
    \quad\quad\quad\quad\quad\quad\quad
    \cdot\,
    \delta^n(\mathcal{W}^1+c_{13}\mathcal{W}^3+c_{14}\mathcal{W}^4)
    \delta^n(\mathcal{W}^2+c_{23}\mathcal{W}^3+c_{24}\mathcal{W}^4)\,.
    \end{split}
\end{align}
This agrees up to a shift of the spectral parameter (and again a
normalization factor) with the deformed amplitude:
\begin{align}
  \label{eq:amp-compare-4site}
  |\Psi_{4,2}(\inhdiff)\rangle
  \propto
  \tilde{\mathcal{A}}_{4,2}(\inhdiff-2s_3)\Big|_{n|0}\,
  \quad
  \text{for}
  \quad
  s_3-\inhdiff=1,2,3,\ldots
\end{align}
In \cite{Ferro:2013dga} $\tilde{\mathcal{A}}_{4,2}(\inhdiff)$ is also used
for generic values of $\inhdiff$. However, at least for our choice of the
integration contours around zero this is problematic due to the branch
cut of the complex power function in \eqref{eq:amp-def-a42}. We want
to stress that in the present formulation
\eqref{eq:amp-delta42-special} is only valid at the special points of
the spectral parameter and the full four-site invariant, i.e.\ the
invariant corresponding to the R-matrix, is given by
\eqref{eq:amp-delta42} involving a hypergeometric function. The
interesting question whether a $\mathfrak{gl}(4|4)$ version of
\eqref{eq:amp-delta42} might be a more appropriate deformation of the
four-point $\text{MHV}$ amplitude $\mathcal{A}_{4,2}$ than
\eqref{eq:amp-def-a42} should definitely be clarified.

\section{Bethe ansatz for Yangian invariants}
\label{sec:bethe}

In section~\ref{sec:osc} we discussed some sample Yangian
invariants. Their relation to (deformed) super Yang-Mills scattering
amplitudes was then established in section~\ref{sec:amp}. We will
proceed to a systematic construction of Yangian invariants based on
their characterization as solutions of the set of eigenvalue equations
\eqref{eq:yi-inv-rmm}, which involves the monodromy matrix elements
$\mon_{ab}(\spec)$. This characterization shows that the invariant
$|\Psi\rangle$ is a special eigenstate of the transfer matrix
\begin{align}
  \label{eq:bethe-trans}
  T(\spec)=\tr\mon(\spec)\,,
\end{align}
where the trace is taken over the auxiliary space
$V_\square=\mathbb{C}^n$. Indeed, \eqref{eq:yi-inv-rmm} implies
\begin{align}
  \label{eq:bethe-inv-eigen}
  T(\spec)|\Psi\rangle=n|\Psi\rangle
\end{align} 
with the fixed eigenvalue $n$. The transfer matrix
\eqref{eq:bethe-trans} can be diagonalized by means of a Bethe ansatz,
see e.g.\ the introduction \cite{Faddeev:1996iy} and
\cite{Kulish:1983rd} for the $\mathfrak{gl}(n)$ case. Therefore a
Yangian invariant $|\Psi\rangle$ is a special Bethe vector. This is
the key observation leading to the construction of $|\Psi\rangle$ by a
\emph{Bethe ansatz for Yangian invariants} in this section.

For simplicity, we first focus on $\mathfrak{gl}(2)$ monodromies with
finite-dimensional highest weight representations in the quantum
space. After a brief reminder of the general algebraic Bethe ansatz
technique for $\mathfrak{gl}(2)$ spin chains in
section~\ref{sec:bethe-gl2-review}, we specialize in
section~\ref{sec:bethe-gl2} to the case of Yangian invariant Bethe
vectors. This leads to a set of functional relations characterizing
Yangian invariants, which are equivalent to a degenerate case of the
Baxter equation \cite{Baxter:1972hz}. These equations determine the
Bethe roots and, in addition, constrain the allowed representation
labels and inhomogeneities of the monodromy. Remarkably, a large class
of explicit solutions of these functional relations can be
obtained. They show an interesting structure which is discussed in
section~\ref{sec:bethe-gl2-sol}. The Bethe roots form \emph{exact}
strings in the complex plane. The positions of these strings depend on
the inhomogeneities of the monodromy. The length of the strings, i.e.\
the number of Bethe roots per string, is determined by the
representation labels. We illustrate this structure using the sample
invariants already known from section \ref{sec:osc}. We also present
solutions to the functional relations corresponding to Baxter lattices
with $\lines$ lines. In particular, this includes lattices where all lines
carry the spin~$\frac{1}{2}$ representation of $\mathfrak{su}(2)$. In
section~\ref{sec:bethe-pba} this special case is shown to reproduce
Baxter's original perimeter Bethe ansatz, cf.\
section~\ref{sec:pba}. Finally, in section~\ref{sec:bethe-gln} we
sketch the generalization of the set of functional relations
characterizing Yangian invariants from the $\mathfrak{gl}(2)$ to the
$\mathfrak{gl}(n)$ case, postponing the details to a future
publication \cite{Frassek:2013}.

\subsection{Algebraic Bethe ansatz for
  \texorpdfstring{$\mathfrak{gl}(2)$}{gl(2)} spin chains}
\label{sec:bethe-gl2-review}

An extensive review of the algebraic Bethe ansatz for
$\mathfrak{gl}(2)$ invariant spin chains can be found e.g.\ in
\cite{Faddeev:1996iy}. Here we only recapitulate the essential ideas
and highlight those features that are of importance for the
construction of Yangian invariants.

The algebraic Bethe ansatz allows to diagonalize transfer matrices
defined as traces of suitable monodromies.  We begin with a monodromy
matrix $\mon(\spec)$ satisfying the RTT-relation
\eqref{eq:yi-ybe-rmm}. In the $\mathfrak{gl}(2)$ case the auxiliary
space is $V_\square=\mathbb{C}^2$. Thus, it is convenient to think
about the monodromy as a $2\times 2$ matrix with operatorial entries
acting on the total quantum space,
\begin{align}
  \label{eq:bethe-gl2-monodromy}
  \mon(\spec)=
  \begin{pmatrix}
    A(\spec)&B(\spec)\\
    C(\spec)&D(\spec)\\
  \end{pmatrix}.
\end{align}
We define a transfer matrix $T(\spec)$ as the trace of the monodromy
\eqref{eq:bethe-gl2-monodromy} over the auxiliary space:
\begin{align}
  \label{eq:bethe-gl2-trans}
  T(\spec)=A(\spec)+D(\spec)\,.
\end{align}
Its diagonalization can then be achieved in an efficient way using the
algebraic relations imposed on \eqref{eq:bethe-gl2-monodromy} by the
RTT-equation \eqref{eq:yi-ybe-rmm}.

We assume the existence of a vacuum state $\bvac$ characterized by the
action of the monodromy elements as
\begin{align}
  \label{eq:bethe-gl2-vacuum}
  A(\spec)\bvac=\alpha(\spec)\bvac\,,
  \quad
  D(\spec)\bvac=\delta(\spec)\bvac\,,
  \quad
  C(\spec)\bvac=0\,.
\end{align}
The operators $A(\spec)$ and $D(\spec)$ act diagonally on the
reference state $\bvac$ and hence $\alpha(\spec)$ and $\delta(\spec)$
are scalar functions. The conditions in \eqref{eq:bethe-gl2-vacuum}
are satisfied for monodromies \eqref{eq:yi-mono} built up from Lax
operators \eqref{eq:yi-lax-fund-xi}, where the $i$-th local quantum
space carries a $\mathfrak{gl}(2)$ representation
$\Xi_i=(\xi_i^{(1)},\xi_i^{(2)})$ with a highest weight state
$|\sigma_i\rangle$ defined by
\begin{align}
  \label{eq:bethe-gl2-hws-loc}
  J^i_{11}|\sigma_i\rangle=\xi_i^{(1)}|\sigma_i\rangle\,,
  \quad  
  J^i_{22}|\sigma_i\rangle=\xi_i^{(2)}|\sigma_i\rangle\,,
  \quad  
  J^i_{12}|\sigma_i\rangle=0\,.
\end{align}
For such a monodromy the reference state is 
\begin{align}
  \label{eq:bethe-gl2-hws-tot}
  \bvac=|\sigma_1\rangle\otimes\cdots\otimes|\sigma_\sites\rangle\,,
\end{align} 
and we immediately obtain
\begin{align}
  \label{eq:bethe-gl2-alphadelta}
  \begin{aligned}
    \alpha(\spec)
    =
    \prod_{i=1}^\sites f_{\Xi_i}(\spec-\inh_i)
    \frac{\spec-\inh_i+\xi_i^{(1)}}{\spec-\inh_i}\,,
    \quad
    \delta(\spec)
    =
    \prod_{i=1}^\sites f_{\Xi_i}(\spec-\inh_i)
    \frac{\spec-\inh_i+\xi_i^{(2)}}{\spec-\inh_i}\,.
  \end{aligned}
\end{align} 
However, here and also in section \ref{sec:bethe-gl2} we do not use
these explicit expressions for $\alpha(\spec)$ and $\delta(\spec)$. It
suffices to demand \eqref{eq:bethe-gl2-vacuum}. To proceed, we make an
ansatz for the eigenstates of the transfer matrix:
\begin{align}
  \label{eq:bethe-gl2-eigenvector}
  |\Psi\rangle=B(\brt_1)\cdots B(\brt_\brts)\bvac\,,
\end{align}
where the $\brts$ complex parameters $\brt_k$ are referred to as Bethe
roots. In general the vector \eqref{eq:bethe-gl2-eigenvector} is not
an eigenvector of the transfer matrix $T(\spec)$. It is, however, if the
Bethe roots satisfy a set of Bethe equations. To derive them, we need
some of the commutation relations between the monodromy elements
encoded in \eqref{eq:yi-rmm-comp}. With the notation introduced in
\eqref{eq:bethe-gl2-monodromy} the relevant commutators are
\begin{align}
  \label{eq:bethe-gl2-commrel-abd}
  \begin{aligned}
    A(\spec)B(\specp)
    &=
    \frac{\spec-\specp-1}{\spec-\specp}
    B(\specp)A(\spec)+\frac{1}{\spec-\specp}B(\spec)A(\specp)\,,\\
    D(\spec)B(\specp)
    &=
    \frac{\spec-\specp+1}{\spec-\specp}
    B(\specp)D(\spec)-\frac{1}{\spec-\specp}B(\spec)D(\specp)\,,\\
    B(\spec)B(\specp)
    &=
    B(\specp)B(\spec)\,.
  \end{aligned}
\end{align}
In the next step we act with the operators $A(\spec)$ and $D(\spec)$
appearing in the transfer matrix \eqref{eq:bethe-gl2-trans} on the
vector \eqref{eq:bethe-gl2-eigenvector}. Using
\eqref{eq:bethe-gl2-commrel-abd} one commutes these operators to the
right and obtains after some algebra, see \cite{Faddeev:1996iy},
\begin{align}
  \label{eq:bethe-gl2-apsi-dpsi}
  \begin{aligned}
    A(\spec)|\Psi\rangle
    &=
    \alpha(\spec)\frac{Q(\spec-1)}{Q(\spec)}|\Psi\rangle-
    \sum_{k=1}^\brts\frac{\alpha(\brt_k)Q(\brt_k-1)}{\spec-\brt_k}B(\spec)
    \prod_{\substack{i=1\\i\neq k}}^\brts\frac{B(\brt_i)}{\brt_k-\brt_i}\bvac\,,\\
    D(\spec)|\Psi\rangle
    &=
    \delta(\spec)\frac{Q(\spec+1)}{Q(\spec)}|\Psi\rangle-
    \sum_{k=1}^\brts\frac{\delta(\brt_k)Q(\brt_k+1)}{\spec-\brt_k}B(\spec)
    \prod_{\substack{i=1\\i\neq k}}^\brts\frac{B(\brt_i)}{\brt_k-\brt_i}\bvac\,.\\
  \end{aligned}
\end{align}
Here we introduced Baxter's Q-function, which is defined as a
polynomial of degree $\brts$ in the spectral parameter and its zeros are
located at the Bethe roots $\brt_i$,
\begin{align}
  \label{eq:bethe-gl2-qfunct}
  Q(\spec)=\prod_{i=1}^\brts(\spec-\brt_i)\,.
\end{align}
For $|\Psi\rangle$ of \eqref{eq:bethe-gl2-eigenvector} to be an
eigenstate of the transfer matrix we have to impose the \emph{Bethe
  equations}
\begin{align}
  \label{eq:bethe-gl2-betheeq}
  \alpha(\brt_k)Q(\brt_k-1)+\delta(\brt_k)Q(\brt_k+1)=0
\end{align}
for $k=1,\ldots,\brts$. These equations assure that the ``unwanted
terms'', namely the sums on the right hand sides of each of the two
equations in \eqref{eq:bethe-gl2-apsi-dpsi}, cancel each other upon
addition of the equations. 

A more common form of \eqref{eq:bethe-gl2-betheeq} is achieved by
solving for the fraction of the two Q-functions, and inserting
\eqref{eq:bethe-gl2-alphadelta} and \eqref{eq:bethe-gl2-qfunct}:
\begin{align}
  \label{eq:bethe-gl2-betheeq-ordinary}
  \prod_{i=1}^\sites\frac{\brt_k-\inh_i+\xi_i^{(1)}}{\brt_k-\inh_i+\xi_i^{(2)}}
  =
  -\prod_{j=1}^\brts\frac{\brt_k-\brt_j+1}{\brt_k-\brt_j-1}\,.
\end{align}
However, it turns out that for those solutions which are of particular
interest in the following sections we would divide by zero in
\eqref{eq:bethe-gl2-betheeq-ordinary}. Therefore, we keep the Bethe
equations in the original form \eqref{eq:bethe-gl2-betheeq}.

The eigenvalue $\tau(\spec)$ of $T(\spec)$ corresponding to the
eigenstate $|\Psi\rangle$ is then given by the \emph{Baxter equation}
\begin{align}
  \label{eq:bethe-gl2-baxtereq}
  \tau(\spec)
  =
  \alpha(\spec)\frac{Q(\spec-1)}{Q(\spec)}
  +\delta(\spec)\frac{Q(\spec+1)}{Q(\spec)}\,.
\end{align}
It is important to notice that, assuming the regularity of
$\tau(\spec)$, $\alpha(\spec)$ and $\delta(\spec)$ at the Bethe roots
$\brt_k$, the Bethe equations \eqref{eq:bethe-gl2-betheeq} are a
consequence of the Baxter equation \eqref{eq:bethe-gl2-baxtereq}. This
is easily seen by taking the residue of \eqref{eq:bethe-gl2-baxtereq}
at $\spec=\brt_k$ and using the form of the Q-function in
\eqref{eq:bethe-gl2-qfunct}.

With this algebraic Bethe ansatz the problem of diagonalizing the
transfer matrix \eqref{eq:bethe-gl2-trans}, i.e\ determining its
eigenvalues \eqref{eq:bethe-gl2-baxtereq} and the corresponding
eigenvectors \eqref{eq:bethe-gl2-eigenvector}, is reduced to solving
the Bethe equations \eqref{eq:bethe-gl2-betheeq}. Although this method
is very powerful, it is in general difficult to obtain solutions of
the Bethe equations analytically and often one relies on
approximations and numerical methods. In the case of Yangian
invariants the situation will turn out to be much more favorable.

\subsection{Bethe ansatz for invariants of Yangian
  \texorpdfstring{$\mathcal{Y}(\mathfrak{gl}(2))$}{Y(gl(2))}}
\label{sec:bethe-gl2}

Let us explicitly spell out the definition \eqref{eq:yi-inv-rmm} of
Yangian invariants for the $\mathfrak{gl}(2)$ case using the notation
\eqref{eq:bethe-gl2-monodromy} for the monodromy elements,
\begin{align}
  \label{eq:bethe-gl2-diag}
  A(\spec)|\Psi\rangle&=|\Psi\rangle\,,&
  D(\spec)|\Psi\rangle&=|\Psi\rangle\,,\\
  \label{eq:bethe-gl2-offdiag}
  B(\spec)|\Psi\rangle&=0\,,&
  C(\spec)|\Psi\rangle&=0\,.
\end{align}
Here we separated the equations into \eqref{eq:bethe-gl2-diag}
involving the diagonal monodromy elements and
\eqref{eq:bethe-gl2-offdiag} with the off-diagonal elements. To
construct Yangian invariants $|\Psi\rangle$ we first solve
\eqref{eq:bethe-gl2-diag} by specializing the Bethe ansatz of
section~\ref{sec:bethe-gl2-review}. In a second step, we show that for
finite-dimensional representations the Bethe vectors $|\Psi\rangle$
obtained in this way automatically obey also
\eqref{eq:bethe-gl2-offdiag}. The result of this procedure yields a
characterization of Yangian invariants in terms of functional
relations that will be summarized at the end of this section.

Let us first concentrate on the diagonal part
\eqref{eq:bethe-gl2-diag}. Usually, cf.\
section~\ref{sec:bethe-gl2-review}, one wants to construct
eigenvectors of the transfer matrix, i.e.\ eigenvectors of the sum
$A(\spec)+D(\spec)$. However, here we additionally require that
$|\Psi\rangle$ is a common eigenvector of $A(\spec)$ and
$D(\spec)$. As in section~\ref{sec:bethe-gl2-review} we make the
ansatz \eqref{eq:bethe-gl2-eigenvector} for the eigenvector and use
the commutation relations \eqref{eq:bethe-gl2-commrel-abd} to derive
\eqref{eq:bethe-gl2-apsi-dpsi}. However, we now need to demand that
the ``unwanted terms'' are identical to zero separately for each of
the two equations in \eqref{eq:bethe-gl2-apsi-dpsi}. This is
guaranteed by
\begin{align}
  \label{eq:bethe-gl2-specbethe}
  \alpha(\brt_k)Q(\brt_k-1)=0\,,
  \quad
  \delta(\brt_k)Q(\brt_k+1)=0\,,
\end{align}
which is the degenerate case of the Bethe equations
\eqref{eq:bethe-gl2-betheeq} where each term vanishes individually.
In order to fix the eigenvalues of $A(\spec)$ and $D(\spec)$ to be $1$,
equation \eqref{eq:bethe-gl2-apsi-dpsi} implies that we have to
require
\begin{align}  
  \label{eq:bethe-gl2-specbaxter}
  1=\alpha(\spec)\frac{Q(\spec-1)}{Q(\spec)}\,,
  \quad
  1=\delta(\spec)\frac{Q(\spec+1)}{Q(\spec)}\,.
\end{align}
This is the degenerate case of the Baxter equation
\eqref{eq:bethe-gl2-baxtereq} where each term on the r.h.s.\ is equal
to $1$. It leads to the required transfer matrix eigenvalue
$\tau(\spec)=2$, which is the rank of $\mathfrak{gl}(2)$. Assuming the
regularity of $\alpha(\spec)$ and $\delta(\spec)$ at the Bethe roots
$\spec=\brt_k$, one shows by taking the residue as in
section~\ref{sec:bethe-gl2-review} that
\eqref{eq:bethe-gl2-specbaxter} implies
\eqref{eq:bethe-gl2-specbethe}. Consequently, the problem of
constructing common solutions of the eigenvalue equations in
\eqref{eq:bethe-gl2-diag} has been reduced to solving
\eqref{eq:bethe-gl2-specbaxter}.

To address \eqref{eq:bethe-gl2-offdiag} involving the off-diagonal
monodromy elements, we use \eqref{eq:bethe-gl2-diag}, which we already
solved. We expand \eqref{eq:bethe-gl2-offdiag} using
\eqref{eq:yi-mono-expanded} to obtain
\begin{align}
  \label{eq:bethe-gl2-gl2-weight}
  \mon_{11}^{(1)}|\Psi\rangle=0\,,
  \quad
  \mon_{22}^{(1)}|\Psi\rangle=0\,.
\end{align}
As discussed in the context of \eqref{eq:yi-rmm-expanded}, the
generators $-\mon^{(1)}_{ab}(\spec)$ form a $\mathfrak{gl}(2)$ algebra
and thus \eqref{eq:bethe-gl2-gl2-weight} means that $|\Psi\rangle$ has
$\mathfrak{gl}(2)$ weight $(0,0)$. The expansion of $C(\spec)\bvac=0$
from \eqref{eq:bethe-gl2-vacuum} implies
$\mon_{21}^{(1)}\bvac=0$. Using the commutation relations
\eqref{eq:yi-rmm-gln-comp} and \eqref{eq:bethe-gl2-apsi-dpsi} one
shows
\begin{align}
  \label{eq:bethe-gl2-m21psi}
  \mon^{(1)}_{21}|\Psi\rangle
  =
  -\sum_{k=1}^\brts(\alpha(\brt_k)Q(\brt_k-1)+\delta(\brt_k)Q(\brt_k+1))
  \prod_{\substack{i=1\\i\neq k}}^\brts\frac{B(\brt_i)}{\brt_k-\brt_i}\bvac
  =
  0\,,
\end{align}
where we needed \eqref{eq:bethe-gl2-specbethe} for the last
equality. As we are dealing with a finite-dimensional
$\mathfrak{gl}(2)$ representation, \eqref{eq:bethe-gl2-gl2-weight} and
\eqref{eq:bethe-gl2-m21psi} imply that $|\Psi\rangle$ is a
$\mathfrak{gl}(2)$ singlet. Hence, also
\begin{align}
  \label{eq:bethe-gl2-m12psi}
  \mon^{(1)}_{12}|\Psi\rangle=0\,.
\end{align}
Finally, we obtain from \eqref{eq:yi-rmm-gln-comp} the relations
\begin{align}
  \label{eq:bethe-gl2-comm}
  [\mon_{12}^{(1)},A(\spec)-D(\spec)]=2B(\spec)\,,
  \quad
  [\mon_{21}^{(1)},D(\spec)-A(\spec)]=2C(\spec)\,.
\end{align}
Acting with these on $|\Psi\rangle$ and using
\eqref{eq:bethe-gl2-diag}, \eqref{eq:bethe-gl2-m21psi} and
\eqref{eq:bethe-gl2-m12psi}, we see that also the off-diagonal
part \eqref{eq:bethe-gl2-offdiag} of the Yangian invariance condition
is satisfied.

In conclusion, we have reduced the problem of constructing invariants
$|\Psi\rangle$ of the Yangian $\mathcal{Y}(\mathfrak{gl}(2))$ to the
problem of solving the functional relations
\eqref{eq:bethe-gl2-specbaxter}. Given a solution
$(\alpha(\spec),\delta(\spec),Q(\spec))$ of
\eqref{eq:bethe-gl2-specbaxter}, where the Q-function is of the form
\eqref{eq:bethe-gl2-qfunct}, and both $\alpha(\spec)$ and
$\delta(\spec)$ are regular at the Bethe roots $\brt_k$, the Bethe vector
$|\Psi\rangle$ given in \eqref{eq:bethe-gl2-eigenvector} is Yangian
invariant. It is convenient to represent the functional relations
\eqref{eq:bethe-gl2-specbaxter} in a slightly different
form. Remarkably, the system of two equations in
\eqref{eq:bethe-gl2-specbaxter} can be decoupled into an equation that
depends only on the eigenvalues \eqref{eq:bethe-gl2-vacuum} of the
monodromy acting on the reference state and not on the Bethe roots,
\begin{align}
  \label{eq:bethe-gl2-ad}
  1=\alpha(\spec)\delta(\spec-1)\,,
\end{align} 
and a further equation also involving the Bethe roots contained in
the Q-function,
\begin{align}
  \label{eq:bethe-gl2-qaq}
  \frac{Q(\spec)}{Q(\spec+1)}=\delta(\spec)\,.
\end{align} 
The main task is to understand the solutions of
\eqref{eq:bethe-gl2-ad}. As $\alpha(\spec)$ and $\delta(\spec)$
contain the representation labels and inhomogeneities, cf.\
\eqref{eq:bethe-gl2-alphadelta}, this equation determines those
monodromies that correspond to a Yangian invariant, i.e.\ for which
\eqref{eq:yi-inv-rmm} admits a solution $|\Psi\rangle$. Once a
suitable solution of \eqref{eq:bethe-gl2-ad} is found, the difference
equation \eqref{eq:bethe-gl2-qaq} can typically be solved with ease
for the Bethe roots $\brt_k$. This is in stunning contradistinction to
the usual situation in most spin chain spectral problems, where the
Bethe equations are very hard to solve. Substituting the Bethe roots
into \eqref{eq:bethe-gl2-eigenvector} yields the Bethe state, and
hence the invariant $|\Psi\rangle$. We term this construction a
\emph{Bethe ansatz for Yangian invariants}.

\subsection{Sample solutions of
  \texorpdfstring{$\mathfrak{gl}(2)$}{gl(2)} functional relations}
\label{sec:bethe-gl2-sol}

At present, we lack a complete understanding of the set of solutions
to the functional relations \eqref{eq:bethe-gl2-ad} and
\eqref{eq:bethe-gl2-qaq}. Gaining it should lead to a classification
of invariants of the Yangian $\mathcal{Y}(\mathfrak{gl}(2))$, clearly
an interesting problem for future research. In this paper we take
first steps and analyze a few sample solutions. We show how the
$\mathfrak{gl}(2)$ versions of our favorite invariants in oscillator
form with representations of type $\s=(s,0)$ and $\bs=(0,-s)$, cf.\
section~\ref{sec:osc}, fit into the framework of the Bethe ansatz for
Yangian invariants. In particular, we again discuss the invariant
$|\Psi_{2,1}\rangle$, which was represented by a Baxter lattice with a
single line, the three-vertices $|\Psi_{3,1}\rangle$ and
$|\Psi_{3,2}\rangle$, as well as the four-vertex (R-matrix)
$|\Psi_{4,2}(\inhdiff)\rangle$. We also consider the invariant
associated with a Baxter lattice of $\lines$ lines. For all these
examples the Bethe roots are given explicitly. They arrange themselves
into strings in the complex plane.

\subsubsection{Line}
\label{sec:bethe-gl2-sol-line}

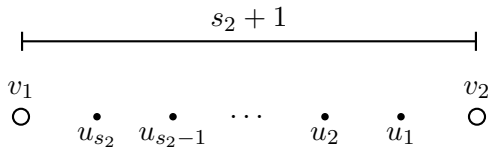
\begin{figure}[!t]
  \begin{center}
    \begin{tikzpicture}
      \draw[thick] (0,0) circle (3pt) node[above=0.1cm]{$\inh_2$};
      \filldraw[thick] (-1,0) circle (1pt) node[below]{$\brt_1$};
      \filldraw[thick] (-2,0) circle (1pt) node[below]{$\brt_2$};
      \node at (-3,0) {\ldots};
      \filldraw[thick] (-4,0) circle (1pt) node[below]{$\brt_{s_2-1}$};
      \filldraw[thick] (-5,0) circle (1pt) node[below]{$\brt_{s_2}$};
      \draw[thick] (-6,0) circle (3pt) node[above=0.1cm]{$\inh_1$};
      \draw[thick,|-|] (-6,1) -- node[midway,above] {$s_2+1$} (0,1);
    \end{tikzpicture}
    \caption{The Bethe roots $\brt_k$ associated with the Yangian invariant
      $|\Psi_{2,1}\rangle$ of section~\ref{sec:osc-line} arrange into 
      a string between the two inhomogeneities $\inh_1$ and $\inh_2$,
      cf.\ \eqref{eq:bethe-gl2-sol-line-roots} in the complex
      plane. This string consists of $s_2$ roots with a uniform real
      spacing of $1$.}
    \label{fig:bethe-gl2-sol-line}
  \end{center} 
\end{figure} 

Let us recall the representation labels and inhomogeneities for the
$\mathfrak{gl}(2)$ case of the invariant $|\Psi_{2,1}\rangle$
discussed in section~\ref{sec:osc-line} and associated with the spin
chain monodromy $\mon_{2,1}(\spec)$ with $\sites=2$ sites, cf.\
\eqref{eq:osc-m21} and \eqref{eq:osc-m21-vs}:
\begin{align}
  \label{eq:bethe-gl2-sol-line-constr}
  \begin{gathered}
    \Xi_1=\bs_1\,,
    \quad
    \Xi_2=\s_2\,,\\
    \inh_1=\inh_2-1-s_2\,,
    \quad
    s_1=s_2\,.
  \end{gathered}
\end{align}
With these relations and the trivial normalization
\eqref{eq:osc-m21-norm} of the monodromy,
\eqref{eq:bethe-gl2-alphadelta} simplifies to
\begin{align}
  \label{eq:bethe-gl2-sol-line-ad-eval}  
  \alpha(\spec)=\frac{\spec-\inh_2+s_2}{\spec-\inh_2}\,,
  \quad
  \delta(\spec)=\frac{\spec-\inh_2+1}{\spec-\inh_2+1+s_2}\,.
\end{align}
In this form one directly sees that the first functional relation
\eqref{eq:bethe-gl2-ad} holds. The remaining relation
\eqref{eq:bethe-gl2-qaq} is solved by
\begin{align}
  \label{eq:bethe-gl2-sol-line-q}  
  Q(\spec)
  =
  \frac{\Gamma(\spec-\inh_2+s_2+1)}{\Gamma(\spec-\inh_2+1)}
  =
  \prod_{k=1}^{s_2}(\spec-\inh_2+k)\,,
\end{align} 
where the freedom of multiplying this solution by a function of period
$1$ in $\spec$ has been fixed by imposing the polynomial form
\eqref{eq:bethe-gl2-qfunct} of the Q-function. Because $s_2$ is a
positive integer, the gamma functions in
\eqref{eq:bethe-gl2-sol-line-q} indeed reduce to a polynomial and we
can read off the Bethe roots as zeros of the Q-function,
\begin{align}
  \label{eq:bethe-gl2-sol-line-roots}  
  \brt_k=\inh_2-k\quad\text{for}\quad k=1,\ldots,s_2\,.
\end{align}
They form a string in the complex plane, see
figure~\ref{fig:bethe-gl2-sol-line}. Note that, as is usual for a
$\mathfrak{gl}(2)$ Bethe ansatz, the labels of the Bethe roots can be
permuted because the operators $B(\spec)$ appearing in the Bethe
vector \eqref{eq:bethe-gl2-eigenvector} commute for different values
of the spectral parameter $\spec$, cf.\
\eqref{eq:bethe-gl2-commrel-abd}. Finally, we want to construct the
Yangian invariant Bethe vector \eqref{eq:bethe-gl2-eigenvector}
corresponding to this solution of the functional relations. For this
purpose we need the reference state \eqref{eq:bethe-gl2-hws-tot} for
the representations specified in
\eqref{eq:bethe-gl2-sol-line-constr}. It is given by a tensor product
of the highest weight states \eqref{eq:osc-hws}:
\begin{align}
  \label{eq:bethe-gl2-sol-line-vac}  
  \bvac
  =
  (\bar\oscb_2^1)^{s_2}
  (\bar\osca_1^2)^{s_2}
  |0\rangle\,.
\end{align}
Then we can evaluate \eqref{eq:bethe-gl2-eigenvector} using
\eqref{eq:bethe-gl2-sol-line-constr},
\eqref{eq:bethe-gl2-sol-line-roots} and
\eqref{eq:bethe-gl2-sol-line-vac}, where we note that because of
\eqref{eq:osc-m21-norm} also the normalization of the operators
$B(\brt_k)$ is trivial. Some details of this straightforward
computation for general $s_2\in\mathbb{N}$ are given in appendix
\ref{sec:twositeproof}. One finds
\begin{align}
  \label{eq:bethe-gl2-sol-line-inv}  
  |\Psi \rangle
  =
  B(\brt_1)\cdots B(\brt_{s_2})\bvac
  =
  (-1)^{s_2}
  (\bar\oscb^1\cdot \bar\osca^2)^{s_2}|0\rangle
  \propto |\Psi_{2,1}\rangle\,.
\end{align}
Thus, our Bethe ansatz for Yangian invariants nicely matches $|\Psi_{2,1}\rangle$ as given in \eqref{eq:osc-psi21}. 

\subsubsection{Three-vertices}
\label{sec:bethe-gl2-sol-three}

\begin{figure}[!t]
  \begin{center}
    \begin{tikzpicture}
      \draw[thick] (0,0) circle (3pt) node[above=0.1cm]{$\inh_1$};
      \filldraw[thick] (1,0) circle (1pt) node[below]{$\brt_1$};
      \filldraw[thick] (2,0) circle (1pt) node[below]{$\brt_2$};
      \node at (3,0) {\ldots};
      \filldraw[thick] (4,0) circle (1pt) node[below]{$\brt_{s_3}$};
      \filldraw[thick] (5,0) circle (1pt) node[below]{$\brt_{s_3+1}$};
      \draw[thick] (5,0) circle (3pt) node[above=0.1cm]{$\inh_3$};
      \filldraw[thick] (6,0) circle (1pt) node[below]{$\brt_{s_3+2}$};
      \node at (7,0) {\ldots};
      \filldraw[thick] (8,0) circle (1pt) node[below]{$\brt_{s_1-1}$};
      \filldraw[thick] (9,0) circle (1pt) node[below]{$\brt_{s_1}$};
      \draw[thick] (10,0) circle (3pt) node[above=0.1cm]{$\inh_2$};
      \draw[thick,|-|] (0,1) -- node[midway,above] {$s_3+1$} (5,1);
      \draw[thick,|-|] (5,1.5) -- node[midway,above] {$s_2$} (10,1.5);
    \end{tikzpicture}
    \caption{The invariant $|\Psi_{3,1}\rangle$ gives rise to a real
      string of $s_1$ uniformly spaced Bethe roots $\brt_k$ in the
      complex plane, see \eqref{eq:bethe-gl2-sol-three1-roots}. They
      lie in between the inhomogeneities $\inh_1$, $\inh_2$ and one root
      coincides with $\inh_3$.}
    \label{fig:bethe-gl2-three1-line}
  \end{center} 
\end{figure}
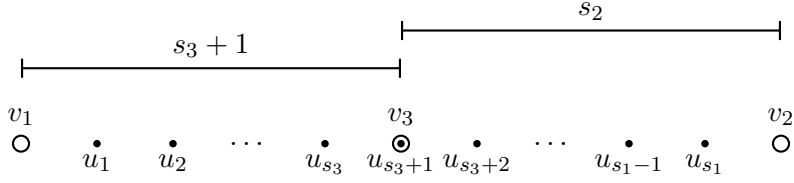 

In section~\ref{sec:osc-3vertices} we discussed two different
three-site invariants. For the $\mathfrak{gl}(2)$ case the monodromy
$\mon_{3,1}(\spec)$ associated with the first invariant
$|\Psi_{3,1}\rangle$ is defined by, cf.\ \eqref{eq:osc-m31} and
\eqref{eq:osc-m31-vs},
\begin{align}
  \label{eq:bethe-gl2-sol-three1-constr}
  \begin{gathered}
    \Xi_1=\bs_1\,,
    \quad
    \Xi_2=\s_2\,,
    \quad
    \Xi_3=\s_3\,,\\
    \inh_2=\inh_1+1+s_2+s_3\,,
    \quad
    \inh_3=\inh_1+1+s_3\,,
    \quad
    s_1=s_2+s_3\,.
  \end{gathered}
\end{align}
With \eqref{eq:bethe-gl2-sol-three1-constr} and the trivial
normalization of the monodromy \eqref{eq:osc-m31-norm}, the
eigenvalues of the monodromy on the reference state of the Bethe
ansatz in \eqref{eq:bethe-gl2-alphadelta} turn into
\begin{align}
  \label{eq:bethe-gl2-sol-three1-ad-eval}  
  \alpha(\spec)=\frac{\spec-\inh_1-1}{\spec-\inh_1-s_1-1}\,,
  \quad
  \delta(\spec)=\frac{\spec-\inh_1-s_1}{\spec-\inh_1}\,.
\end{align} 
Obviously, they obey \eqref{eq:bethe-gl2-ad}. The other functional
relation \eqref{eq:bethe-gl2-qaq} is uniquely solved by
\begin{align}
  \label{eq:bethe-gl2-sol-three1-q}  
  Q(\spec)
  =
  \frac{\Gamma(\spec-\inh_1)}{\Gamma(\spec-\inh_1-s_1)}
  =
  \prod_{k=1}^{s_1}(\spec-\inh_1-k)\,,
\end{align}
because the Q-function is of the form \eqref{eq:bethe-gl2-qfunct}. The
zeros of \eqref{eq:bethe-gl2-sol-three1-q} yield the Bethe roots
\begin{align}
  \label{eq:bethe-gl2-sol-three1-roots}  
  \brt_k=\inh_1+k\quad\text{for}\quad k=1,\ldots,s_1\,.
\end{align}
For this invariant the Bethe roots again form a string in the complex
plane, see figure~\ref{fig:bethe-gl2-three1-line}. We now turn to the
construction of the associated Bethe vector. With \eqref{eq:osc-hws}
the reference state \eqref{eq:bethe-gl2-hws-tot} for the Bethe ansatz
with the representation labels found in
\eqref{eq:bethe-gl2-sol-three1-constr} becomes
\begin{align}
  \label{eq:bethe-gl2-sol-three1-vac}  
  \bvac
  =
  (\bar\oscb_2^1)^{s_2+s_3}
  (\bar\osca_1^2)^{s_2}
  (\bar\osca_1^3)^{s_3}|0\rangle\,.
\end{align}
Notice that one Bethe root is identical to an inhomogeneity,
$\brt_{s_3+1}=\inh_3$. Consequently, the Lax operator
$R_{\square\,{\s}_3}(\brt_{s_3+1}-\inh_3)$, cf.\
\eqref{eq:osc-lax-fund-s}, contributing to $B(\brt_{s_3+1})$ in the
Bethe vector \eqref{eq:bethe-gl2-eigenvector} diverges. Nevertheless,
we obtain a finite Bethe vector using an ad hoc prescription, which we
verified for small values of $s_2$ and $s_3$: First, all
non-problematic Bethe roots are inserted into
\eqref{eq:bethe-gl2-eigenvector}, while $\brt_{s_3+1}$ is kept
generic. In the resulting expression the divergence at
$\brt_{s_3+1}=\inh_3$ disappears. Hence, in a second step, we can
safely insert the last root, leading to
\begin{align}
  \label{eq:bethe-gl2-sol-three1-inv}  
  |\Psi\rangle=B(\brt_1)\cdots B(\brt_{s_1})\bvac
  =
  (-1)^{s_2+s_3}
  (\bar\oscb^1\cdot\bar\osca^2)^{s_2}
  (\bar\oscb^1\cdot\bar\osca^3)^{s_3}
  |0\rangle
  \propto
  |\Psi_{3,1}\rangle\,.
\end{align}
Therefore, we have obtained also the three-site Yangian invariant
$|\Psi_{3,1}\rangle$ presented in \eqref{eq:osc-psi31} from a Bethe
ansatz. A derivation of \eqref{eq:bethe-gl2-sol-three1-inv} for
general $s_2,s_3\in\mathbb{N}$ and a better understanding of the
divergence should be possible, perhaps in generalization of the method
detailed in appendix \ref{sec:twositeproof}.

So-called ``singular solutions'' of the Bethe equations leading
naively to divergent Bethe vectors are well known for the homogeneous
$\mathfrak{su}(2)$ spin~$\frac{1}{2}$ chain, see e.g.\ the recent
discussion \cite{Nepomechie:2013mua}, \cite{Baxter:2001sx} and the
references therein. Such solutions were already known to Bethe himself
\cite{Bethe:1931hc} and appeared also early on in the planar
$\mathcal{N}=4$ super Yang-Mills spectral problem
\cite{Beisert:2003xu}. There are several ways to treat them properly,
cf.\ \cite{Nepomechie:2013mua}, which might also be applicable for the
inhomogeneous spin chain with mixed representations needed for the
three-site invariant $|\Psi_{3,1}\rangle$.

\begin{figure}[!t]
  \begin{center}
    \begin{tikzpicture}
      \draw[thick] (0,0) circle (3pt) node[above=0.1cm]{$\inh_2$};
      \filldraw[thick] (1,0) circle (1pt) node[below]{$\brt_{s_3}$};
      \filldraw[thick] (2,0) circle (1pt) node[below]{$\brt_{s_3-1}$};
      \node at (3,0) {\ldots};
      \filldraw[thick] (4,0) circle (1pt) node[below]{$\brt_{s_1+2}$};
      \filldraw[thick] (5,0) circle (1pt) node[below]{$\brt_{s_1+1}$};
      \draw[thick] (5,0) circle (3pt) node[above=0.1cm]{$\inh_1$};
      \filldraw[thick] (6,0) circle (1pt) node[below]{$\brt_{s_1}$};
      \node at (7,0) {\ldots};
      \filldraw[thick] (8,0) circle (1pt) node[below]{$\brt_2$};
      \filldraw[thick] (9,0) circle (1pt) node[below]{$\brt_1$};
      \draw[thick] (10,0) circle (3pt) node[above=0.1cm]{$\inh_3$};
      \draw[thick,|-|] (0,1) -- node[midway,above] {$s_2$} (5,1);
      \draw[thick,|-|] (5,1.5) -- node[midway,above] {$s_1+1$} (10,1.5);
    \end{tikzpicture}
    \caption{The string of Bethe roots $\brt_k$ belonging to the
      invariant $|\Psi_{3,2}\rangle$. The roots lie between the
      inhomogeneities $\inh_2$ and $\inh_3$. One of them coincides with
      $\inh_1$.}
    \label{fig:bethe-gl2-three2-line}
  \end{center} 
\end{figure}
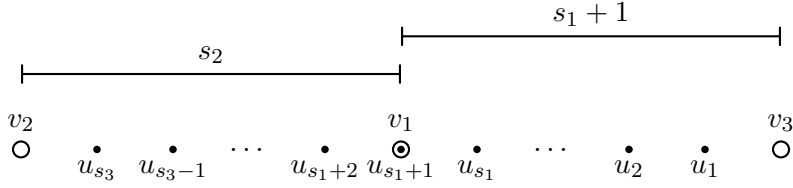 

The $\mathfrak{gl}(2)$ version of the second three-site invariant
discussed in section~\ref{sec:osc-3vertices}, $|\Psi_{3,2}\rangle$, is
characterized by the monodromy $\mon_{3,2}(\spec)$ defined in
\eqref{eq:osc-m32} and \eqref{eq:osc-m32-vs},
\begin{align}
  \label{eq:bethe-gl2-sol-three2-constr}
  \begin{gathered}
    \Xi_1=\bs_1\,,
    \quad
    \Xi_2=\bs_2\,,
    \quad
    \Xi_3=\s_3\,,\\
    \inh_1=\inh_3-1-s_1\,,
    \quad
    \inh_2=\inh_3-1-s_1-s_2\,,
    \quad
    s_3=s_1+s_2\,.
  \end{gathered}
\end{align}
The trivial normalization of this monodromy, cf.\
\eqref{eq:osc-m32-norm}, together with
\eqref{eq:bethe-gl2-sol-three2-constr} implies that
\eqref{eq:bethe-gl2-alphadelta} simplifies to
\begin{align}
  \label{eq:bethe-gl2-sol-three2-ad-eval}  
  \alpha(\spec)=\frac{\spec-\inh_3+s_3}{\spec-\inh_3}\,,
  \quad
  \delta(\spec)=\frac{\spec-\inh_3+1}{\spec-\inh_3+1+s_3}\,,
\end{align} 
which is a solution of the functional relation
\eqref{eq:bethe-gl2-ad}. The second relation \eqref{eq:bethe-gl2-qaq}
is then solved by 
\begin{align}
  \label{eq:bethe-gl2-sol-three2-q}  
  Q(\spec)
  =
  \frac{\Gamma(\spec-\inh_3+s_3+1)}{\Gamma(\spec-\inh_3+1)}
  =
  \prod_{k=1}^{s_3}(\spec-\inh_3+k)\,.
\end{align}
Demanding this solution to be of the form \eqref{eq:bethe-gl2-qfunct}
guarantees its uniqueness and allows us to read off the Bethe roots
\begin{align}
  \label{eq:bethe-gl2-sol-three2-roots}  
  \brt_k=\inh_3-k\quad\text{for}\quad k=1,\ldots,s_3\,.
\end{align}
Once again, they form a string, see
figure~\ref{fig:bethe-gl2-three2-line}. To obtain the corresponding
Bethe vector, we first evaluate the reference state
\eqref{eq:bethe-gl2-hws-tot} with \eqref{eq:osc-hws} and the
representations labels given in
\eqref{eq:bethe-gl2-sol-three2-constr}. This leads to
\begin{align}
  \label{eq:bethe-gl2-sol-three2-vac}  
  \bvac
  =
  (\bar\oscb_2^1)^{s_1}
  (\bar\oscb_2^2)^{s_2}
  (\bar\osca_1^3)^{s_1+s_2}|0\rangle\,.
\end{align}
Just like the other three-site invariant, the operators $B(\brt_{s_1+1})$
diverges because $\brt_{s_1+1}=\inh_1$. With the same ad hoc prescription as
above, we obtain again a finite Bethe vector that, for small values of
$s_1$ and $s_2$, has the explicit form
\begin{align}
  \label{eq:bethe-gl2-sol-three2-inv}  
  |\Psi\rangle=B(\brt_1)\cdots B(\brt_{s_1})\bvac
  =(-1)^{s_1+s_2}
  (\bar\oscb^1\cdot\bar\osca^3)^{s_1}
  (\bar\oscb^2\cdot\bar\osca^3)^{s_2}
  |0\rangle
  \propto
  |\Psi_{3,2}\rangle\,.
\end{align}
This matches the form of the three-site invariant $|\Psi_{3,2}\rangle$
given in \eqref{eq:osc-psi32}.

\subsubsection{Four-vertex}
\label{sec:bethe-gl2-sol-four}

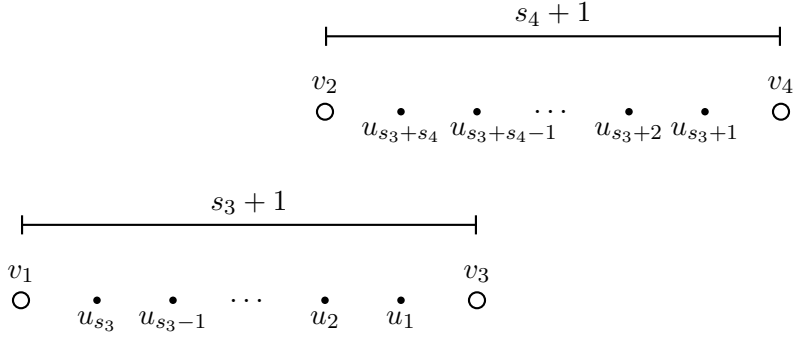
\begin{figure}[!t]
  \begin{center}
    \begin{tikzpicture}
      \begin{scope}
        \draw[thick] (0,0) circle (3pt) node[above=0.1cm]{$\inh_3$};
        \filldraw[thick] (-1,0) circle (1pt) node[below]{$\brt_1$};
        \filldraw[thick] (-2,0) circle (1pt) node[below]{$\brt_2$};
        \node at (-3,0) {\ldots};
        \filldraw[thick] (-4,0) circle (1pt) node[below]{$\brt_{s_3-1}$};
        \filldraw[thick] (-5,0) circle (1pt) node[below]{$\brt_{s_3}$};
        \draw[thick] (-6,0) circle (3pt) node[above=0.1cm]{$\inh_1$};
        \draw[thick,|-|] (-6,1) -- node[midway,above] {$s_3+1$} (0,1);
      \end{scope}
      \begin{scope}[shift={(4,2.5)}]
        \draw[thick] (0,0) circle (3pt) node[above=0.1cm]{$\inh_4$};
        \filldraw[thick] (-1,0) circle (1pt) node[below]{$\brt_{s_3+1}$};
        \filldraw[thick] (-2,0) circle (1pt) node[below]{$\brt_{s_3+2}$};
        \node at (-3,0) {\ldots};
        \filldraw[thick] (-4,0) circle (1pt) 
        node[below]{\hspace{20pt}$\brt_{s_3+s_4-1}$};
        \filldraw[thick] (-5,0) circle (1pt) node[below]{$\brt_{s_3+s_4}$};
        \draw[thick] (-6,0) circle (3pt) node[above=0.1cm]{$\inh_2$};
        \draw[thick,|-|] (-6,1) -- node[midway,above] {$s_4+1$} (0,1);
      \end{scope}
    \end{tikzpicture}
    \caption{The Bethe roots $\brt_k$ corresponding to the four site
      invariant $|\Psi_{4,2}(\inhdiff)\rangle$, i.e.\ to the R-matrix
      $R_{\s_3\,\s_4}(\inhdiff)$, arrange into two real strings in the
      complex plane. The strings consist of $s_3$ and $s_4$ roots,
      respectively. The difference of their endpoints
      $\inhdiff:=\inh_3-\inh_4$, cf.\ \eqref{eq:osc-diffinh}, is the
      spectral parameter of the R-matrix.}
    \label{fig:bethe-gl2-four-line}
  \end{center} 
\end{figure} 

The $\mathfrak{gl}(2)$ version of the four site invariant
$|\Psi_{4,2}(\inh_3-\inh_4)\rangle$ of section \ref{sec:osc-4vertex} is
characterized by a monodromy matrix $\mon_{4,2}(\spec)$ that is specified by,
cf.\ \eqref{eq:osc-m42} and \eqref{eq:osc-m42-vs},
\begin{align}
  \label{eq:bethe-gl2-sol-four-constr}
  \begin{gathered}
    \Xi_1=\bs_1\,,
    \quad
    \Xi_2=\bs_2\,,
    \quad
    \Xi_3=\s_3\,,
    \quad
    \Xi_4=\s_4\,,\\
    \inh_1=\inh_3-1-s_3\,,
    \quad
    \inh_2=\inh_4-1-s_4\,,
    \quad
    s_1=s_3\,,
    \quad
    s_2=s_4\,.
  \end{gathered}
\end{align}
For this monodromy the overall normalization
\eqref{eq:osc-m42-norm} is once again trivial and with
\eqref{eq:bethe-gl2-sol-four-constr} the eigenvalues
\eqref{eq:bethe-gl2-alphadelta} become
\begin{align}
  \label{eq:bethe-gl2-sol-four-ad-eval}
 \alpha(\spec)
 =
 \frac{\spec-\inh_3+s_3}{\spec-\inh_3}\,
 \frac{\spec-\inh_4+s_4}{\spec-\inh_4}\,,
 \quad
 \delta(\spec)
 =
 \frac{\spec-\inh_3+1}{\spec-\inh_3+1+s_3}\,
 \frac{\spec-\inh_4+1}{\spec-\inh_4+1+s_4}\,.
\end{align} 
They obey the functional relation \eqref{eq:bethe-gl2-ad}. A
solution of \eqref{eq:bethe-gl2-qaq} is given by
\begin{align}
  \label{eq:bethe-gl2-sol-four-q}  
  \begin{aligned}
    Q(\spec)
    =
    \frac{\Gamma(\spec-\inh_3+s_3+1)}{\Gamma(\spec-\inh_3+1)}\,
    \frac{\Gamma(\spec-\inh_4+s_4+1)}{\Gamma(\spec-\inh_4+1)}
    =
    \prod_{k=1}^{s_3}(\spec-\inh_3+k)\prod_{k=1}^{s_4}(\spec-\inh_4+k)\,.
  \end{aligned}
\end{align}
Because of \eqref{eq:bethe-gl2-qfunct} this solution is unique. The
Bethe roots
\begin{align}
  \label{eq:bethe-gl2-sol-four-roots}  
  \begin{aligned}
    \brt_k&=\inh_3-k\quad\text{for}\quad k=1,\ldots,s_3\,,\\
    \brt_{k+s_3}&=\inh_4-k\quad\text{for}\quad k=1,\ldots,s_4\,,
  \end{aligned}
\end{align}
which we read off as the zeros of \eqref{eq:bethe-gl2-sol-four-q},
form two strings, see figure~\ref{fig:bethe-gl2-four-line}. To
construct the Bethe vector \eqref{eq:bethe-gl2-eigenvector} we need
the reference state \eqref{eq:bethe-gl2-hws-tot} for the
representation labels found in \eqref{eq:bethe-gl2-sol-four-constr}:
\begin{align}
  \label{eq:bethe-gl2-sol-four-vac}  
  \bvac
  =
  (\bar\oscb_2^1)^{s_3}(\bar\osca_1^3)^{s_3}
  (\bar\oscb_2^2)^{s_4}(\bar\osca_1^4)^{s_4}
  |0\rangle\,.
\end{align}
Then the explicit evaluation of \eqref{eq:bethe-gl2-eigenvector} for
small values of $s_3$ and $s_4$ yields
\begin{align}
  \label{eq:bethe-gl2-sol-four-inv}  
  \begin{aligned}
    |\Psi\rangle
    &=
    B(\brt_1)\cdots B(\brt_{s_3})B(\brt_{s_3+1})\cdots 
    B(\brt_{s_3+s_4})\bvac\\
    &=
    (-1)^{s_3+s_4}s_3!s_4!
    \!\!
    \prod_{l=1}^{\Min(s_3,s_4)}
    \!\!\!
    (\inh_3-\inh_4+s_4-l+1)^{-1}
    \!\!
    \sum_{k=0}^{\Min(s_3,s_4)}
    \!\!\!
    \frac{1}
    {(s_3-k)!(s_4-k)!k!}
    \\
    &\quad
    \cdot\,
    \!\!
    \prod_{l=k+1}^{\Min(s_3,s_4)}
    \!\!\!
    (\inh_3-\inh_4-s_3+l)\;
    (\bar\oscb^1\cdot\bar\osca^3)^{s_3-k}
    (\bar\oscb^2\cdot \bar\osca^4)^{s_4-k}
    (\bar\oscb^2\cdot \bar\osca^3)^{k}
    (\bar\oscb^1\cdot \bar\osca^4)^{k}
    |0\rangle\\
    &\propto|\Psi_{4,2}(\inh_3-\inh_4)\rangle\,,
  \end{aligned}
\end{align}
which matches the expression for $|\Psi_{4,2}(\inhdiff)\rangle$ from
\eqref{eq:osc-psi42} with \eqref{eq:osc-phi}, \eqref{eq:osc-diffinh}
and \eqref{eq:osc-coeff}. As the invariant
$|\Psi_{4,2}(\inhdiff)\rangle$ can be understood as the R-matrix
$R_{\s_3\,\s_4}(\inhdiff)$, we might say that this R-matrix is a
special Bethe vector.

\subsubsection{Baxter lattice with $\lines$ lines}
\label{sec:bethe-gl2-sol-lattice}

We know from section \ref{sec:osc} that the invariants
$|\Psi_{2,1}\rangle$ and $|\Psi_{4,2}(\inhdiff)\rangle$ can be
understood as a Baxter lattice with, respectively, one and two lines
carrying conjugate symmetric representations. Here we work out the
solution to the functional relations \eqref{eq:bethe-gl2-ad} and
\eqref{eq:bethe-gl2-qaq} for a Baxter lattice consisting of $\lines$
lines of this type. In this case the monodromy has $\sites=2\lines$
sites. According to the $\mathfrak{gl}(2)$ version of
\eqref{eq:osc-line-bs}, the $k$-th line of the Baxter lattice with
endpoints $\epe_k<\epb_k$, the representation $\Lambda_k=\bs_{\epe_k}$
and a spectral parameter $\rap_k$ gives rise to the two spin chain
sites
\begin{align}
  \label{eq:bethe-gl2-sol-lattice-constr}
  \begin{gathered}
  \Xi_{\epe_k}=\bs_{\epe_k}\,,
  \quad
  \Xi_{\epb_k}=\s_{\epb_k}\,,\\
  \inh_{\epe_k}=\rap_k\,,
  \quad
  \inh_{\epb_k}=\rap_k+s_{\epe_k}+1\,,
  \quad
  s_{\epe_k}=s_{\epb_k}\,.
  \end{gathered}
\end{align}
This turns the monodromy eigenvalues \eqref{eq:bethe-gl2-alphadelta}
into
\begin{align}
  \label{eq:bethe-gl2-sol-lattice-ad-eval}
  \begin{aligned}
    \alpha(\spec)
    &=
    \prod_{k=1}^\lines
    f_{\bs_{\epe_k}}(\spec-\inh_{\epe_k})f_{\s_{\epb_k}}(\spec-\inh_{\epb_k})
    \frac{\spec-\inh_{\epb_k}+s_{\epb_k}}{\spec-\inh_{\epb_k}}
    =\prod_{k=1}^\lines\frac{\spec-\inh_{\epb_k}+s_{\epb_k}}{\spec-\inh_{\epb_k}}\,,\\
    \delta(\spec)
    &=
    \prod_{k=1}^\lines
    f_{\bs_{\epe_k}}(\spec-\inh_{\epe_k})f_{\s_{\epb_k}}(\spec-\inh_{\epb_k})
    \frac{\spec-\inh_{\epe_k}-s_{\epe_k}}{\spec-\inh_{\epe_k}}
    =\prod_{k=1}^\lines\frac{\spec-\inh_{\epb_k}+1}{\spec-\inh_{\epb_k}+1+s_{\epb_k}}\,.
  \end{aligned}
\end{align}
For the last equality in both equations we used that each factor of
the products corresponds to one line of the Baxter lattice. Using
\eqref{eq:bethe-gl2-sol-lattice-constr} the normalization factors
belonging to each of these lines reduce to $1$ analogously to the case
of a single line explained before~\eqref{eq:osc-m21-norm}.  Obviously,
the eigenvalues in \eqref{eq:bethe-gl2-sol-lattice-ad-eval} satisfy
\eqref{eq:bethe-gl2-ad}. The relation \eqref{eq:bethe-gl2-qaq} is
solved by
\begin{align}
  \label{eq:bethe-gl2-sol-lattice-q}  
  \begin{aligned}
    Q(\spec)
    =
    \prod_{k=1}^\lines
    \frac{\Gamma(\spec-\inh_{\epb_k}+s_{\epb_k}+1)}{\Gamma(\spec-\inh_{\epb_k}+1)}
    =
    \prod_{k=1}^\lines\prod_{l=1}^{s_{\epb_k}}(\spec-\inh_{\epb_k}+l)\,,
  \end{aligned}
\end{align} 
which is the unique solution because we also demand the Q-function to
be of the form \eqref{eq:bethe-gl2-qfunct}. We read off the Bethe
roots as zeros of \eqref{eq:bethe-gl2-sol-lattice-q},
\begin{align}
  \label{eq:bethe-gl2-sol-lattice-roots}  
  \begin{aligned}
    \brt_k&
    =
    \inh_{\epb_1}-k\quad 
    \text{for}\quad 
    k=1,\ldots,s_{\epb_1}\,,\\
    \brt_{k+s_{\epb_1}}&
    =
    \inh_{\epb_2}-k\quad 
    \text{for}\quad 
    k=1,\ldots,s_{\epb_2}\,,\\
    &\hspace{6pt}\vdots\\
    \brt_{k+s_{\epb_{\lines-1}}}&
    =
    \inh_{\epb_\lines}-k\quad 
    \text{for}\quad 
    k=1,\ldots,s_{\epb_\lines}\,.
  \end{aligned}
\end{align}
They arrange into $\lines$ strings. The $k$-th line of the Baxter
lattice with representation $\Lambda_k=\bs_{\epe_k}$ leads to one string
of $s_{\epe_k}=s_{\epb_k}$ Bethe roots with a uniform real spacing of $1$
lying between the inhomogeneities $\inh_{\epe_k}$ and $\inh_{\epb_k}$. The
arrangement of these strings in the complex plane is determined by the
spectral parameters $\rap_k=\inh_{\epe_k}$ of the lines. Next, we
concentrate on the associated Bethe vector. With the form of the
highest weight states \eqref{eq:osc-hws} and
\eqref{eq:bethe-gl2-sol-lattice-constr} the reference state
\eqref{eq:bethe-gl2-hws-tot} turns into
\begin{align}
  \label{eq:bethe-gl2-sol-lattice-vac}  
  \bvac
  =
  \prod_{k=1}^\lines
  (\bar\oscb_2^{\epe_k})^{s_{\epb_k}}(\bar\osca_1^{\epb_k})^{s_{\epb_k}}|0\rangle\,.
\end{align}
The Yangian invariant is then given by the Bethe vector
\eqref{eq:bethe-gl2-eigenvector}. Note that as in the special cases of
one- and two-line Baxter lattices discussed, respectively, in
section~\ref{sec:bethe-gl2-sol-line} and
section~\ref{sec:bethe-gl2-sol-four}, for generic values of
$\rap_k=\inh_{\epe_k}$ no Bethe root coincides with an
inhomogeneity. Consequently, these Bethe vectors with an even number
of spin chain sites are manifestly non-divergent.

We finish with a remark on the general structure of the set of
solutions to the functional relations \eqref{eq:bethe-gl2-ad} and
\eqref{eq:bethe-gl2-qaq}. Notice that the solution of these relations
defined by \eqref{eq:bethe-gl2-sol-lattice-ad-eval} and
\eqref{eq:bethe-gl2-sol-lattice-q} is actually the product of $\lines$ line
solutions of the type discussed in
section~\ref{sec:bethe-gl2-sol-line}. More generally, given two
solutions $(\alpha_1(\spec),\delta_1(\spec),Q_1(\spec))$ and
$(\alpha_2(\spec),\delta_2(\spec),Q_2(\spec))$ of the functional
relations, a new one is obtained as the product
\begin{align}
  \label{eq:bethe-gl2-superpos}
  (\alpha_1(\spec)\alpha_2(\spec),
  \delta_1(\spec)\delta_2(\spec),
  Q_1(\spec)Q_2(\spec))\,.
\end{align}
Using this method one can construct new Yangian invariants by
``superposing'' known ones. For example, it should be possible to
combine line solutions with the three-vertices discussed in
section~\ref{sec:bethe-gl2-sol-three}.

\subsection{Relation to perimeter Bethe ansatz}
\label{sec:bethe-pba}

In the previous section~\ref{sec:bethe-gl2-sol-lattice} we analyzed
the solution to the functional relations \eqref{eq:bethe-gl2-ad} and
\eqref{eq:bethe-gl2-qaq} that corresponds to a Baxter lattice with
$\lines$ lines. Here we show that a special case of it reproduces the
perimeter Bethe ansatz of section~\ref{sec:pba}. Therefore, we first
use special properties of the $\mathfrak{gl}(2)$ Lax operators. Then
the Baxter lattice is specialized to the case where all lines carry
the conjugate of the fundamental, i.e.\ the antifundamental,
representation. This allows us to express the associated Yangian
invariant $|\Psi\rangle$, which was discussed in
section~\ref{sec:bethe-gl2-sol-lattice} in the algebraic formulation
of the Bethe ansatz, in terms of a coordinate Bethe ansatz wave
function. The resulting expression matches the perimeter Bethe ansatz
formula \eqref{eq:pba-partition-wave} for the partition function
$\mathcal{Z}(\graph,\rapset,\stateset)$.

In order to obtain the special properties of the Lax operators, we
employ a relation between representations $\s$ and $\bs$, which is
valid in the $\mathfrak{gl}(2)$ case but not for $\mathfrak{gl}(n)$ in
general. The generators \eqref{eq:osc-gen-s-bs} and the highest weight
states \eqref{eq:osc-hws} of both representations are linked by
\begin{align}
  \label{eq:bethe-gl2-special-rep}
  UJ_{ab}U^{-1}=\bar J_{ab}\big|_{\oscb_a\mapsto\osca_a}\!\!\!+s\delta_{ab}\,,
  \quad
  U|\sigma\rangle=(-1)^s|\bar\sigma\rangle\big|_{\oscb_a\mapsto\osca_a}\,,
\end{align}
where the unitary operator
\begin{align}
  \label{eq:bethe-gl2-special-v}
  U=e^{\frac{\pi}{2}(\bar\osca_1\osca_2-\bar\osca_2\osca_1)}
  \quad
  \text{obeys}
  \quad
  U|0\rangle=|0\rangle\,,
  \quad
  \bar\osca_1 U=U\bar\osca_2\,,
  \quad
  \bar\osca_2 U=-U\bar\osca_1\,.  
\end{align}
To avoid spurious divergencies in the following, we introduce Lax
operators with a different normalization than before,
\begin{align}
  \label{eq:bethe-gl2-special-lax}
  \tilde{R}_{\square\,\s}(\spec-\inht)
  =
  (\spec-\inht)1+\sum_{a,b=1}^2e_{ab}\bar\osca_b\osca_a\,.
\end{align}
They build up a monodromy with inhomogeneities denoted by $\inht_i$,
\begin{align}
  \label{eq:bethe-gl2-special-mono}
  \tilde{M}(\spec)
  =
  \tilde{R}_{\square\,\s_1}(\spec-\inht_1)
  \cdots
  \tilde{R}_{\square\,\s_\sites}(\spec-\inht_\sites)\,.
\end{align}
Using \eqref{eq:bethe-gl2-special-rep}, the ordinary Lax operators
\eqref{eq:osc-lax-fund-s} for the representation $\s$ and
\eqref{eq:osc-lax-fund-bs} for $\bs$ can be expressed in terms of
\eqref{eq:bethe-gl2-special-lax} as
\begin{align}
  \label{eq:bethe-gl2-special-lax-symm}
  R_{\square\,\s}(\spec)
  =
  \frac{f_{\s}(\spec)}{\spec}
  \tilde{R}_{\square\,\s}(\spec)\,,
  \quad
  R_{\square\,\bs}(\spec)\big|_{\oscb_a\mapsto\osca_a}
  =
  \frac{f_{\bs}(\spec)}{\spec}U
  \tilde{R}_{\square\,\s}(\spec-s)U^{-1}\,.
\end{align}
These properties at hand, any $\mathfrak{gl}(2)$ monodromy $M(\spec)$
consisting of $R_{\square\,\s}(\spec)$ and $R_{\square\,\bs}(\spec)$
can be reformulated as $\tilde{\mon}(\spec)$ that solely comprises Lax
operators of the type $\tilde{R}_{\square\,\s}(\spec)$. 

We will apply this observation to the monodromy specified in
\eqref{eq:bethe-gl2-sol-lattice-constr} which is associated with a
Baxter lattice with $\lines$ lines. To connect with the perimeter
Bethe ansatz, we first need to specialize to lattices where each line
carries the antifundamental representation,
\begin{align}
  \label{eq:bethe-gl2-special-reps}
  \repset=(\overline{(1,0)},\ldots,\overline{(1,0)})\,,
\end{align}
cf.\ \eqref{eq:yi-baxter-data} for the notation. From
\eqref{eq:bethe-gl2-special-reps} together with
\eqref{eq:bethe-gl2-sol-lattice-constr} we have
$s_{\epe_k}=s_{\epb_k}=1$. Hence, the strings of Bethe roots
\eqref{eq:bethe-gl2-sol-lattice-roots} degenerate into individual
points in the complex plane,
\begin{align}
  \label{eq:bethe-gl2-special-roots}
  \brt_k=\rap_k+1
  \quad
  \text{for}
  \quad
  k=1,\ldots,\lines\,.
\end{align}
This pattern of Bethe roots is identical to that of the perimeter
Bethe ansatz in \eqref{eq:pba-inhomo-rap}. Using
\eqref{eq:bethe-gl2-special-lax-symm}, we express the monodromy
defined by \eqref{eq:bethe-gl2-sol-lattice-constr} and
\eqref{eq:bethe-gl2-special-reps} as
\begin{align}
  \label{eq:bethe-gl2-special-trafo-mono}
  \mon(\spec)\big|_{\oscb_a^{\epe_k}\mapsto\osca_a^{\epe_k}}
  =
  \prod_{i=1}^{2\lines}\frac{1}{\spec-\inh_i}
  W\tilde{\mon}(\spec)W^{-1}
  \quad
  \text{with}
  \quad
  W=\prod_{k=1}^\lines U^{\epe_k}\,,
\end{align}
where the normalizations of the Lax operators cancel as explained
after \eqref{eq:bethe-gl2-sol-lattice-ad-eval}.  The unitary
$U^{\epe_k}$ acts on site $\epe_k$. Thus $W$ transforms all conjugate
sites. The parameters of $\tilde{\mon}(\spec)$ are
\begin{align}
  \label{eq:bethe-gl2-special-rep-inhomo}
  \s_i=(1,0)\,,
  \quad
  \inht_{\epe_k}=\rap_k+1\,,
  \quad
  \inht_{\epb_k}=\rap_k+2\,,
\end{align}
where the inhomogeneities $\inht_{\epe_k}$ originating from the
conjugate sites of $\mon(\spec)$ are shifted by $1$ with respect to
the $\inh_{\epe_k}$ in \eqref{eq:bethe-gl2-sol-lattice-constr}. The
inhomogeneities in \eqref{eq:bethe-gl2-special-rep-inhomo} agree with
those of the perimeter Bethe ansatz in \eqref{eq:pba-inhomo-rap}. To
obtain the highest weight state $|\tilde{\Omega}\rangle$ in the total
quantum space of $\tilde{\mon}(\spec)$, we apply
\eqref{eq:bethe-gl2-special-rep} to $\bvac$ in
\eqref{eq:bethe-gl2-sol-lattice-vac},
\begin{align}
  \label{eq:bethe-gl2-special-refstate}
  |\tilde{\Omega}\rangle
  =
  (-1)^\lines W^{-1}\bvac\big|_{\oscb_a^i\mapsto\osca_a^i}
  =
  \bar\osca_1^1\cdots\bar\osca_1^{\sites}|0\rangle\,.
\end{align}
Equations \eqref{eq:bethe-gl2-special-trafo-mono} and
\eqref{eq:bethe-gl2-special-refstate} allow us to express also the
Bethe vector \eqref{eq:bethe-gl2-eigenvector}, i.e.\ the Yangian
invariant $|\Psi\rangle$ of the Baxter lattice, as a Bethe vector that
is constructed using the matrix elements
$\tilde{\mon}_{12}(\spec)=\tilde{B}(\spec)$ of the new monodromy,
\begin{align}
  \label{eq:bethe-gl2-special-trafo-vector}
  |\Psi\rangle\big|_{\oscb_a^{\epe_k}\mapsto\osca_a^{\epe_k}}
  =
  (-1)^\lines\prod_{k=1}^\lines\prod_{i=1}^{2\lines}\frac{1}{\brt_k-\inh_i}
  W|\tilde{\Psi}\rangle
  \quad
  \text{with}
  \quad
  |\tilde{\Psi}\rangle
  =
  \tilde{B}(\brt_1)\cdots \tilde{B}(\brt_\lines)
  |\tilde{\Omega}\rangle\,.
\end{align}

Next, the algebraic Bethe ansatz vector $|\tilde{\Psi}\rangle$ is
represented by coordinate Bethe ansatz wave functions. For monodromies
$\tilde{\mon}(\spec)$ of the type \eqref{eq:bethe-gl2-special-mono}
with $\s_i=(1,0)$ at all sites one has, see e.g.\
\cite{Ovchinnikov:2010vb} and appendix~3.E of
\cite{Essler:2010},\footnote{See \cite{Avdeev:1985cx} for a proof of
  the corresponding relation in case of more general representations
  $\s_i=(s,0)$ but no inhomogeneities, $\inht_i=0$.}
\begin{align}
  \label{eq:bethe-gl2-special-vector}
  \begin{aligned}
  |\tilde{\Psi}\rangle
  =
  \tilde{B}(\brt_1)\cdots \tilde{B}(\brt_\brts)|\tilde{\Omega}\rangle
  =
  \sum_{1\leq \magn_1<\cdots <\magn_\brts\leq \sites}
  \!\!\!\!\!\!
  \Phi(\inhtset,\brtset,\magnset)\,
  J_{21}^{\magn_1}\cdots J_{21}^{\magn_\brts}|\tilde{\Omega}\rangle\,,
  \end{aligned}
\end{align}
with generators $J_{ab}^i=\bar\osca_a^i\osca_b^i$ and the wave
function $\Phi(\inhtset,\brtset,\magnset)$ in \eqref{eq:pba-psi}. The
arguments $\inhtset$, $\brtset$ and $\magnset$ denote respectively the
inhomogeneities $\inht_i$, Bethe roots $\brt_k$ and magnon positions
$\magn_k$, cf.\ \eqref{eq:pba-psi-param}. We apply
\eqref{eq:bethe-gl2-special-vector} in
\eqref{eq:bethe-gl2-special-trafo-vector} with $\sites=2\lines$ sites
and $\brts=\lines$ Bethe roots.

To obtain a partition function from the Yangian invariant vector
$|\Psi\rangle$, recall \eqref{eq:yi-partition-psi}:
\begin{align}
  \label{eq:bethe-gl2-special-z-psi}
  \mathcal{Z}(\graph,
  \repset,\rapset,\stateset)
  \propto
  \langle\stateset|\Psi\rangle\,.
\end{align}
For a Baxter lattice with the representations
\eqref{eq:bethe-gl2-special-reps} the possible states are
$|\stateset\rangle =
|\alpha_1\rangle\otimes\cdots\otimes|\alpha_{2\lines}\rangle$ with
$\alpha_i=1,2$. After the replacement
$|\stateset\rangle\big|_{\oscb_a^{\epe_k}\mapsto\osca_a^{\epe_k}}$,
the state at each site is either $|1\rangle=\bar\osca_1^{i}|0\rangle$
or $|2\rangle=\bar\osca_2^{i}|0\rangle$.  The computation of the
scalar product \eqref{eq:bethe-gl2-special-z-psi} reduces using
\eqref{eq:bethe-gl2-special-trafo-vector} to that with each term of
the sum in \eqref{eq:bethe-gl2-special-vector}. This turns out to be
only non-zero if the state labels $\stateset$ obey the ice rule
\eqref{eq:pba-ice-rule-global}, and if in addition $\magnset$ is
determined in terms of $\graph$ and $\stateset$ by
\eqref{eq:pba-alpha-position}. Then we have
\begin{align}
  \label{eq:bethe-gl2-special-scalarprod}
  \langle\stateset|\big|_{\oscb_a^{\epe_k}\mapsto\osca_a^{\epe_k}}
  W
  J_{21}^{\magn_1}\cdots J_{21}^{\magn_\lines}|\tilde{\Omega}\rangle
  =
  (-1)^{\mathcal{K}(\graph,\stateset)}
\end{align}
with $\mathcal{K}(\graph,\stateset)$ defined in
\eqref{eq:pba-partition-wave-exp}. The factors of $-1$ stem from sites
transformed by $W$.

Finally, the combination of \eqref{eq:bethe-gl2-special-trafo-vector},
\eqref{eq:bethe-gl2-special-vector} and
\eqref{eq:bethe-gl2-special-scalarprod} leads to an expression for the
partition function \eqref{eq:bethe-gl2-special-z-psi}. Again, it is
only non-zero if the state labels $\stateset$ satisfy
\eqref{eq:pba-ice-rule-global}. In this case
\begin{align}
  \label{eq:bethe-gl2-special-partitionfct}
  \mathcal{Z}(\graph,
  \repset,
  \rapset,
  \stateset)
  \propto
  \langle\stateset|\Psi\rangle
  =
  (-1)^\lines
  \prod_{k=1}^\lines\prod_{i=1}^{2\lines}\frac{1}{\brt_k-\inh_i}\,
  (-1)^{\mathcal{K}(\graph,\stateset)}
  \Phi(\inhtset,\brtset,\magnset)\,.
\end{align}
Here $\repset$ is fixed in \eqref{eq:bethe-gl2-special-reps}, and the
arguments $\inhtset$, $\brtset$, $\magnset$ of the wave function are
determined by the variables $\graph$, $\rapset$, $\stateset$ of the
partition function with \eqref{eq:pba-alpha-position} and
\eqref{eq:pba-inhomo-rap}. Up to an $\stateset$-independent
normalization factor, the l.h.s.\ of
\eqref{eq:bethe-gl2-special-partitionfct} is the perimeter Bethe
ansatz formula \eqref{eq:pba-partition-wave}.

However, this factor cannot be directly determined by the Bethe
ansatz. To show that its choice in \eqref{eq:pba-partition-wave}
guarantees the agreement with the partition function
\eqref{eq:pba-partition}, note the following. With the normalization
of the Boltzmann weights in \eqref{eq:pba-r-matrix} it is easy to see
that for the particular state labels $\stateset_0=(1,\ldots,1)$ the
partition function \eqref{eq:pba-partition} equals
$\mathcal{Z}(\graph,\rapset, \stateset_0)=1$. The
$\stateset$-independent normalization in \eqref{eq:pba-partition-wave}
trivially guarantees that also this expression is equal to $1$ for
$\stateset=\stateset_0$. As we already know from
\eqref{eq:bethe-gl2-special-partitionfct} that the
$\stateset$-dependent part of \eqref{eq:pba-partition-wave} is
proportional to the partition function \eqref{eq:pba-partition}, this
concludes our derivation of \eqref{eq:pba-partition-wave}. It shows
that the perimeter Bethe ansatz as reviewed in section~\ref{sec:pba}
is a special case of the Bethe ansatz for Yangian invariants.

\subsection{Outline of \texorpdfstring{$\mathfrak{gl}(n)$}{gl(n)}
  functional relations}
\label{sec:bethe-gln}

In section \ref{sec:bethe-gl2} we discussed in detail how the Bethe
ansatz for $\mathfrak{gl}(2)$ spin chains can be specialized in such a
way that the resulting Bethe vector $|\Psi\rangle$ is Yangian
invariant. This leads to functional relations
\eqref{eq:bethe-gl2-specbaxter} which restrict the allowed
representations and inhomogeneities of the monodromy and determine the
Bethe roots. The derivation was based on the observation
\eqref{eq:bethe-inv-eigen} that a Yangian invariant $|\Psi\rangle$ is
a special eigenvector of a transfer matrix. Of course, this
observation is also valid more generally for invariants of the Yangian
of $\mathfrak{gl}(n)$. In this $\mathfrak{gl}(n)$ case the nested
algebraic Bethe ansatz for monodromies with a finite-dimensional
highest weight representation at each site can be found e.g.\ in
\cite{Kulish:1983rd}. In generalization of the discussion of the
$\mathfrak{gl}(2)$ situation in section \ref{sec:bethe-gl2}, it can be
specialized to the case where the Bethe vectors are Yangian
invariant. The details of this calculation will be presented in a
separate publication \cite{Frassek:2013}.

Here we only state one of the main results, the set of functional
relations determining the representation labels, inhomogeneities and
Bethe roots of Yangian invariants in the $\mathfrak{gl}(n)$ case:
\begin{align}
  \label{eq:bethe-gln-specbaxter}
  \begin{aligned}
    1&=\mu_1(\spec)
    \frac{Q_1(\spec-1)}{Q_1(\spec)}\,,\\
    1&=\mu_2(\spec)
    \frac{Q_1(\spec+1)}{Q_1(\spec)}\,
    \frac{Q_2(\spec-1)}{Q_2(\spec)}\,,\\
    1&=\mu_3(\spec)
    \frac{Q_2(\spec+1)}{Q_2(\spec)}\,
    \frac{Q_3(\spec-1)}{Q_3(\spec)}\,,\\
    &\;\;\vdots\\
    1&=\mu_{n-1}(\spec)
    \frac{Q_{n-2}(\spec+1)}{Q_{n-2}(\spec)}\,
    \frac{Q_{n-1}(\spec-1)}{Q_{n-1}(\spec)}\,,\\
    1&=\mu_{n}(\spec)
    \frac{Q_{n-1}(\spec+1)}{Q_{n-1}(\spec)}\,.
  \end{aligned}
\end{align}
Here $\mu_1(\spec),\ldots,\mu_n(\spec)$ are the eigenvalues of the
monodromy elements $\mon_{11}(\spec),\ldots,\mon_{nn}(\spec)$ on the
pseudo vacuum of the Bethe ansatz, cf.\ \eqref{eq:bethe-gl2-vacuum}
for the $\mathfrak{gl}(2)$ case. For a monodromy \eqref{eq:yi-mono},
which is composed out of the Lax operators \eqref{eq:yi-lax-fund-xi}
with a finite-dimensional $\mathfrak{gl}(n)$ representation of highest
weight $\Xi_i=(\xi_{i}^{(1)},\ldots,\xi_{i}^{(n)})$ at the local
quantum space of the $i$-th site, these eigenvalues are given by
\begin{align}
  \label{eq:bethe-gln-mu}
  \mu_{a}(\spec)
  =
  \prod_{i=1}^{\sites}f_{\Xi_i}(\spec-\inh_i)
  \frac{\spec-\inh_{i}+\xi_{i}^{(a)}}{\spec-\inh_{i}}\,.
\end{align} 
The Bethe roots are encoded into the Q-functions
\begin{align}
  \label{eq:bethe-gln-q}
  Q_k(\spec)=\prod_{i=1}^{\brts_k}(\spec-\brt_{i}^{(k)})\,,
\end{align}
where $k=1,\ldots,n-1$ is the nesting level with $\brts_k$ Bethe roots
$\brt_i^{(k)}$. Obviously, for $n=2$
equation~\eqref{eq:bethe-gln-specbaxter} reduces to the functional
relations \eqref{eq:bethe-gl2-specbaxter}. As one can see from the
Baxter equation for $\mathfrak{gl}(n)$, see e.g.\
\cite{Bazhanov:2010jq}, \eqref{eq:bethe-gln-specbaxter} is compatible
with the fixed eigenvalue in \eqref{eq:bethe-inv-eigen}. More
precisely, each term in the Baxter equation is equal to one.

Interestingly, the functional relations
\eqref{eq:bethe-gln-specbaxter} can also be written in the form
\begin{align}
  \label{eq:bethe-gln-separated-mu}
  1&=\;\prod_{\mathclap{a=1}}^n\;\mu_a(\spec-a+1)\,,\\
  \label{eq:bethe-gln-separated-qk}
  \frac{Q_k(\spec)}{Q_k(\spec+1)}
  &=
  \;\prod_{\mathclap{a=k+1}}^n\;\mu_a(\spec-a+k+1)
\end{align}
for $k=1,\ldots,n-1$. The first equation
\eqref{eq:bethe-gln-separated-mu} does not involve the Bethe roots and
only constrains the representation labels and inhomogeneities of the
monodromy. Each of the remaining equations
\eqref{eq:bethe-gln-separated-qk} only involves the Bethe roots of one
nesting level $k$. The equations \eqref{eq:bethe-gln-separated-mu} and
\eqref{eq:bethe-gln-separated-qk} generalize \eqref{eq:bethe-gl2-ad}
and \eqref{eq:bethe-gl2-qaq}, respectively, to the $\mathfrak{gl}(n)$
case.

\section{Conclusion and outlook}
\label{sec:concl}

In this paper we have proposed a systematic approach to the
construction of Yangian invariants by means of the quantum inverse
scattering method (QISM). Our motivation is two-fold. The first is
mathematical. It appears that the possibility to construct such
invariants for a given algebra and representation in a methodical
fashion has not yet been explored. This is clearly a rich field. The
second is physical. Following \cite{Drummond:2009fd}, Yangian
invariance appears as the hallmark of integrability in the form of a
hidden symmetry of the tree-level scattering problem of planar
$\mathcal{N}=4$ super Yang-Mills theory. This opens the exciting
possibility to directly construct such amplitudes by the techniques of
integrability, such as the various versions of the Bethe ansatz.

The present work is complementary to
\cite{Ferro:2012xw,Ferro:2013dga}, where spectral parameter
deformations of Yangian invariants in general, and of scattering
amplitudes in particular were proposed. Here we can look at the
spectral parameters $\inhdiff$ in \cite{Ferro:2012xw,Ferro:2013dga}
from a slightly different perspective: In the above, these appear as
(differences of) inhomogeneities of some auxiliary spin chain
monodromies. The latter contain in turn a spectral parameter $\spec$,
which is a very useful quantity in the QISM. However, the Yangian
invariants and thus the amplitudes do not depend on this spectral
parameter. We should also point the reader to the recent works
\cite{Chicherin:2013sqa,Chicherin:2013ora}, which bear some
similarities with our approach.

There is a large number of open problems. The first concerns
completing the exploratory study of the $\mathfrak{gl}(n)$ invariants
begun in this paper. Clearly it remains to construct the general
$\sites$-site invariants, and to analyze the freedom in assigning the
inhomogeneities (and thus the spectral parameters in the sense of
\cite{Ferro:2012xw,Ferro:2013dga}). Furthermore, the attentive reader
will have noticed that we essentially derived the $2,3,4$-site
invariants directly from \eqref{mon_ev1}, and subsequently proved that
the Bethe ansatz equations are satisfied. We would really prefer to
proceed in the opposite fashion: First solve the Bethe equations,
which should always be fairly trivial, as all roots are expected to
assemble into exact strings. Then construct the invariants as the
corresponding on-shell Bethe states. Bethe wave functions are in
general very complicated. However, here it should help that the roots
are so simple.

The second open problem concerns the replacement of compact
representations of $\mathfrak{gl}(n)$ by the non-compact
representations of $\mathfrak{gl}(4|4)$ appropriate for the study of
the $\mathcal{N}=4$ scattering amplitudes. The goal would clearly be
to derive the Yangian invariant tree-level amplitudes from an
appropriate ``Bethe ansatz''. We suspect that functional methods will
be important here, as the solution presumably involves considering
infinite sets of Bethe roots. Q-operator methods
\cite{Bazhanov:2010jq,Frassek:2010ga,Frassek:2011aa,Frassek:2012mg}
might be helpful here.

The third and obviously most exciting open problem is the derivation
of higher loop corrections to the tree-level amplitudes from a
Bethe-like ansatz. Here there is a crucial open conceptual problem:
What is the precise fate of Yangian invariance beyond one loop? See
e.g.\ the discussion in \cite{Beisert:2010jq}. The main trouble is that
the infrared divergences of loop amplitudes naively break conformal
symmetry and thus also Yangian symmetry. In
\cite{Ferro:2012xw,Ferro:2013dga} it was proposed that parametric
deformations of loop-level on-shell diagrams might regulate the
divergences. Vexingly, however, exact Yangian invariance seems to
clash with convergence. On the other hand, Yangian invariance appears
to be a key feature of the on-shell diagrammatic approach of
\cite{ArkaniHamed:2012nw}. If it is true that the integrands of the
higher-loop amplitudes may be constructed in a Yangian-invariant way,
these integrands should definitely be constructible by an extension of
the methods proposed in the present paper.

\section*{Acknowledgments}
We thank Zoltan Bajnok, Ludvig Faddeev, Frank Göhmann, Vladimir Mitev,
Nicolai Reshetikhin, and especially Livia Ferro, Tomek {\L}ukowski,
Carlo Meneghelli and Jan Plefka for very useful discussions. N.K.\
thanks the organizers of the XXIst International Conference on Integrable
Systems and Quantum Symmetries (ISQS-21) for the possibility to
present the above results in June 2013 in Prague.  R.F.\ thanks the
organizers of IGST 2013 for the invitation to present our work at this
conference in Utrecht in August 2013. This research is supported in
part by the SFB 647 \emph{``Raum-Zeit-Materie. Analytische und
  Geometrische Strukturen''} and the Marie Curie network GATIS
(\texttt{\href{http://gatis.desy.eu}{gatis.desy.eu}}) of the European
Union’s Seventh Framework Programme FP7/2007-2013/ under REA Grant
Agreement No 317089. N.K.\ is supported by a
\emph{Promotionsstipendium} of the \emph{Studienstiftung des Deutschen
  Volkes}, and receives partial support by the GK 1504 \emph{``Masse,
  Spektrum, Symmetrie''}.  Two of us (N.K.\ and M.S.) thank the Kavli
IPMU for hospitality while working on parts of the manuscript, and
acknowledge the support of the Marie Curie International Research
Staff Exchange Network UNIFY.

\appendix

\section{Some oscillator algebra representations}
\label{sec:bargmann}

In this appendix we substantiate the representations of the oscillator
algebra introduced formally in \eqref{eq:amp-osc-delta}. It is easily
seen that this algebra can be represented on holomorphic functions of
one complex variable. The creation operator is realized as
multiplication by this variable and the annihilation operator
corresponds to differentiation. This was made precise by Bargmann
\cite{Bargmann:1961gm} who provided an inner product guaranteeing that
both operators are Hermitian conjugates of each other. We review his
construction in section~\ref{sec:bargmann-barg}.

In section~\ref{sec:amp} we reformulated the Yangian invariants of
section~\ref{sec:osc} in a way that is reminiscent of planar
$\mathcal{N}=4$ super Yang-Mills scattering amplitudes. For this we
also used a representation of the oscillator algebra which is
``conjugate'' to that of Bargmann in the sense that the role of the
operators is exchanged: The creation operator acts by differentiation
and the annihilation operator as multiplication. Furthermore, the Fock
vacuum is realized as a delta function of a complex argument.

Such a representation already appeared previously in rather different
contexts, see e.g. \cite{Fan:1984,Ashtekar:1991vz,Ribeiro:2008}, and
it is even traced back in \cite{Fan:1994} to work by Dirac
\cite{Dirac:1943}. In contrast to these references, we explain this
conjugate Bargmann representation in
section~\ref{sec:bargmann-conj-barg} completely within the Bargmann
framework.

\subsection{Bargmann representation}
\label{sec:bargmann-barg}

We start by reviewing the \emph{Bargmann representation}
\cite{Bargmann:1961gm}, which is also called \emph{holomorphic
  representation}, see e.g.\ \cite{Takhtajan:2008,ZinnJustin:2005} for
recent expositions. The oscillator algebra, the Hermiticity condition
and the characterization of the Fock vacuum,
\begin{align}
  \label{eq:bargmann-osca}
  [\osca,\bar\osca]=1\,,
  \quad
  \bar\osca^\dagger=\osca\,,
  \quad
  \osca|0\rangle=0\,,
\end{align}
are realized in terms of a complex variable $\mathcal{W}$ by
\begin{align}
  \label{eq:bargmann-holo}
  \bar\osca\mathrel{\widehat{=}}\mathcal{W}\,,
  \quad
  \osca\mathrel{\widehat{=}}\partial_\mathcal{W}\,,
  \quad
  |0\rangle\mathrel{\widehat{=}}1\,.
\end{align}
In this representation a state translates into a holomorphic function,
\begin{align}
  \label{eq:bargmann-state}
  |\Sigma\rangle=\Sigma(\bar\osca)|0\rangle
  \mathrel{\widehat{=}}
  \Sigma(\mathcal{W})\,.
\end{align}
The inner product of two states is defined as
\begin{align}
  \label{eq:bargmann-holo-inner}
  \langle \Theta|\Sigma\rangle
  =
  \int_{\mathbb{C}}
  \frac{\D\!\overline{\mathcal{W}}\D\!\mathcal{W}}{2\pi i}\,
  e^{-\mathcal{W}\overline{\mathcal{W}}}\,
  \overline{\Theta(\mathcal{W})}
  \Sigma(\mathcal{W})\,,
\end{align}
where the integral is to be understood as a two-dimensional real
integral with $\D\!\overline{\mathcal{W}}\D\!\mathcal{W}=2
i\D\!\text{Re}\mathcal{W}\D\!\text{Im}\mathcal{W}$. Because of the
exponential function in the measure, the creation and annihilation
operators are indeed related by Hermitian conjugation, i.e.\
$\mathcal{W}^\dagger=\partial_{\mathcal{W}}$. This is easily verified
using partial integration. States with finite norm with respect to the
inner product form a Hilbert space with an orthonormal basis
\begin{align}
  \label{eq:bargmann-basis}
  |k\rangle=\frac{\bar\osca^k|0\rangle}{\sqrt{k!}}
  \mathrel{\widehat{=}}
  \frac{\mathcal{W}^k}{\sqrt{k!}}\,.
\end{align}

Likewise, one defines an \emph{antiholomorphic representation}, where
a family of oscillators
\begin{align}
  \label{eq:bargmann-oscb}
  [\oscb,\bar\oscb]=1\,,
  \quad
  \bar\oscb^\dagger=\oscb\,,
  \quad
  \oscb|0\rangle=0\,,
\end{align}
is realized in terms of a complex
conjugate variable $\overline{\mathcal{W}}$ as
\begin{align}
  \label{eq:bargmann-antiholo}
  \bar\oscb\mathrel{\widehat{=}}\overline{\mathcal{W}}\,,
  \quad
  \oscb\mathrel{\widehat{=}}\partial_{\overline{\mathcal{W}}}\,,
  \quad
  |0\rangle\mathrel{\widehat{=}}1\,.
\end{align}
Here the inner product is
\begin{align}
  \label{eq:bargmann-antiholo-inner}
  \langle \Theta|\Sigma\rangle
  =
  \int_{\mathbb{C}}\frac{\D\!\overline{\mathcal{W}}\D\!\mathcal{W}}{2\pi i}\,
  e^{-\mathcal{W}\overline{\mathcal{W}}}\,
  \overline{\Theta(\overline{\mathcal{W}})}
  \Sigma(\overline{\mathcal{W}})\,.
\end{align}

Let us employ both representations \eqref{eq:bargmann-holo} and
\eqref{eq:bargmann-antiholo} in case of a simple example. Consider the
operator
\begin{align}
  \label{eq:bargmann-example-op}
  \mathcal{O}_\Psi
  =
  \sum_{k,l=0}^\infty
  \mathcal{O}_{kl}(\bar\osca^2)^k|0\rangle\langle 0|(\oscb^1)^l
\end{align}
mapping from a Fock space $V_1$ with oscillators $\bar\oscb^1$,
$\oscb^1$ into $V_2$ with $\bar\osca^2$, $\osca^2$. It should be
thought of as a simple analogue of the Yangian invariants discussed in
section~\ref{sec:osc}, see e.g.\ \eqref{eq:osc-opsi21}. Of course, for
generic coefficients $\mathcal{O}_{kl}$ and with only one family of
oscillators per space, it is not an actual invariant of the Yangian
$\mathcal{Y}(\mathfrak{gl}(n))$. We use the representation
\eqref{eq:bargmann-holo} for the oscillators in $V_2$ and
\eqref{eq:bargmann-antiholo} for those in $V_1$. Then the action of
the operator on a ``test state'' $|f\rangle\in V_1$ becomes
\begin{align}
  \label{eq:bargmann-example-action}
  \mathcal{O}_\Psi|f\rangle
  =
  \int_{\mathbb{C}}\frac{\D\!\overline{\mathcal{W}}^1\D\!\mathcal{W}^1}{2\pi i}\,
  e^{-\mathcal{W}^1\overline{\mathcal{W}}^1}
  \mathcal{O}_\Psi(\mathcal{W}^2,\mathcal{W}^1)
  f(\overline{\mathcal{W}}^1)
\end{align}
with the kernel
\begin{align}
  \label{eq:bargmann-example-action-kernel}
  \mathcal{O}_\Psi(\mathcal{W}^2,\mathcal{W}^1)
  =
  \sum_{k,l=0}^\infty\mathcal{O}_{kl}(\mathcal{W}^2)^k(\mathcal{W}^1)^l\,.
\end{align}
In this article we mostly work with the vector version
$|\Psi\rangle=\mathcal{O}_\Psi^{\dagger_1}$ of operators like
\eqref{eq:bargmann-example-op}, see e.g.\
\eqref{eq:osc-psi21}. Written in terms of this vector,
\eqref{eq:bargmann-example-action} turns into
\begin{align}
  \label{eq:bargmann-example-testfct-opvec}
  \mathcal{O}_\Psi|f\rangle
  =
  \overline{\langle f|\Psi\rangle}^{\,_1}\,.
\end{align}
Note that the inner product is only in the space $V_1$ and not in
$V_2$.  The complex conjugation affects also only $V_1$.

\subsection{Conjugate Bargmann representation}
\label{sec:bargmann-conj-barg}

Motivated by \eqref{eq:bargmann-example-testfct-opvec} of this example
we employ the representation \eqref{eq:bargmann-antiholo} to study the
inner product of two states $|\Sigma\rangle$ and $|f\rangle$, where
the latter will play the role of a test state. Choosing
$|\Sigma\rangle=|0\rangle$ to be the Fock vacuum we write
\begin{align}
  \label{eq:bargmann-conj-delta}
  \langle f|0\rangle=\overline{f(0)}
  =
  \int_{\mathbb{C}}\frac{\D\!\overline{\mathcal{W}}\D\!\mathcal{W}}{2\pi i}\,
  \overline{f}(\mathcal{W})\delta(\mathcal{W})
  \quad
  \text{with}
  \quad
  \delta(\mathcal{W})
  :=
  e^{-\mathcal{W}\overline{\mathcal{W}}}\,.
\end{align}
Here we interpreted the exponential function of the measure in
\eqref{eq:bargmann-antiholo-inner} as a ``delta function of a complex
argument'', cf.\ \cite{ZinnJustin:2005}. This delta function does
\emph{not} coincide with the reproducing kernel, which usually plays
the role of a delta function in the Bargmann representation. However,
this interpretation of the exponential function is essential for our
purpose, see below. For a general state $|\Sigma\rangle$ we obtain
\begin{align}
  \label{eq:bargmann-conj-genstate}
  \langle f|\Sigma\rangle
  =
  \int_{\mathbb{C}}\frac{\D\!\overline{\mathcal{W}}\D\!\mathcal{W}}{2\pi i}\,
  \overline{f}(\mathcal{W})\hat{\Sigma}(\mathcal{W})\,,
  \quad
  \text{with}
  \quad
  \hat{\Sigma}(\mathcal{W})
  :=
  \Sigma(-\partial_\mathcal{W})\delta(\mathcal{W})\,.
\end{align}
The action of the oscillators \eqref{eq:bargmann-antiholo} translates
into
\begin{align}
  \label{eq:bargmann-conj-action}
  \begin{aligned}
    \langle f|\bar\oscb|\Sigma\rangle
    &=
    \int_{\mathbb{C}}\frac{\D\!\overline{\mathcal{W}}\D\!\mathcal{W}}{2\pi i}\,
    \overline{f}(\mathcal{W})(-\partial_\mathcal{W})\hat{\Sigma}(\mathcal{W})\,,\\
    \langle f|\oscb|\Sigma\rangle
    &=
    \int_{\mathbb{C}}\frac{\D\!\overline{\mathcal{W}}\D\!\mathcal{W}}{2\pi i}\,
    \overline{f}(\mathcal{W})\mathcal{W}\hat{\Sigma}(\mathcal{W})\,.
  \end{aligned}
\end{align}
Also the inner product \eqref{eq:bargmann-antiholo-inner} can be
expressed in terms of $\hat{\Sigma}(\mathcal{W})$ and
$\hat{\Theta}(\mathcal{W})$,\footnote{As similar inner product was
  introduced in \cite{Ashtekar:1991vz}.}
\begin{align}
  \label{eq:bargmann-conj-inner}
  \langle\Theta|\Sigma\rangle
  =
  \sum_{n=0}^\infty
  \overline{\langle n|\Theta\rangle}
  \langle n|\Sigma\rangle
  =
  \int_{\mathbb{C}}\frac{\D\!\overline{\mathcal{Y}}\D\!\mathcal{Y}}{2\pi i}
  \int_{\mathbb{C}}\frac{\D\!\overline{\mathcal{W}}\D\!\mathcal{W}}{2\pi i}\,
  e^{+\overline{\mathcal{Y}}\mathcal{W}}\,
  \overline{\hat{\Theta}(\mathcal{Y})}\hat{\Sigma}(\mathcal{W})\,.
\end{align}
With this one easily verifies
$(-\partial_{\mathcal{W}})^\dagger=\mathcal{W}$ after identifying
$\mathcal{W}\leftrightarrow\mathcal{Y}$. Notice the positive sign in
the exponential function in \eqref{eq:bargmann-conj-inner}, whereas it
is negative in the measure of \eqref{eq:bargmann-antiholo-inner}. In
particular, this leads to a finite norm of
$\hat{\Sigma}(\mathcal{W})=\delta(\mathcal{W})$, while that of
$\hat{\Sigma}(\mathcal{W})=1$ diverges.

In conclusion, \eqref{eq:bargmann-conj-delta} and
\eqref{eq:bargmann-conj-action} together with
\eqref{eq:bargmann-conj-inner} constitute a realization of the
oscillators \eqref{eq:bargmann-oscb} which we call \emph{conjugate
  Bargmann representation}:
\begin{align}
  \label{eq:bargmann-conj}
  \bar\oscb\mathrel{\widehat{=}}-\partial_{\mathcal{W}}\,,
  \quad
  \oscb\mathrel{\widehat{=}}\mathcal{W}\,,
  \quad
  |0\rangle\mathrel{\widehat{=}}\delta(\mathcal{W})\,.
\end{align}
The fact that we can realize the creation operator as differentiation
and the annihilation operator as multiplication (and not the other way
around) depends crucially on the reinterpretation of the exponential
factor in the measure as a delta function, cf.\
\eqref{eq:bargmann-conj-delta}. In this sense the representation
\eqref{eq:bargmann-conj} can also be thought of as
\eqref{eq:bargmann-antiholo} ``in disguise''. With
\eqref{eq:bargmann-conj} a general state, cf.\
\eqref{eq:bargmann-conj-genstate}, and the orthonormal states from
above respectively take the form
\begin{align}
  \label{eq:bargmann-conj-states-basis}
  |\Sigma\rangle
  \mathrel{\widehat{=}}
  \hat{\Sigma}(\mathcal{W})=
  \Sigma(-\partial_\mathcal{W})\delta(\mathcal{W})\,,
  \quad
  |k\rangle
  \mathrel{\widehat{=}}
  \frac{(-\partial_{\mathcal{W}})^k\delta(\mathcal{W})}{\sqrt{k!}}\,.
\end{align}
Given a state $|\Sigma\rangle$, the representative
$\Sigma(\mathcal{W})$ in the Bargmann representation
\eqref{eq:bargmann-holo} and $\hat{\Sigma}(\mathcal{W})$ in its
conjugate \eqref{eq:bargmann-conj} are formally related by a
complex generalization of the Fourier transform.\footnote{A definition
  of a complex Fourier transform can be found e.g.\ in
  \cite{Dubinskii:1991}.}

We return to the example from the end of section
\ref{sec:bargmann-barg}. Realizing the oscillators in space $V_1$ by
\eqref{eq:bargmann-conj} and those in $V_2$ by
\eqref{eq:bargmann-holo}, the vector version of the operator
\eqref{eq:bargmann-example-op} becomes
\begin{align}
  \label{eq:bargmann-example-vectorkernel}
  |\Psi\rangle=\mathcal{O}_\Psi^{\dagger_1}
  =
  \sum_{k,l=0}^\infty
  \mathcal{O}_{kl}(\bar\osca^2)^k
  (\bar\oscb^1)^l|0\rangle
  \mathrel{\widehat{=}}
  \mathcal{O}_\Psi(\mathcal{W}^2,-\partial_{\mathcal{W}^1})
  \delta(\mathcal{W}^1)\,.
\end{align}
This example illustrates the use of these oscillator algebra
representations for the Yangian invariants in
section~\ref{sec:amp}. There the Bargmann representation
\eqref{eq:bargmann-holo} is employed for oscillators at sites carrying
a totally symmetric representation $\s$ of $\mathfrak{gl}(n)$ and the
conjugate Bargmann representation \eqref{eq:bargmann-conj} appears at
sites with a conjugate $\bs$ of a totally symmetric $\mathfrak{gl}(n)$
representation. Equation~\eqref{eq:bargmann-example-vectorkernel} is
also an example of how our expressions for Yangian invariants in terms
of delta functions are related to the kernels, cf.\
\eqref{eq:bargmann-example-action-kernel}, of the corresponding
intertwiners.

\section{Derivation of two-site invariant}
\label{sec:twositeproof}

Here we present, by way of example, the derivation of
\eqref{eq:bethe-gl2-sol-line-inv}, i.e.~the construction of the
two-site invariant $|\Psi_{2,1}\rangle$ for a spin $\frac{s}{2}$
representation of $\mathfrak{gl}(2)$ using the algebraic Bethe ansatz,
starting from the reference state $|\Omega\rangle$. Let
$s:=s_1=s_2$. The ``generalized lowering operators'' are for two sites
\begin{align}
  \label{eq:Bop}
  B(u)=
  \left(1-\frac{1}{u-v_1}\,\bar \oscb_1 \oscb_1\right) 
  \frac{1}{u-v_2}\,\bar \osca_2 \osca_1
  -\left(1+\frac{1}{u-v_2}\,\bar \osca_2 \osca_2\right) 
  \frac{1}{u-v_1}\,\bar \oscb_1 \oscb_2\,.
\end{align}
For convenience we dropped the (in this case) redundant upper site
indices.  Using the expressions in section
\ref{sec:bethe-gl2-sol-line} for the Bethe roots and inhomogeneities,
one finds for $k=1, \ldots, s$
\begin{align}
  \label{eq:Bopk}
  B(u_k)=\alpha_k+\beta_k\,,
\end{align}
where we have defined the operators
\begin{align}
  \label{eq:alphabeta}
  \alpha_k:=-\left(1-\frac{1}{s+1-k}\,\bar \oscb_1 \oscb_1\right) 
  \frac{1}{k}\,\bar \osca_2 \osca_1\,,\quad
  \beta_k=-\left(1-\frac{1}{k}\,\bar \osca_2 \osca_2\right) 
  \frac{1}{s+1-k}\,\bar \oscb_1 \oscb_2\,.
\end{align}
We thus need to calculate
\begin{align}
  \label{eq:bruteforce1}
  |\Psi_{2,1}\rangle=\prod_{k=1}^sB(u_k)\,|\Omega\rangle
  =\prod_{k=1}^s\left(\alpha_k+\beta_k\right)\, 
  (\bar \oscb_2)^s (\bar \osca_1)^s\, |0\rangle\,.
\end{align}
Let us expand the product $\prod_{k=1}^s\,
\left(\alpha_k+\beta_k\right)$, denoting by $m$ the number of times
$\beta_k$ appears:
\begin{align}
  \label{eq:bruteforce2}
  \sum_{m=0}^s\;\sum_{1 \leq j_1 < \ldots < j_m \leq s}
  \!\!\!\!\!\!\!
  \alpha_1\cdots\alpha_{j_1-1}
  \,\beta_{j_1}\,
  \alpha_{j_1+1}\cdots\alpha_{j_2-1}
  \,\beta_{j_2}\,
  \alpha_{j_2+1}
  \cdots
  \alpha_{j_m-1}
  \,\beta_{j_m}\,
  \alpha_{j_m+1}\cdots\alpha_s\,.
\end{align}
For fixed $m$ and insertion positions $j_1, \ldots, j_m$ the number
operators $\bar \oscb_1 \oscb_1$ and $\bar \osca_2 \osca_2$ in
\eqref{eq:alphabeta} take fixed values which are easily read off. Also
taking into account the combinatorial factors when acting with $\bar
\osca_2 \osca_1$ and $\bar \oscb_1 \oscb_2$ on the reference state,
and canceling all common factors, one easily obtains, up to a trivial
factor of $(-1)^s$, for $|\Psi_{2,1}\rangle$
\begin{align}
  \label{eq:bruteforce3}
  \sum_{m=0}^s\;\sum_{1 \leq j_1 < \ldots < j_m \leq s}
  \!\!\!\!\!\!\!\!
  \frac{\prod_{k=1}^m(2 j_k-s+m-k)}{m!}
  (\bar \oscb_1 \bar \osca_1)^m (\bar \oscb_2 \bar \osca_2)^{s-m} |0\rangle\,.
\end{align}
By a curious identity this then simplifies to
\begin{align}
  \label{eq:bruteforce4}
  |\Psi_{2,1}\rangle=
  (-1)^s\,\sum_{m=0}^s \binom{s}{m}
  (\bar \oscb_1 \bar \osca_1)^m (\bar \oscb_2 \bar \osca_2)^{s-m} |0\rangle
  =(-1)^s\,\left(\bar \oscb_1 \bar \osca_1 + 
    \bar \oscb_2 \bar \osca_2\right)^s |0\rangle\,,
\end{align}
and \eqref{eq:bethe-gl2-sol-line-inv} is proven. Incidentally, the
just mentioned curious identity means that the operator in
\eqref{eq:bruteforce2} may be replaced by the ordered expression
\begin{align}
  \label{eq:curious}
  \sum_{m=0}^s\,\alpha_1 \cdots \alpha_{s-m} \beta_{s-m+1} \cdots \beta_s
\end{align}
when acting on the reference state $|\Omega\rangle$. All other orders
of insertions of the $\beta_k$ operators either yield zero or cancel
against another ordering.

\bibliography{literature}{}
\bibliographystyle{utphys}

\end{document}